\documentclass[times,twocolumn,5p,final]{elsarticle}
\usepackage{framed,multirow}
\journal{Computer Aided Geometric Design}

\usepackage{amssymb}
\usepackage{latexsym}

\usepackage[colorlinks]{hyperref}

\usepackage[switch,pagewise]{lineno} %Required by command \linenumbers below
\usepackage{tabularx}
\usepackage{amsmath,amssymb,amsfonts,bm,fixmath}
\usepackage{subcaption}
\graphicspath{ {./figures/}}

\newif\iflowres
\newcommand{\norm}[1]{\left\lVert\,#1\,\right\rVert}

\newcommand{\set}[1]{{\mathcal{#1}}}
\newcommand{\setsize}[1]{{n_{\mathcal{#1}}}}

\newcommand{\etal}{{et al.}}
\newcommand{\matlab}{\textsc{Matlab}}
\newcommand{\abaqus}{\textsc{Abaqus}}
\newcommand{\blender}{\textsc{Blender}}
\DeclareMathOperator{\subto}{s.t.}

\DeclareMathOperator{\atantwo}{atan2}

\begin{document}
%\verso{Preprint submitted for review}
%\linenumbers

\begin{frontmatter}

\title{Pose to Seat: Automated Design of Body-Supporting Surfaces}

\author[1]{Kurt {Leimer}\corref{cor1}}
\cortext[cor1]{Corresponding author.}
\emailauthor{kurt.leimer@tuwien.ac.at}{K. Leimer}
%\ead{example@email.com}
\author[1]{Andreas {Winkler}}
\author[1]{Stefan {Ohrhallinger}}
\author[1,2]{Przemyslaw {Musialski}}

\address[1]{Technische Universit\"at Wien (TU Wien)}
\address[2]{New Jersey Institute of Technology (NJIT)}

%\received{1 February 2017}
%\received{20 March 2020}
%%%% Do not use the below for submitted manuscripts
%\finalform{28 March 2017}
%\accepted{2 April 2017}
%\availableonline{21 March 2207}
%\communicated{S. Sarkar}

\begin{abstract}
%%%
The design of functional seating furniture is a complicated process which often requires extensive manual design effort and empirical evaluation. We propose a computational design framework for pose-driven automated generation of body-supports which are optimized for comfort of sitting. 
Given a human body in a specified pose as input, our method computes an approximate pressure distribution that also takes frictional forces and body torques into consideration which serves as an objective measure of comfort. 
Utilizing this information to find out where the body needs to be supported in order to maintain comfort of sitting, our algorithm can create a supporting mesh suited for a person in that specific pose. This is done in an automated fitting process, using a template model capable of supporting a large variety of sitting poses. The results can be used directly or can be considered as a starting point for further interactive design. 
%%%%
\end{abstract}

\end{frontmatter}

\section{Introduction}

\begin{figure*}[t]
	\centering
	\iflowres
	\includegraphics[width=0.19\textwidth, trim=0.4cm 1.4cm 0cm 1.4cm, clip]{./renders/teaser/f11_empty}
	\includegraphics[width=0.19\textwidth, trim=0.4cm 1.4cm 0cm 1.4cm, clip]{./renders/teaser/f11_empty}
	\includegraphics[width=0.19\textwidth, trim=0.4cm 1.4cm 0cm 1.4cm, clip]{./renders/teaser/f11_empty}
	\includegraphics[width=0.19\textwidth, trim=0.4cm 1.4cm 0cm 1.4cm, clip]{./renders/teaser/f11_empty}
	\includegraphics[width=0.19\textwidth, trim=0.4cm 1.4cm 0cm 1.4cm, clip]{./renders/teaser/f11_empty}
	\else
	\includegraphics[width=0.19\textwidth, trim=0.4cm 1.4cm 0cm 1.4cm, clip]{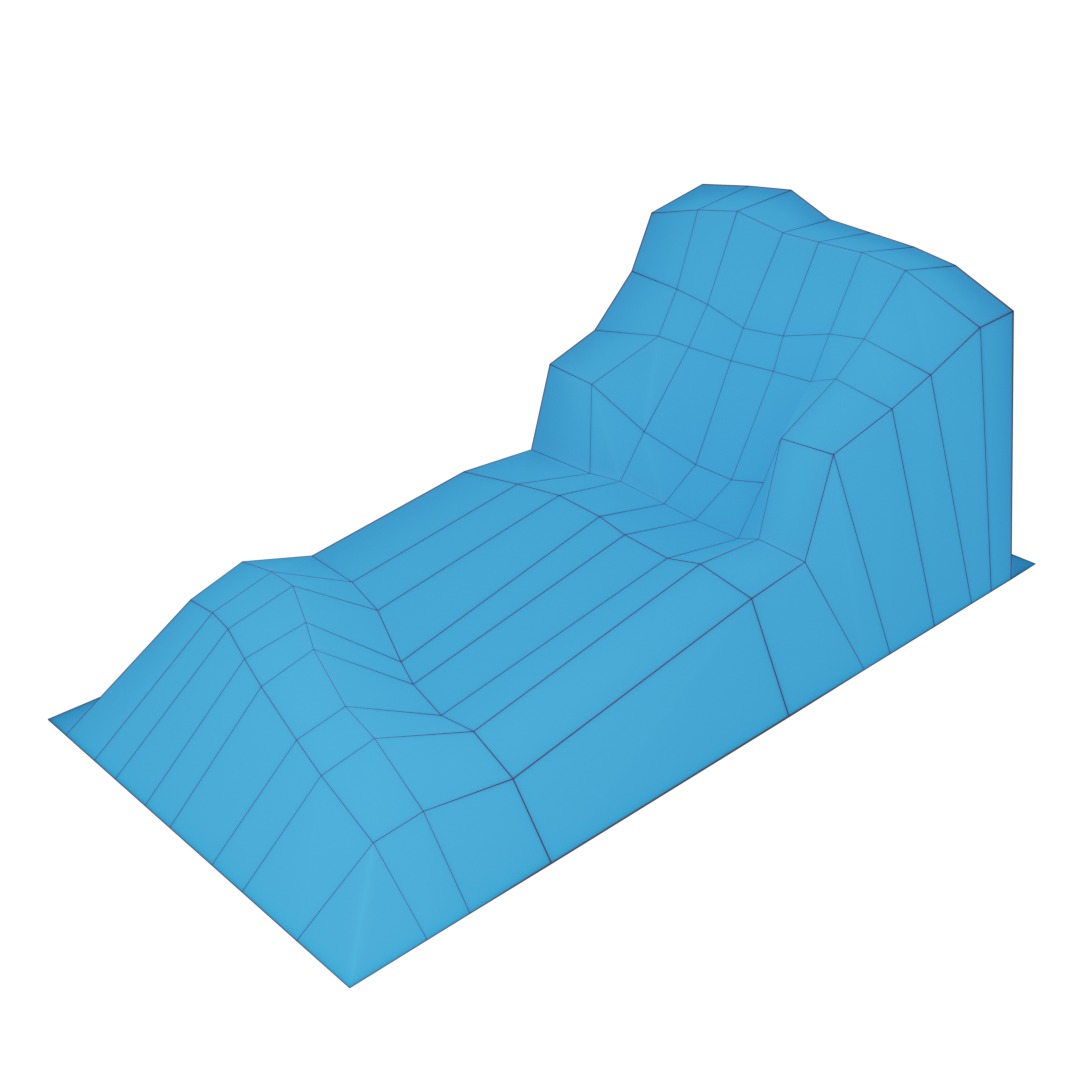}
	\includegraphics[width=0.19\textwidth, trim=0.8cm 2.8cm 0cm 2.8cm, clip]{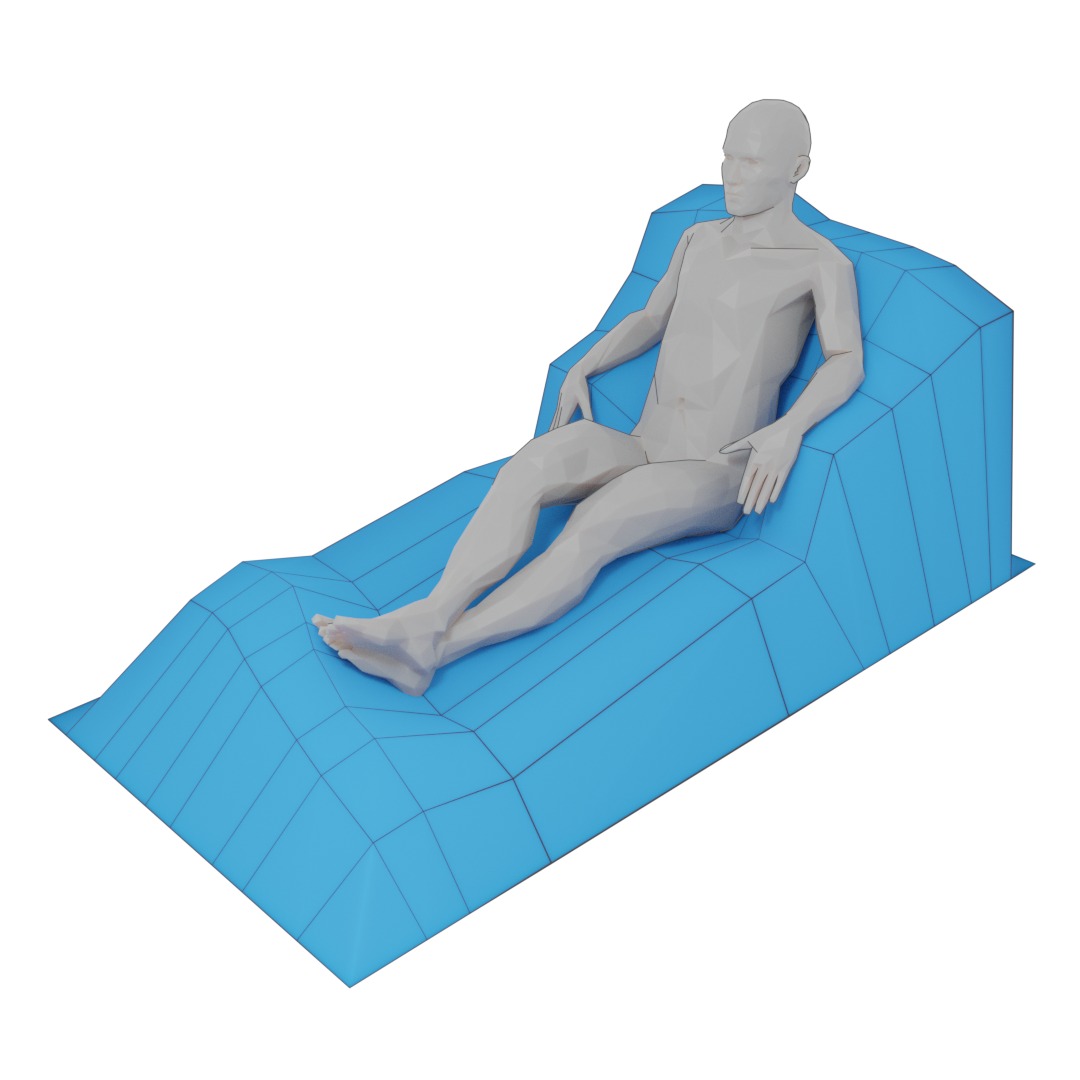}
	\includegraphics[width=0.19\textwidth, trim=0.4cm 1.4cm 0cm 1.4cm, clip]{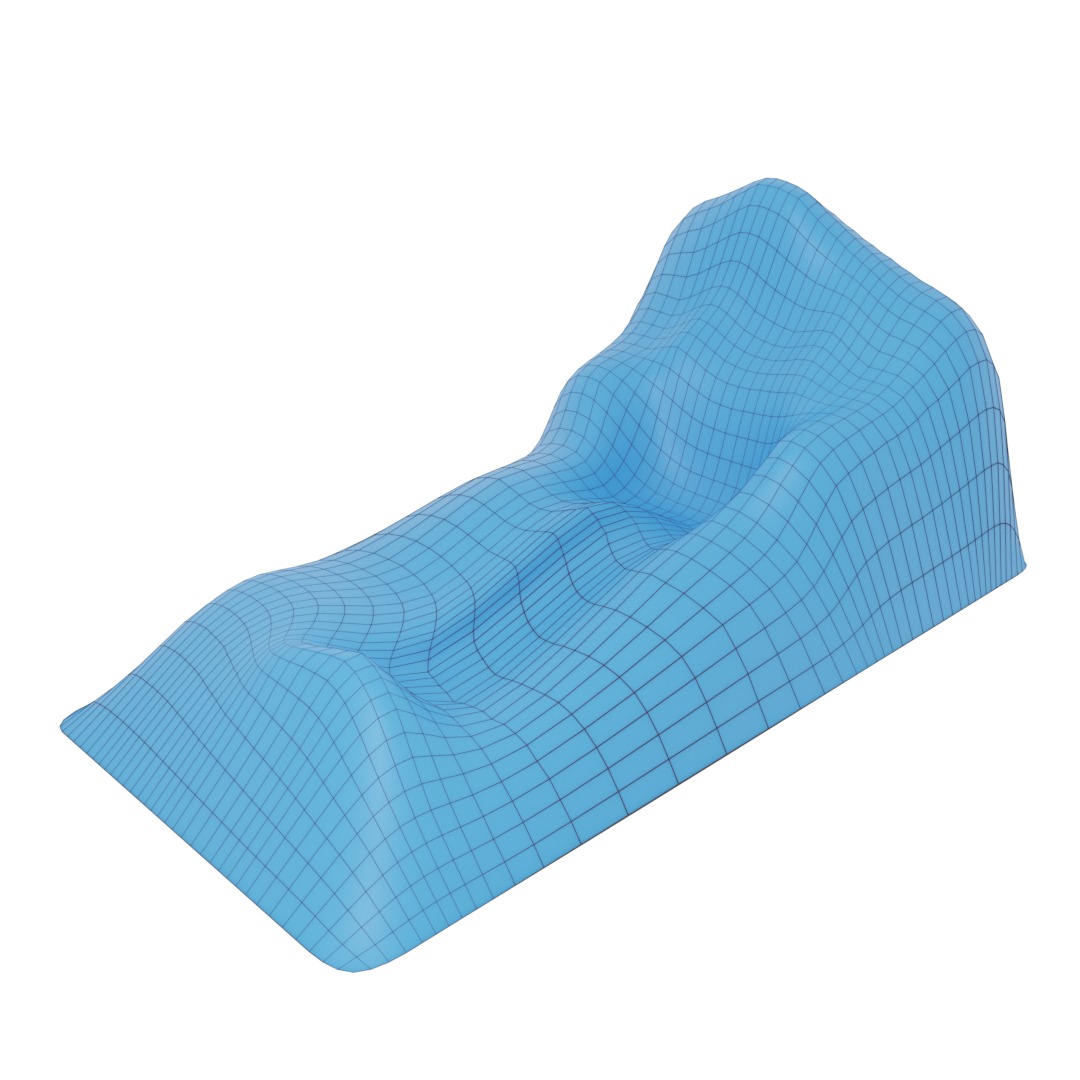}
	\includegraphics[width=0.19\textwidth, trim=0.8cm 2.8cm 0cm 2.8cm, clip]{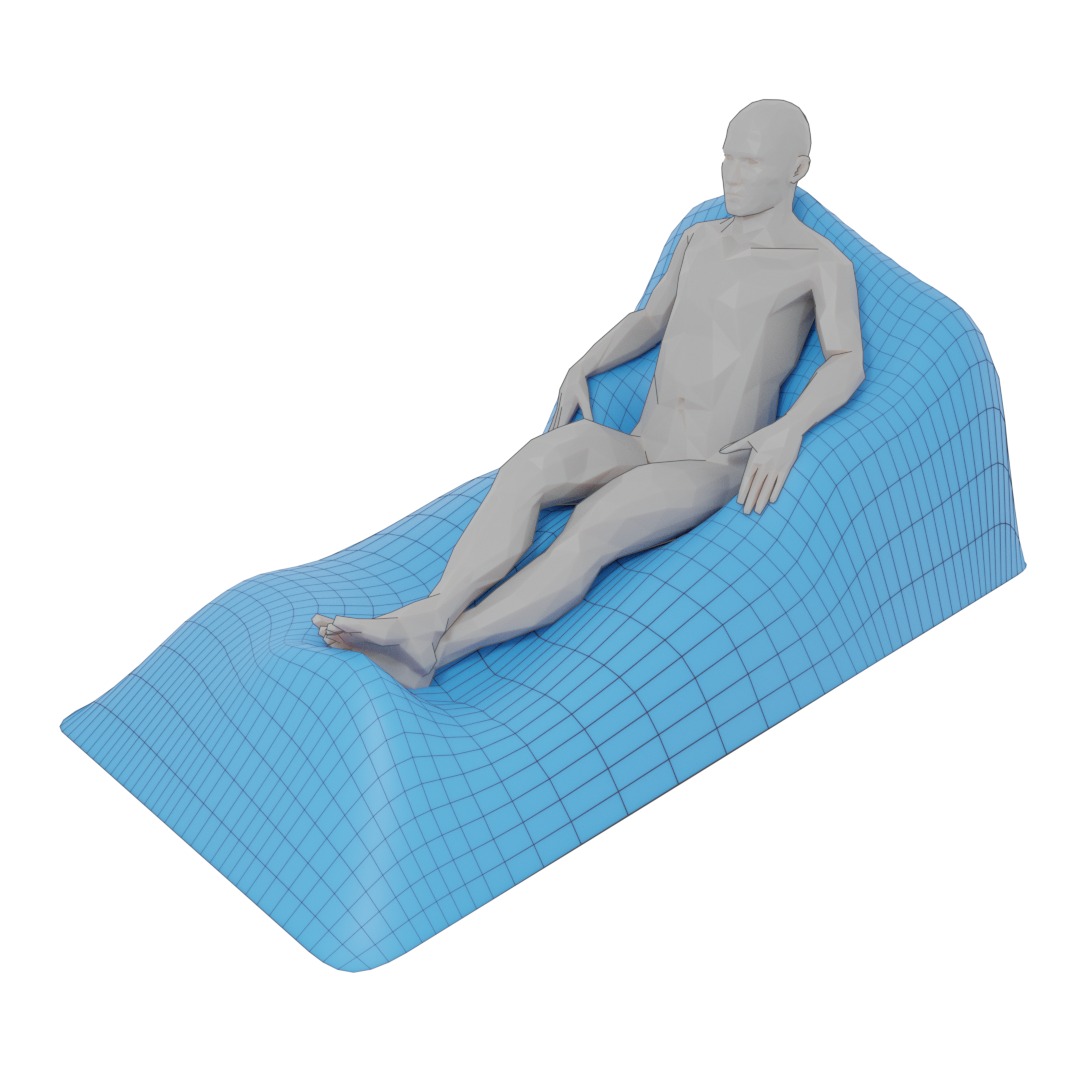}	
	\includegraphics[width=0.19\textwidth, trim=0.4cm 1.4cm 0cm 1.4cm, clip]{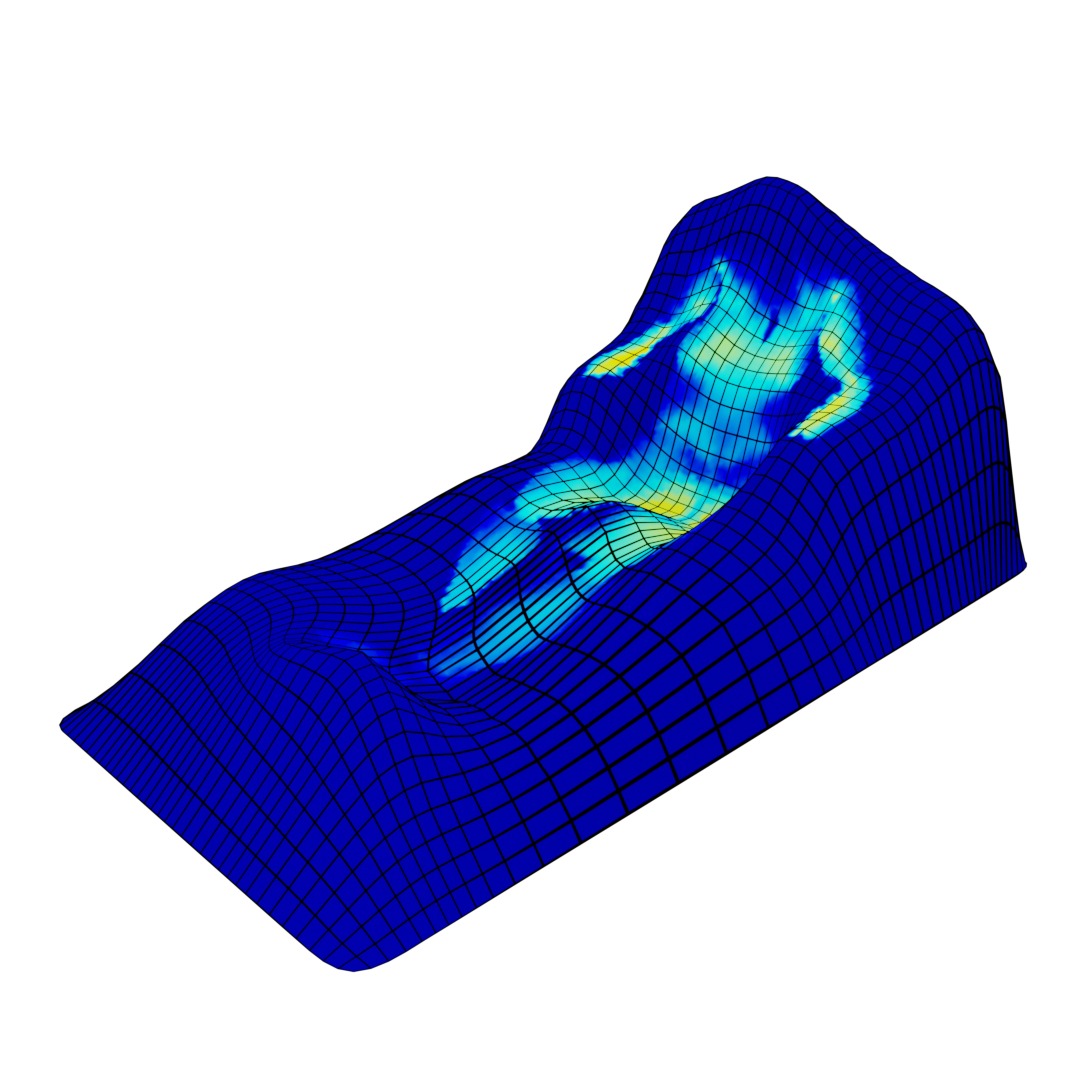}
	\fi
	\caption{Left: a control mesh created by our automated body support-template fitting algorithm. Note that the mesh is generated to in order to optimally support the body in a given pose.  Right: the surface fitting algorithm of Leimer \etal~\cite{leimer2018sit} applied to the generated control mesh, again using the body model as guidance. Finally, the pressure distribution on the body indicated on the surface.  }
	\label{fig:teaser}
\end{figure*}

Todays furniture industry can be roughly separated into two main groups: on the one side there is high-end customization with expensive design and functionality aspects in the foreground, on the other,  
standardized mass production aiming at efficiently meeting the needs of the mainstream. 
While other home industries have already recognized the potential of mass customization arising due to the the growth of digital fabrication capabilities, this field is still barely covered in furniture design.

One of the reasons is that the customization task is particularly challenging in terms of furniture design, especially seating furniture. 
Indeed, traditional design of custom seating furniture is a costly process. It usually involves a number of iterations, where physical prototypes need to be produced in one-to-one scale in order to determine how functional they actually are. In practice, it is difficult to predict how comfortable a final product will eventually be if used by humans.

Nowadays, products are typically designed using advanced CAD software with their aesthetic, structural, ergonomic, and economic aspects in mind. However, most of these aspects are left to the judgments of the designer and her or his experience and expertise. For instance, the ergonomics of seats has been researched for a long time  \cite{zheng2016ergonomics,brintrup2008ergonomic} and there exist sets of rules and guidelines which can be applied during the design process. 
Nonetheless, the prototypes of products still require further testing to determine if they meet the desired criteria, like comfort of sitting.

Another option is to perform physical simulations, which again is more involved and interrupts the pure design flow when working with CAD software. Additionally, it also poses further technical requirements on the design team and thus  increases the costs. In practice, the product development pipelines in the mid-level industry are still very awkward and require manufacturing of many physical prototypes.

%\przem{move to discussion?} 
%However, there exists a trade off between general usability and individualization. On the two extremes of this spectrum one may find a simple flat board that anyone can sit on and the seat of a race car cockpit that is custom made to perfectly fit the driver. While the former is significantly lacking in comfort, it is easy to mass produce and a general solution that works for any individual. The latter may provide optimal comfort to its intended user, but the level of comfort is reduced significantly when used by someone else, making it unsuitable for mass production. Because of this, commercial furniture is usually designed with general usability and mass production in mind, offering less comfort for the individual. Our design approach based on poses and body types aims for increased individualization compared to commercial furniture while still maintaining general usability.
%

In order to address these limitations, we propose a computational design framework which aims at automated design of body-supports which are ensured to be comfortable---at least to the extents which can be measured quantitatively by the pressure distribution on the body.  
This quantity indicates where the body should be supported in order to easily hold a particular pose, which is one of the measures of comfortable sitting as defined in the ergonomics literature \cite{lueder1983seat,de2003sitting}. 

Our method is meant for computational design of personalized furniture and can be used by both inexperienced users and by professionals to quickly create unique designs. The results can be used directly or can be interactively modified in order to explore potential design variations. The advantage of the designs is that they automatically account for the human-body given in desired pose, and maximize its support, which is one of the objective measures of comfort \cite{de2003sitting}.

Our contributions can be divided into main components: 
%The central goal of this paper is to provide an improvement into that pipeline by introduction support at two different stages: 
\begin{itemize}
	\item We propose a novel computational human-body model for the approximation of the comfort of sitting in a given pose, based on both pressure distribution of the body on the seat and on the moments (torques) acting on the limbs of the body. Our model is driven by physical assumptions and extends previously proposed models. Nonetheless, it is simplified to a system of linear equations in order to account for interactive rates. 
	%A suitable representation of a human body shape and pose is required in order to create personalized seating surfaces. 
	%Details of this model are described in Sec. \ref{sec:compmodel}. 
	
	\item Moreover, we propose a generic body-support template which delivers a control mesh that can be used for further interactive design and refinement. Our model is capable of supporting a large variety of poses and body shapes, and we demonstrate its applicability by using it to derive control meshes for subdivision surfaces which fulfill the functional requirements. 
	%Our template seating mesh can automatically support individual parts of a person's body.
	%A RANSAC based surface fitting algorithm is developed to ensure an optimal fit between the surface model and the body shape.
	%Additionally, in order to maintain globally appealing aesthetics, an optimization algorithm is applied that is capable of improving the model's visual quality while preserving its functionality as supporting surface. Details are described in Sec. \ref{sec:template_model}. 	
\end{itemize}

In the following section we review related work and in Section \ref{sec:overview} we provide an overview of the components of the framework. In Section \ref{sec:compmodel} we provide the details of the computational body-model, in Section \ref{sec:template_model} the details of the support-templates, and in Section \ref{sec:results} we present and evaluate our results. Finally, we discuss and conclude the work in Sections \ref{sec:limitations} and \ref{sec:conclusions}.

\begin{figure*}[t]
	\includegraphics[width=0.99\textwidth]{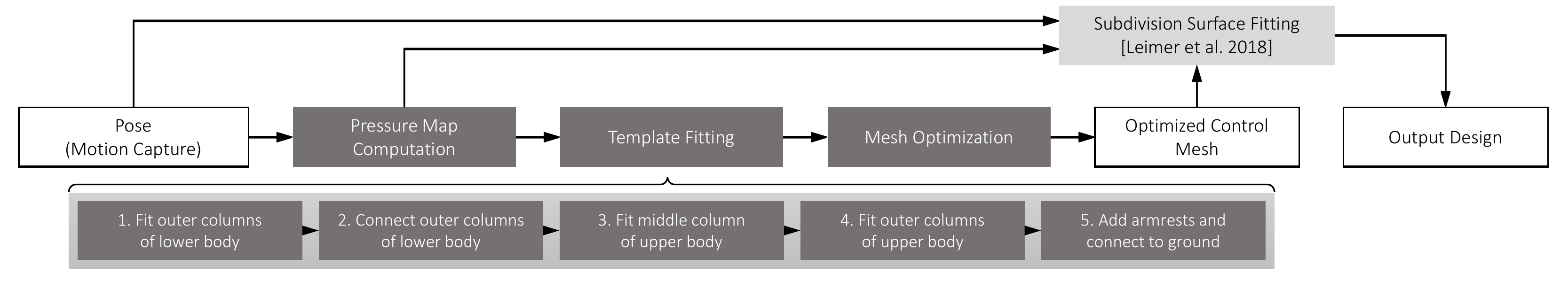}
	\centering
	\caption{Overview of our automated pose-driven furniture design method. Given a human pose as input, we first compute a pressure distribution under the assumption that the body is perfectly supported. We then use this pressure distribution as an importance map to fit a template geometry thats supports the body in its given pose. The support mesh is then further optimized. The resulting control mesh can then serve as an initial design candidate that can be edited with conventional modeling tools, or used with the fitting algorithm of Leimer \etal~\cite{leimer2018sit} to create a surface with an even better optimized fit.}
	\label{fig:overview}
\end{figure*}

\section{Related Work}\label{sec:related}

\paragraph{Furniture Design}
From a general perspective, the central goal of this paper is to provide an automated computational design system for usable seating furniture. Furniture creation is a very broad task with a rich history in a variety of fields including wood working, product design or medicine. 
An important question is whether a designed seating surface is aimed for a general application or to be used in a specific situation only.
In any public place or transport, the use of a one-size-fits-all solution is inevitable. With modern design methods, for instance using 3d scans of human body shapes \cite{smulders_aircraftseats}, a large range of body sizes can be covered. 

In a human centered design process the attention is shifted to the needs and requirements of a human person.
The goal is to find a seating surface that optimally matches the requirements of a human person, ranging from physical properties such as shape or size~\cite{reed2008modeling} to semantic constraints.
Research in function driven design aims to find seating surfaces which match a general class of poses or guidelines~\cite{zheng2016ergonomics,brintrup2008ergonomic}. 
In \textit{pose driven} design the goal is to fulfill much stricter human requirements. Research in this area considers a given pose as optimal for a specific situation and aims to design a seating surface to match a person in that pose as close as possible~\cite{fu2017pose,leimer2018sit}.

Interactivity is also an important factor in personalized furniture design.
For example, Saul et al.~\cite{saul2011sketchchair} created a furniture design system intended for end users, which allows them to design chairs from free-form shapes. Interactivity is a core element in their research.
Lee et al.~\cite{lee2016posing} designed their system around VR technology to allow users to personalize furniture via poses and voice commands.
Umetani~\etal~\cite{Umetani2012b} proposed a system for computational design of shelves using a physical model which supported the users during the design such that only structurally stable models where created. 
Other research focuses on automated systems aimed to design fitting furniture in an automated process. User interaction is mostly limited to customizing input data and parameters~\cite{zheng2016ergonomics,verhaert2011use}.
Researchers in furniture design have also used hybrid approaches for their systems, where the furniture shapes are created by an automated system, but users can steer or manipulate the design process in various stages~\cite{fu2017pose}.

\paragraph{Comfort}
Comfort is an important measure to evaluate the functional requirements of furniture. Historically, the most elementary way to determine comfort or discomfort of a seat is to keep note of the subjective feelings of its users \cite{hertzberg1958seat,jones69}.
Subjective measures are the most direct and reliable indicators of comfort, however, in most furniture design applications, objective measures would be advantageous compared to subjective ratings~\cite{de2003sitting}.
Therefore, researchers have aimed to find a relation from subjective feelings to objective measures for comfort and discomfort.
De Looze et al.~\cite{de2003sitting} identified a variety of objective measures for comfort or discomfort from literature in medicine and ergonomics and concluded that pressure distribution showed the most clear association with the subjective ratings. Similar findings have also been shown in many other studies \cite{kamijo1982evaluation,yun1992using,LOPEZTORRES2008123,ZENK2012290,NORO2012308}.

Related to this is the field of Biomechanics, which studies mechanical effects on human bodies \cite{Hall1995,Robertson}---like forces and moments. In this paper we base our calculations on a link-segment skeleton model of a human body which is also often used for inverse dynamics~\cite{Dumas2004,Robertson,Fluit2014}. Our method is based on a linear least-squares approach of inverse dynamics as often used for such purposes~\cite{Kuo1998,Dumas2004,Vlietstra2014}. 
However, we extend the model by combining the skeleton with a skin of the human using linear blend skinning~\cite{Baran2007a} as commonly used in computer graphics. In order to make our model physically more plausible, we assign each limb a weight and a center of mass that have been determined empirically by Plagenhoef~\etal~\cite{Plagenhoef1983}. In contrast to physical simulations, in our method we propose a linearized model of the transmission of forces from bones to the skin in order to determine an optimal distribution of reaction forces on the body in such a way that the moments acting in the given posture are minimized.

\paragraph{Pose-based Design}
In 2017, Fu et al.~\cite{fu2017pose} introduced a shape synthesis approach with the goal of creating hybrid shapes usable by humans. 
While this work is not limited to furniture shapes, it serves as an example for pose-driven design. 
Lee et al.~\cite{lee2016posing} proposed a novel user centric furniture design process, making digital design interfaces accessible for casual users by using poses and gestures, speech commands and augmented reality technology. 
Zheng et al.~\cite{zheng2016ergonomics} introduced an interactive system that selects and adapts seating furniture for user-specified human body and input poses.
An entirely different approach at personalized furniture design is presented by Wu et al. ~\cite{Wu:2018:AAP:3173574.3174132}:
ActiveErgo is a monitored workplace environment that dynamically adjusts its parameters like desk height or chair position in accordance with ergonomic guidelines, adjusted to the user.  
In 2018, Leimer et al.~\cite{leimer2018sit} presented Sit\&Relax, a pose-driven, interactive furniture design approach.
However, the accuracy of their results is mostly dependent on the quality of the control meshes used as input. In this paper we address this problem.

\section{Overview}\label{sec:overview}
The workflow of our proposed method can be seen in Figure \ref{fig:overview}. 
The input is a human body model in a specific pose, given by a 21-joint skeleton defining the body structure and pose, a triangle mesh forming the body geometry, and a mapping between them. 

We used the dataset of poses provided by Leimer \etal \cite{leimer2018sit}, who uses a \blender{} plugin \cite{Bastioni2018} for the generation of body meshes with varying attributes, like gender, mass, size, stature, etc. The meshes are skinned and rigged to a skeleton and the poses can be adjusted either manually or can be created with a motion capturing device (for instance \textsc{Perception Neuron} system~\cite{Neuron2018}). 
% and use a set of \przem{64??} poses which are shown in Figure~\ref{fig:poses}. 

In the first step after selection of a pose (Section \ref{sec:compmodel}), we compute a pressure distribution on the human body under the assumption that the body is supported everywhere. 
In the second step  (Section \ref{sec:template_model}), we use the pressure map as an importance map for the synthesis of the basic geometry of the body-support by fitting a template-geometry, which further optimized in order to meet certain quality criteria. 

The resulting geometry can then be treated as an initial design step that can be edited manually with conventional modeling applications. It is also possible to directly apply the subsurface fitting algorithm of Leimer \etal~\cite{leimer2018sit} to create an even further optimized fit between the body and the surface. Figure \ref{fig:teaser} shows an example design created using this method. In Section \ref{sec:results} we provide more results of the method and compare them to previous work of Leimer \etal~\cite{leimer2018sit}.

\section{Computational Model of Sitting}\label{sec:compmodel}

In this section we propose our simplified physical model of sitting for the computation of pressure distribution, the moments (torques) acting on the joints as well as friction forces acting on the body. 
%This pressure distribution is used as an importance map in the furniture synthesis stage (see Section \ref{sec:template_model}) to indicate where the human body needs to be supported.
In the recent work of Leimer \etal~\cite{leimer2018sit}, a similar simplified computation model was proposed, however, our method has three advantages over theirs:
\begin{enumerate}
	\item  Our model consists of individual body segments instead of a single rigid body, allowing us to consider a more realistic distribution of body mass, as well as the moments acting on the joints which are caused by the transfer of forces between body segments. 
	%We also use empirically determined masses of individual body parts as proposed by \cite{Plagenhoef1983}. 
	\item  Using our model, we can also compute friction forces which are not available in their approach. % \przem{more dicussion?}
	\item Finally, our algorithm yields physical pressure values instead of only a relative distribution, which we show to be in realistic range by comparing to FEM simulation.
\end{enumerate}

\subsection{Human Body Model}

We propose a novel human body model that combines a skeleton with a surface  that allows us to compute the moments acting on the joints and the pressure distribution on the surface of the body. 

\paragraph{Skeleton Model}

\begin{figure}[t]
	\centering
	\includegraphics[width=1.0\columnwidth]{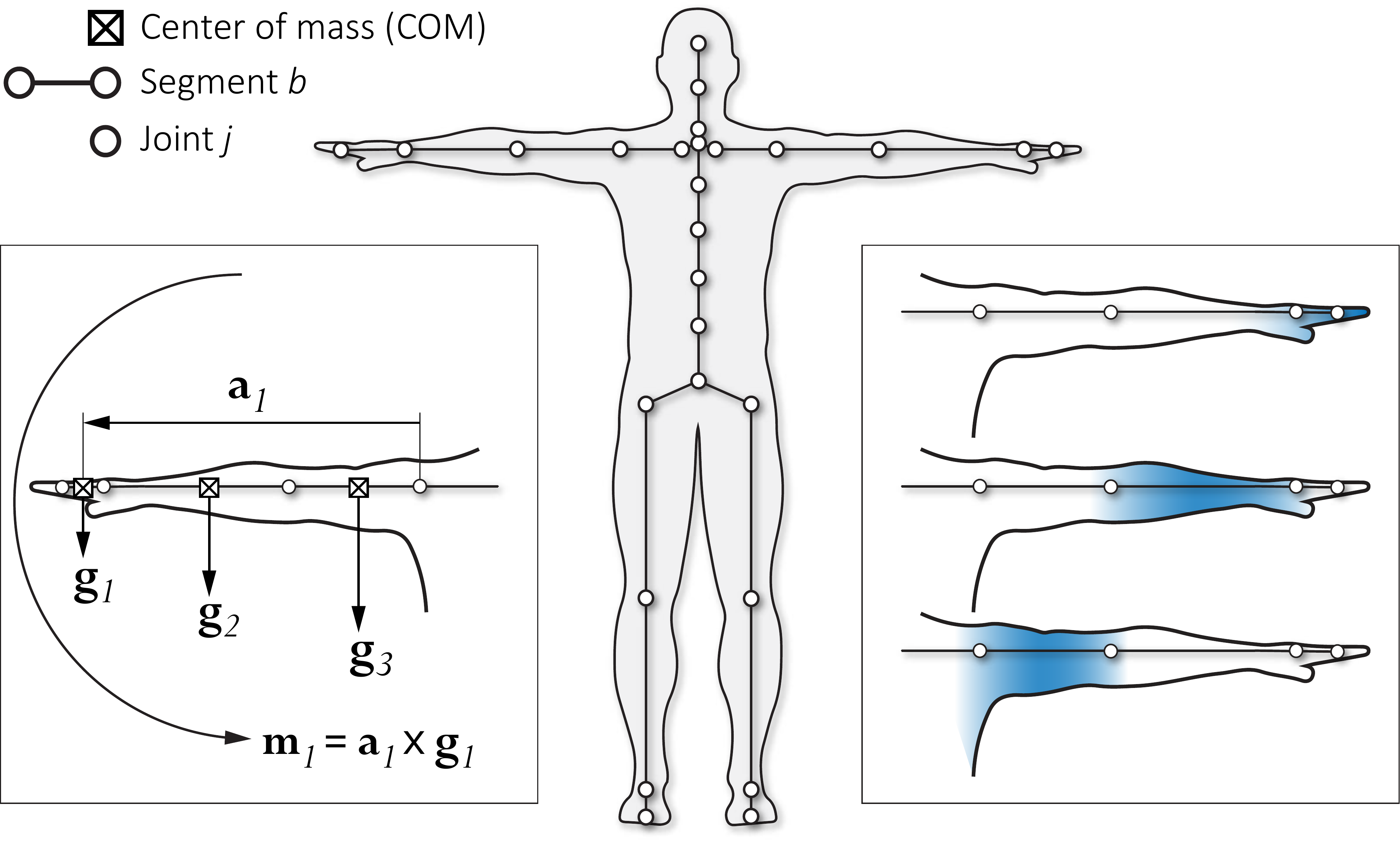}
	\caption{Left: a moment $\textbf{m}_1$ as a cross product of the moment arm $\textbf{a}_1$ and the force $\textbf{g}_1$. Body segments have anatomical values (e.g., $\textbf{g}_1$) assigned from~\cite{Plagenhoef1983}. Middle: a link-segment-skeleton with 21 segments and a polygonal surface mesh. Right: The mesh is rigged using linear blend skinning~\cite{Magnenat-Thalmann1989}.   }
	\label{fig:skinning}
\end{figure}

The skeleton is modeled using a link-segment-model with 21 segments as depicted in Figure~\ref{fig:skinning}. Such a model consists of segments that represent parts of the human body which are connected by joints that allow movement of the segments with varying rotational degrees of freedom. Each segment has its own mass concentrated at the center of mass (COM) and can be influenced by external forces such as gravity or contact with other surfaces. 
The mass and the locations of the center of mass of each segment are based on the data by Plagenhoef \etal~\cite{Plagenhoef1983}, which were determined empirically on experiments with human cadavers. 

The joints themselves are assumed to have no mass and also to not be affected by external forces. They can, however, transfer forces and moments from one segment to another. We model this as two opposing forces (or moments) acting on the joint, one for each segment linked by the joint (cf. Fig.~\ref{fig:bodymodel} for detailed depiction).

Segment-link-models are commonly used in biomechanics to examine the moments acting at joints during certain actions or movements~\cite{Hall1995,Robertson}. 
%Since we use the same data set of poses that is used in the work of Leimer \etal~\cite{leimer2018sit}, which was created using the Perception Neuron motion capturing system~\cite{Neuron2018}, our segment-link model contains 21 segments.  
Please refer to Figures~\ref{fig:skinning} and~\ref{fig:bodymodel} for a depiction.  
%\przem{Explain the link-segment model better here?}
%\przem{Make a figure of a moment-arm!}

\paragraph{Skinning}
We register the skeleton with a human body model given by a triangle mesh. For the generation of body meshes we have used the software provided by Manuel Bastioni~\cite{Bastioni2018} which allows the generation of human bodies with varying parameters, like gender, mass, size, stature, etc. %We roughly match these settings with those of the probands we have captured with the motion capturing device. 
We rig the mesh with the skeleton using  linear blend skinning~\cite{Magnenat-Thalmann1989}, in particular all vertices are defined by their weighted linear combinations: 
\begin{equation*}
\textbf{v} = \sum_b \alpha_{b,v}\textbf{T}_b\textbf{v}^0 \,,
\end{equation*} 
where $\textbf{v}^0$ are the initial and $\textbf{v}$ are the new vertex positions respective,  $\alpha_{b,v}$ are the weights which associate the vertex $v$ to the segment $b$, and $\textbf{T}_b$ are the transformations of the assigned segments $b$. We use the algorithm of \cite{Baran2007a} implemented in \blender.

We further use the weights of the skinning to propagate forces from the segments to the surface vertices and vice versa (cf. Figure~\ref{fig:skinning} and Figure~\ref{fig:bodymodel}). 

\begin{figure}[t]
	\centering
	\includegraphics[width=1.0\columnwidth]{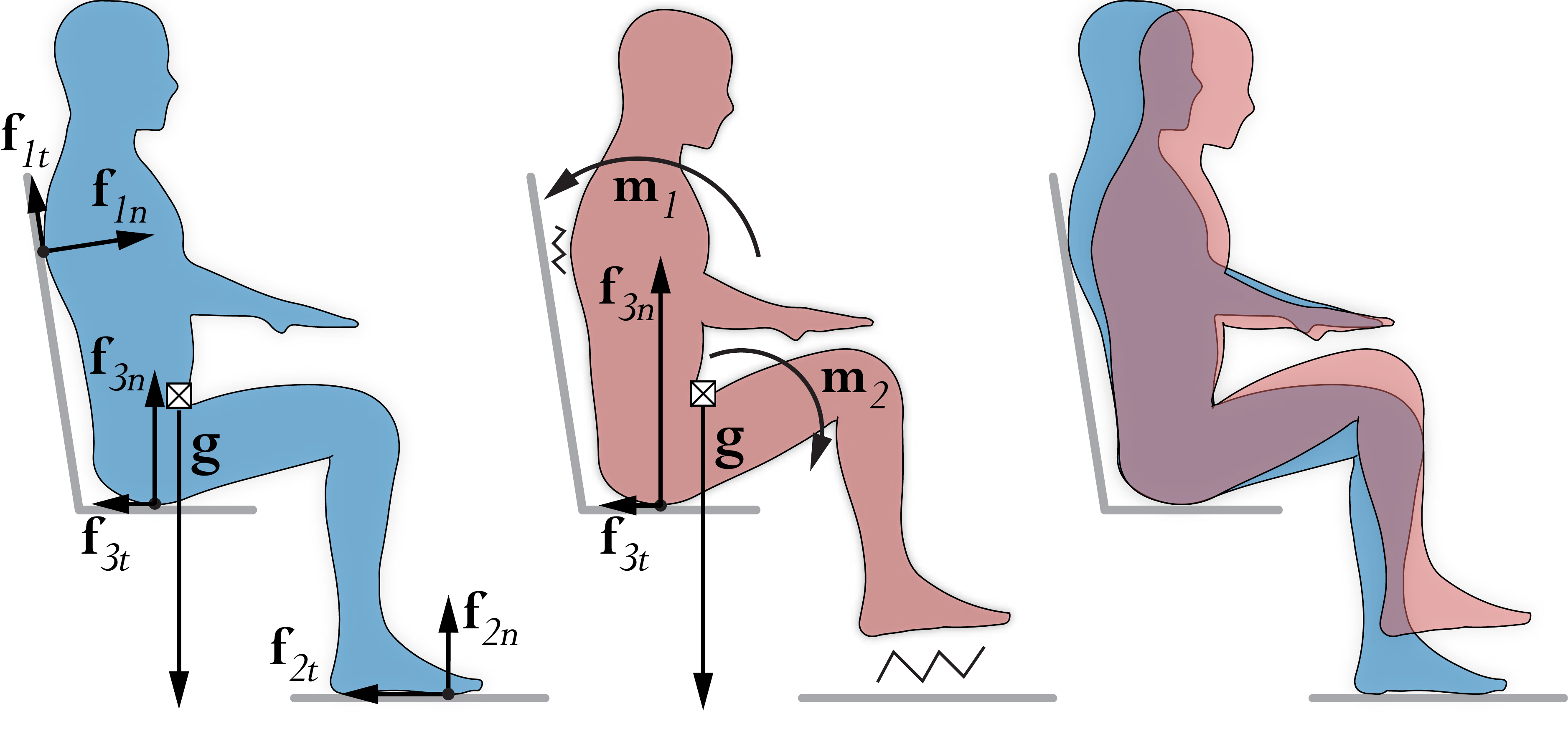}
	\caption{Physics of sitting: if contact with a support surface is given on the buttocks, back, and feet, the moments of the body are minimized and the forces are in equilibrium. If the contact on the feet is lost, the contact to the back is lost automatically due to the missing friction force on the feet.  }
	\label{fig:sitting}
\end{figure}

\paragraph{Friction Model}
In the mechanics of sitting, friction plays an important role. 
Consider the example shown in Figure~\ref{fig:sitting}, left, where a body has three contact points: on the buttocks, the back, and the feet. 

The tangential reaction force $\textbf{f}_{1t}$ that is supporting the back is dependent on the normal force $\textbf{f}_{1n}$ at the same location. This normal force can only exist due to an opposing force $\textbf{f}_{2t}$ existing at the feet since all forces must sum to zero to maintain equilibrium. Therefore, if we lose the contact of the feet to the ground (e.g., by lifting the legs), the force $\textbf{f}_{2t}$ disappears and we also (almost) lose contact on the back and the reaction forces acting there unless an additional force is introduced, for example by pressing the thighs against the seat which requires significantly more muscle activity. 

In consequence, the back is no longer supported and the overall contact area becomes much smaller, resulting in a higher force ($\textbf{f}_3$) on the remaining contact points. Additionally, higher moments ($\textbf{m}_1$, $\textbf{m}_2$) act on the joints, requiring more muscle forces to maintain the pose (cf. Figure~\ref{fig:sitting}, center). 

We use the Coulomb model in which the frictional component of a reaction force depends only linearly on the normal component of the reaction force (refer to Figure~\ref{fig:bodymodel} and to Eq.~\ref{eq:friction} later on).  We choose this simplified model, since due to our assumptions, the by far biggest force is the gravity which implies that forces in any other direction tend to be much smaller. 
% 
%Using an individual friction coefficient for each contact would require knowledge of contacting materials at each contact point and coefficients for each combination of materials that can be in contact, making our model too sophisticated for interactive design. 

\subsection{Reaction Forces}\label{sec:model:optimizaton}

Our goal is now to find a physically plausible distribution of reaction forces that supports the human body with as little need to use additional muscle forces to maintain its current pose as possible. 
Usually, such distribution would be found using a sophisticated finite elements simulation which is very time consuming. 

Since our goal is to achieve interactive rates, we propose a model where we assume the human composed of rigid segments combined by joints, where the surface vertices are related to the segments of the bodies surface by linear combinations. 
%In order to account for the softness of the human body, we introduce a small penetration tolerance. 
This allows us to formulate it as a Pareto-optimization problem where we balance the minimization of the moments acting in the body with the uniformity of the distribution of the reaction forces. 
%\przem{combine with previous section? move to motivation?}

In this section we first describe how we estimate the friction and normal forces on each vertex of the surface, and further we describe to details of the linear optimization problem. Finally, we compare our results to a rigid-body FEM simulation in order to validate our results.

\begin{figure}[t]
	\centering
	\includegraphics[width=1.0\columnwidth]{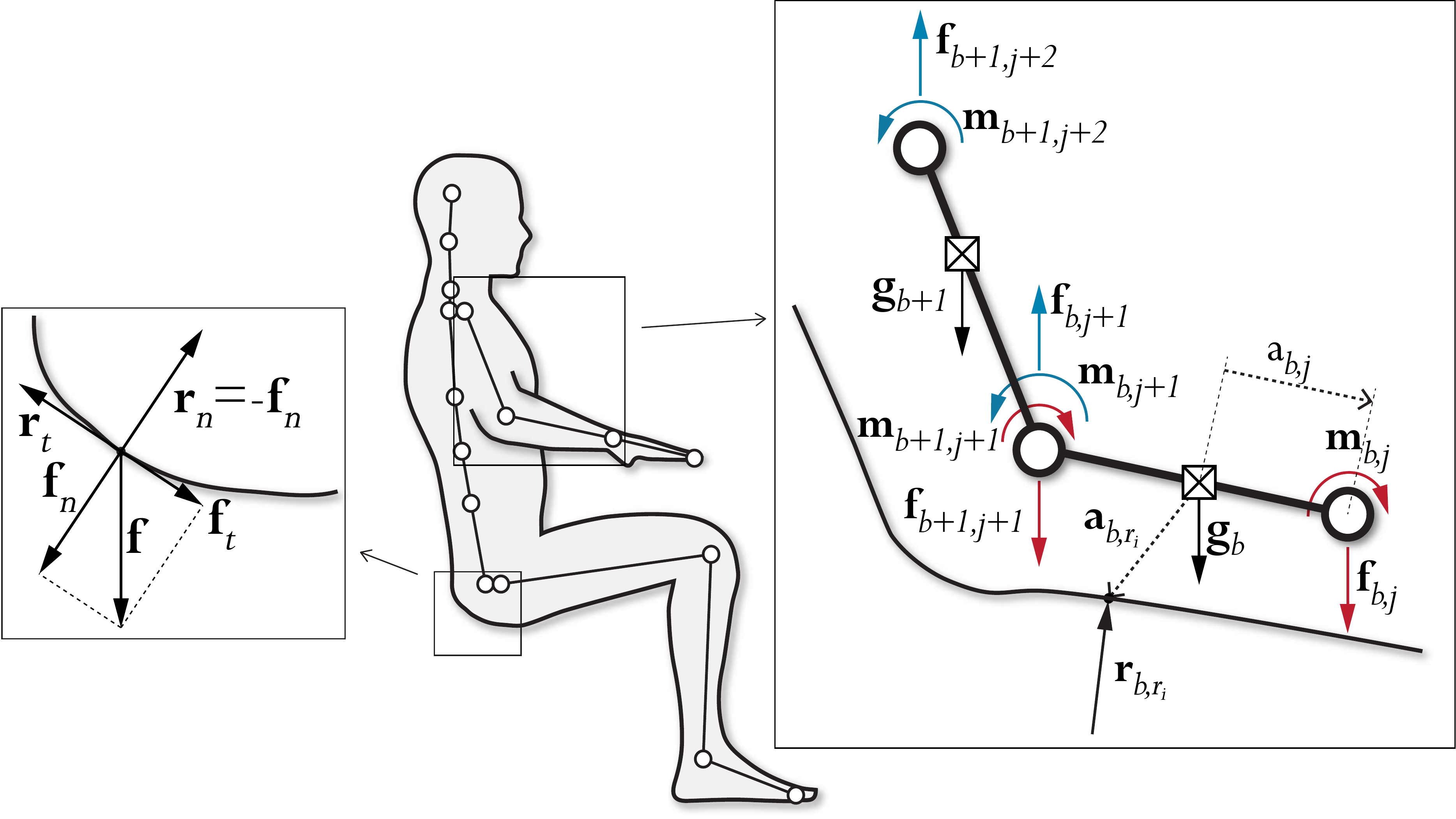}
	\setlength{\abovecaptionskip}{-3pt}
	\setlength{\belowcaptionskip}{-11pt}
	\caption{Computational human body model. Left: simplified friction model, right: free-body diagram of the skeleton model. Please refer to Section~\ref{sec:model:compmodel} for the details.  }
	\label{fig:bodymodel}
\end{figure}

\paragraph{Local Reaction Weights} \label{sec:force:local}
In order to compute the optimal reaction forces for the entire system, we first introduce the local reaction and friction force model, which we utilize for the derivation of \textit{local reaction weights}. In essence, we compute the maximum reaction forces that can occur if a local force $\mathbf{f} = (0,-1,0)^T$ acts on an isolated vertex. We first split $\mathbf{f}$ into its normal component $\mathbf{f}_n$ along the surface normal and its tangential components $\mathbf{f}_{t_1}$ and $\mathbf{f}_{t_2}$. Making use of the well-known friction pyramid of the Coulomb model~\cite{Popov2010}, we have
\begin{equation}
\mathbf{r}_n = - \mathbf{f}_n, \quad\quad
\mathbf{r}_{t_{1}} = -\frac{\mathbf{f}_{t_1}}{ \norm{\mathbf{f}_{t_1}}} \min\left(||\mathbf{f}_{t_1}||, \mu ||\mathbf{f}_n||\right) ,
\label{eq:friction}
\end{equation}
and $\mathbf{r}_{t_2}$ defined analogically, with the friction coefficient $\mu \geq 0$. The total reaction force is then $\mathbf{r} = \mathbf{r}_n + \mathbf{r}_{t_1} + \mathbf{r}_{t_2}$ (cf. Figure~\ref{fig:bodymodel}, left box).

In other words, the magnitude of the friction force must be smaller or equal to the magnitude of the normal force multiplied with the friction coefficient $\mu$, which depends mainly on the roughness of the surface material, hence, in our experiment we use $\mu=0.5$ which is a common default value if the  material  is not known.

Since the distribution of the reaction forces on the body depends on the overall forces acting on the system, which are not known in advance, looking only at each vertex individually is not sufficient. But we can use this information to introduce a weight vector $\mathbf{w} = (w_{n},w_{t_1},w_{t_2})^T$ per vertex with
\begin{equation}
w_{n} = \frac{1}{\norm{ \mathbf{r} }}, \quad\quad
w_{t_1} = \frac{1}{\norm{ \mathbf{r}_{n} + \mathbf{r}_{t_1} }}, \quad\quad
w_{t_2} = \frac{1}{\norm{ \mathbf{r}_{n} + \mathbf{r}_{t_2} }} \,,
\label{eq:react_weight}
\end{equation}
which serves us later to indicate the actual contribution of each individual reaction force to their global distribution during the optimization (cf. Eq.~\ref{eq:obj_press}).

\paragraph{Computation of Reaction Forces}\label{sec:model:compmodel}

%\subsubsection{Assumptions} 
%Assumptions:
%- Segments of the human body are rigid. 
%- We use a simplified friction model, which is currently uniform across the body. 
%- We assume a uniform mass distribution of each body segment. The centers of mass of each segment are taken from a table. 
%- Blender skinning
%In this section we present our computational model we have developed in order to determine the optimal distribution of reaction forces on the body to support the human in the current pose. 
Given the weights, we propose a linear model to determine the optimal distribution of reaction forces on the body to support the human in the current pose. 
Please refer to Figure~\ref{fig:bodymodel} for a depiction of the components.

%\subsubsection{Unknowns} 
First, we compose the vector $\textbf{x}$ of unknowns of the following physical entities: 
\begin{itemize}
	\item $\mathbf{f}_{b,j} \ldots$ the forces acting on the joints $j$ in each body segment $b$. There are $2$ such forces per body segment, except for the hands, feet, and head since they are connected to only $1$ joint, 
	\item $\mathbf{m}_{b,j} \ldots$ the moments acting on the joints $j$ in each body segment $b$. Again there are $2$ such moments per body segment, except for the hands, feet, and head since they are connected to only $1$ joint, 
	\item $\mathbf{r}_v \ldots$ the reaction forces acting at each body vertex $\mathbf{v}$, caused by contact with an external surface. %These are related to the body segments by the linear blend skinning weights $\alpha_{b,v}$, which express how much the force $\mathbf{r}_v$ is influenced by the segment $b$ (and vice versa). 
\end{itemize}
The vector of unknowns $\mathbf{x}$ is the $3(\setsize{F} + \setsize{M} + \setsize{R})$ column vector 
\begin{equation*}
\mathbf{x} = \begin{bmatrix}
\mathbf{f}_{b,j} \\%[-2mm]
%\vdots \\[2mm]
\mathbf{m}_{b,j} \\%[-2mm]
%\vdots \\[2mm]
\mathbf{r}_v \\%[-2mm]
%\vdots
\end{bmatrix} \,,
\end{equation*}
where $\setsize{F}, \setsize{M}, \setsize{R}$ denote the cardinality of the sets for joint forces $\set{F}$, moments $\set{M}$, and reaction forces $\set{R}$ respective. 

\begin{figure*}[t]
	\centering
%	\iflowres
%	\includegraphics[width=1.0\textwidth]{/comparison/Figure/FiniteElements-Matlab-Comparison_v2_copy_lowres}
%	\else
	\includegraphics[width=1.0\textwidth]{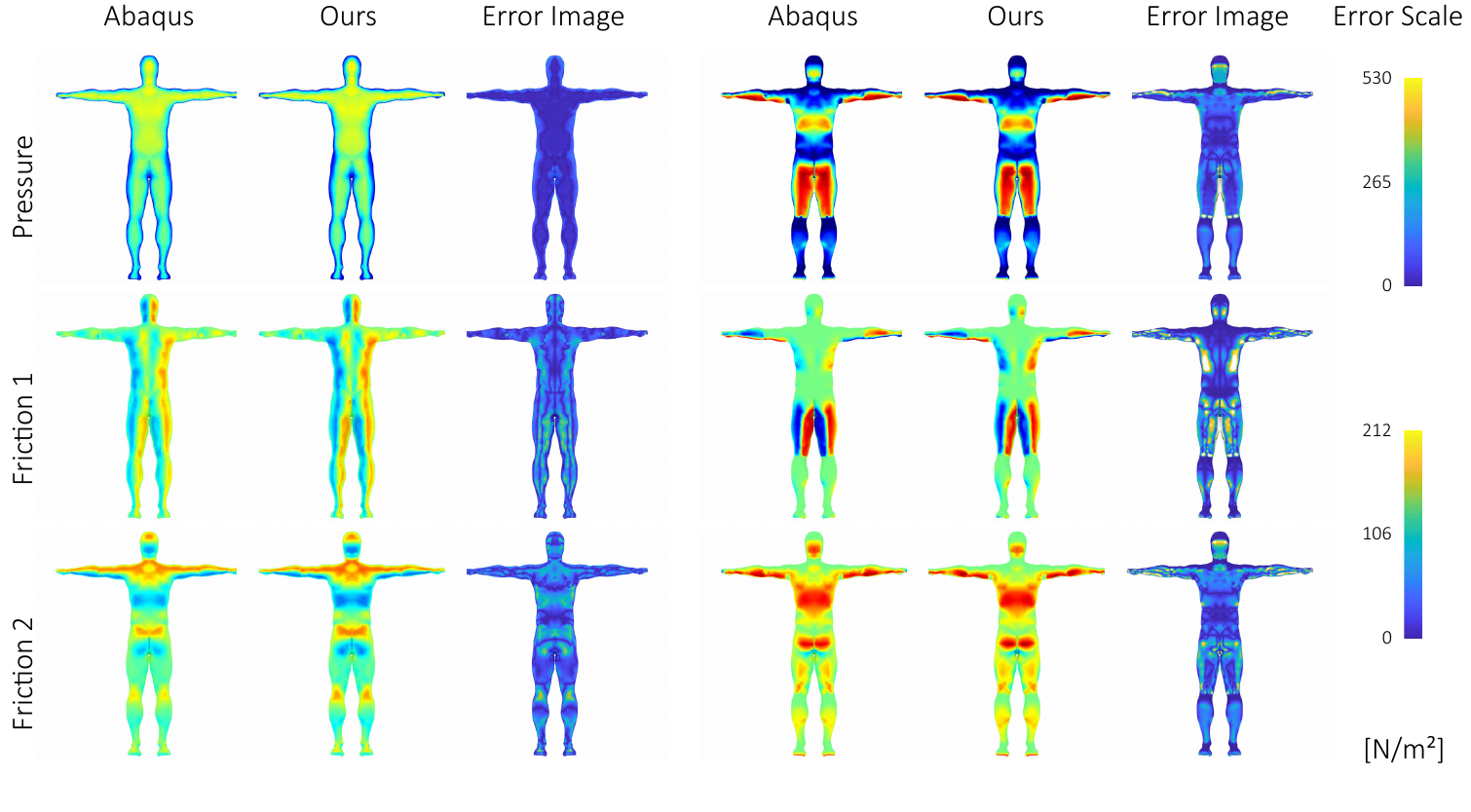}
%	\fi
	\caption{Comparison to the FEM (cf. Section~\ref{sec:model:calibration}). Left: results of a lying T-pose. Right: a sitting pose. Please note that we plot the pressure and shear distributions of the sitting pose on a T-pose mesh for better visualization purpose. The distributions show excellent agreement between our method and the physical simulation in most areas. The largest differences occur at the borders of the contact region where accuracy is not as important.}
	\label{fig:comp_fem}
\end{figure*}

Our main constraint is that the human body must be in static equilibrium, meaning that it must be physically able to maintain its current pose through contact forces, friction, and acting moments (i.e., muscle strength), so that there is no translational or rotational movement of any segment. According to the equations of motion~\cite{Goldstein2002}, a body is in equilibrium if the sum of all acting forces and moments sum to $0$. Applied to our link-segment-model, this includes the following forces:
\begin{itemize}
	\item $\mathbf{g}_b \ldots$ the gravity acting on the center of mass (COM) of the body segment $b$,
	\item $\mathbf{r}_{v} \ldots$ the reaction forces at each vertex of the body segment caused by contact with an external surface,
	\item $\mathbf{f}_{b,j} \ldots$ the forces caused by other body segments transmitted through the joints $j$, 
\end{itemize}
and the following moments:
\begin{itemize}
	%\item $\mathbf{v}_{j,b} \times \mathbf{g}_b \ldots$ the moment acting on joint $j$ caused by gravity, with $\mathbf{v}_{j,b}$ being the vector pointing from joint $j$ to the center of mass of body segment $b$.
	\item $\mathbf{a}_{b,j} \times \mathbf{f}_{b,j} \ldots$ the moments acting on the COM of body segment $b$ caused by the forces from other body segments transmitted through joint $j$, with the moment arm $\mathbf{a}_{b,j}$ being the vector pointing from the COM  to joint $j$.
	\item $\mathbf{a}_{b,r} \times \mathbf{r}_{v} \ldots$ the moment acting on the COM of body segment $b$ caused by the reaction force through contact with an external surface, with the moment arm $\mathbf{a}_{b,r}$ being the vector pointing from the COM to the contact point.
	\item $\mathbf{m}_{b,j} \ldots$ the moments caused by other body segments transmitted through the joints $j$. 
\end{itemize}
Please note that reaction forces $\mathbf{r}_{v}$ at the vertices $v$ are connected to the body segments $b$ by the linear blend skinning weights $\alpha_{b,v}$.  %for each reaction force $\mathbf{r}_{v}$  which assigns its  on a particular body segment $b$. 
This leads to the following constraints for each body part $b$:
\begin{gather}
\sum_{j \in \set{J}_b} \mathbf{f}_{b,j} + \sum_{v \in \set{V}_b} \left(\alpha_{b,v} \mathbf{r}_{v}\right) - \mathbf{g}_b = \textbf{0}, \label{eq:const11}\\
\sum_{j \in \set{J}_b} \mathbf{a}_{b,j} \times \mathbf{f}_{b,j} + \sum_{v \in \set{V}_b} \mathbf{a}_{b,r} \times \left(\alpha_{b,v}\mathbf{r}_{v}\right) + \mathbf{m}_{b,j} = \textbf{0},
\label{eq:const12}
\end{gather}
with $\set{J}_b$ being the set of joints connected to body segment $b$ and $\set{V}_b$ being the set of vertices of body segment $b$. Naturally, the sum of forces, as well as the sum of moments, acting on a joint must also equal $0$, i.e.:
\begin{equation}
\sum_{b \in \set{B}_j} \mathbf{f}_{b,j} = \textbf{0} \quad \text{and} \quad \sum_{b \in \set{B}_j} \mathbf{m}_{b,j} = \textbf{0},
\label{eq:const2}
\end{equation}
with $\set{B}_j$ being the set of body segments connected to joint $j$.

These constraints can be formulated as a system of linear equations $\mathbf{Cx} = \mathbf{z}$. The matrix $\mathbf{C}$ is a $(6 \setsize{B} + 6 \setsize{J}) \times 3(\setsize{F} + \setsize{M} + \setsize{R})$ matrix---$6$ rows for each body segment and each joint ($3$ for the forces and $3$ for the moments), as well as $3$ columns for each unknown force, moment and, reaction force in a body segment. 
More details abut the structure of this matrix can be found in the supplemental material.

To ensure that the resulting reaction forces do not point out of the body (which would be physically equivalent to gluing the body to a surface), we also require inequality constraints
\begin{equation}
-r_{y} \leq 0 \,,
\label{eq:const3}
\end{equation}
with $y$ being the up-direction of the global coordinate system. 

Finally, since the weights computed in Eq.~\ref{eq:react_weight} are used in the objective function and are therefore soft-constraints, we additionally restrict the 
magnitudes of the friction forces based on the normal force with hard constraints. 

To do so, we consider the reaction force vector $\mathbf{\bar{r}}$ in the tangent space of vertex $\mathbf{v}$, with the first coordinate being the normal force and the second and third components being the friction forces, and %To account for friction, we use the commonly used linearized Coulomb friction model that 
limit the magnitude of the latter in relation to the normal force using
\begin{equation}
\bar{r}_{y} \leq \mu\, \bar{r}_{x} \quad \text{and} \quad \bar{r}_{z} \leq \mu\, \bar{r}_{x} \,.
\label{eq:const4}
\end{equation}
We denote the matrix containing these inequality constraints as  $\mathbf{D}$ whose structure is explained in further detail in supplemental material.

We can now formulate the objective function as
\begin{equation}
E_{\text{pres}} = \sum_{i=1}^{\setsize{M}} \norm{ \mathbf{m}_i } ^{2} + \lambda \sum_{i=1}^{\setsize{V}} \frac{1}{A_i}  \norm{ \mathbf{w}_i \circ \mathbf{\bar{r}}_{i} }^{2},
\label{eq:obj_press}
\end{equation}
where $\circ$ denotes the Schur-product, $\mathbf{w}_i$ are the reaction weights (cf. Eq.~\ref{eq:react_weight}) and $\mathbf{\bar{r}}$ is the reaction force at the vertex $\mathbf{v}_i$ in its tangent space. 
Note, that we need to divide the reaction force by the (Voronoi) area $A_i$ of each vertex since we want the forces to be distributed equally over the surface regardless of mesh resolution.

Minimization of the function in Eq.~\ref{eq:obj_press} with constraints in  Eq.~\ref{eq:const11}, \ref{eq:const12}, \ref{eq:const2}, \ref{eq:const3}, and \ref{eq:const4} leads to a  system of linear equations with equality and inequality constraints, which we formulate in matrix form as
\begin{equation*}
\begin{aligned}
& \min_x & \parallel \mathbf{Ax} \parallel ^{2} \\
& \subto & \mathbf{Cx} = \mathbf{z} \\
& & \mathbf{Dx} \leq \mathbf{0}
\end{aligned}\;,
\label{eq:problem1}
\end{equation*}
and solve it using \matlab{}'s \texttt{lsqlin} function. The details of how the matrices $\mathbf{A}$, $\textbf{C}$, and $\textbf{D}$ are constructed can be found in supplemental material.

%\subsubsection{Balancing the Terms}\label{sec:model:lambda}
The free parameter we introduce in Eq.~\ref{eq:obj_press} is the value of $\lambda$. Intuitively, it is a weight which allows to balance between the terms which minimize the moments in the body and which distribute the reaction forces on the surface. 

Physically, we can interpret this parameter as the 'stiffness' of the joints. If it is $0$, no muscle force can be expended to maintain the pose. If it is infinite, the entire human body can be treated as completely rigid. Realistically, we cannot set the parameter to $0$ because we only have a finite number of reaction forces acting at predetermined locations, making it either impossible to fulfill the equilibrium constraints or resulting in a physically implausible solution for most poses. %lower its value, the more own muscle force needs to be expended by the human to hold the posture.  
In empirical experiments, we determined a default value of $\lambda=0.013$, which we have further used in our applications. 

In order to actually compute the pressure distribution of a pose which is used as an importance map in the next step of our approach, we need to know which vertices of the body surface are in contact with the support surface. For this we assume that the given body is supported everywhere, meaning that we consider every vertex of the body surface to be in contact.
%proceed like Leimer \etal~\cite{leimer2018sit} and

%%--------------------------------------------------------------------------------
\subsection{Comparison to Finite Elements Simulation}\label{sec:model:calibration}

In order to evaluate our computational model, we compare it to a FEM simulation using the professional physical simulation software \abaqus{}~\cite{Smith2009}. We select 2 poses for this purpose---a lying pose and a sitting pose. For each pose, we create 2 parts in \abaqus{}, one being the body with the geometry of the original mesh, the other being a shell generated from the original geometry which serves as the contact surface. To create this shell, we first include all mesh faces whose normal is not perpendicular or opposite of the gravity direction, and then manually reduce this set by deleting isolated faces or faces where we do not want to support the body (e.g. under the armpits). 

We create a volume mesh of the body using the \abaqus{} meshing algorithm such that the body consists of roughly equally sized tetrahedrons. We use the same element size to subdivide the shell such that the surfaces of body and shell are still perfectly aligned. We assign both body and shell a Young's Modulus of $2.1e+18$ and Poisson's Ratio of $0.3$, thus making both parts close to rigid. The shell is completely locked in place by boundary constraints, while the body is moved downward by forces totaling $735N$ (roughly equivalent to a body weight of $75kg$). The forces are applied per vertex (both on the surface and the inside of the body), their distribution computed from the weight of each body segment and the skinning weights that determine which vertex belongs to which body part (for each inner vertex we simply copy the weights of the closest surface vertex).

To model the contact between body and shell we use a linear pressure-overclosure relationship with contact stiffness of $8e+12$. The tangential contact behavior is modeled with an isotropic friction coefficient of $\mu = 0.5$ and an elastic slip of $1e-10$. We determined these parameter values empirically, as other values would often result in unstable contact conditions and a physically implausible pressure distribution with immense pressure peaks at some isolated vertices and no contact at all at other vertices.

In our model, we select the set of vertices at which reaction forces are computed by choosing those vertices of the body mesh that are also included in the corresponding contact surface shell to make sure that the contact surface is the same in both methods. We furthermore do not optimize for the joint moments ($\lambda = \infty$), which is equivalent to making the body completely rigid, as is also the case in the \abaqus{} simulation.

For the lying pose, the \abaqus{} simulation took a total of 32 minutes and 5 seconds (18 minutes and 12 seconds for preprocessing and 13 minutes and 53 seconds for actual simulation), while our system takes 1.6 seconds on average. For the sitting pose, the \abaqus{} simulation took a total of 141 minutes and 50 seconds (78 minutes and 8 seconds for preprocessing and 63 minutes and 42 seconds for actual simulation), while our system takes 1.5 seconds on average. The results of both methods can be compared in Figure~\ref{fig:comp_fem}. Note that we use the same color scale for the visualization of the results of both methods, but different scales for the pressure and shear values.

\section{Body-Support Synthesis} \label{sec:template_model}

%Introduction:
%In this section, the structure and design of the template furniture model is explained in detail. %This generic model is the main contribution of this thesis and core resource in the design process.
%The following sections summarize and further clarify the goals and requirements for this model.
%Furthermore, the methodology behind the generic template model is presented in detail.
The general goal of this stage is to automatically create a support-template for the seating surface that closely fits a human body in a specific pose. 
The template model utilizes a hierarchy of non-planar quads for this purpose, chosen for simplicity as well as suitability for the task. 
As a secondary optimization goal, we aim to produce a visually pleasing piece of furniture. Therefore, we impose rough guidelines on the geometric shape of the seating surface regarding planarity and regularity \cite{Liu:2011:GPQ:2070781.2024174,doi:10.1111/j.1467-8659.2010.01776.x}.

\subsection{Template Model} \label{sub:fitting}
The general design concept for the template model is to find a suitable structure of quadrilateral faces which can be fit to the human body in a specific pose according to the comfort measures (represented by the importance map computed in the previous stage) under the defined constraints. For the proposed framework we decided on using a 3x7 grid of faces for the main body shape, excluding the person's arms and head, which are treated separately. Going forward, we refer to the faces along the height direction of the body as \textit{rows} and the faces along the width as \textit{columns}.

\begin{figure}[t]
	\centering	
	\includegraphics[width=0.48\textwidth]{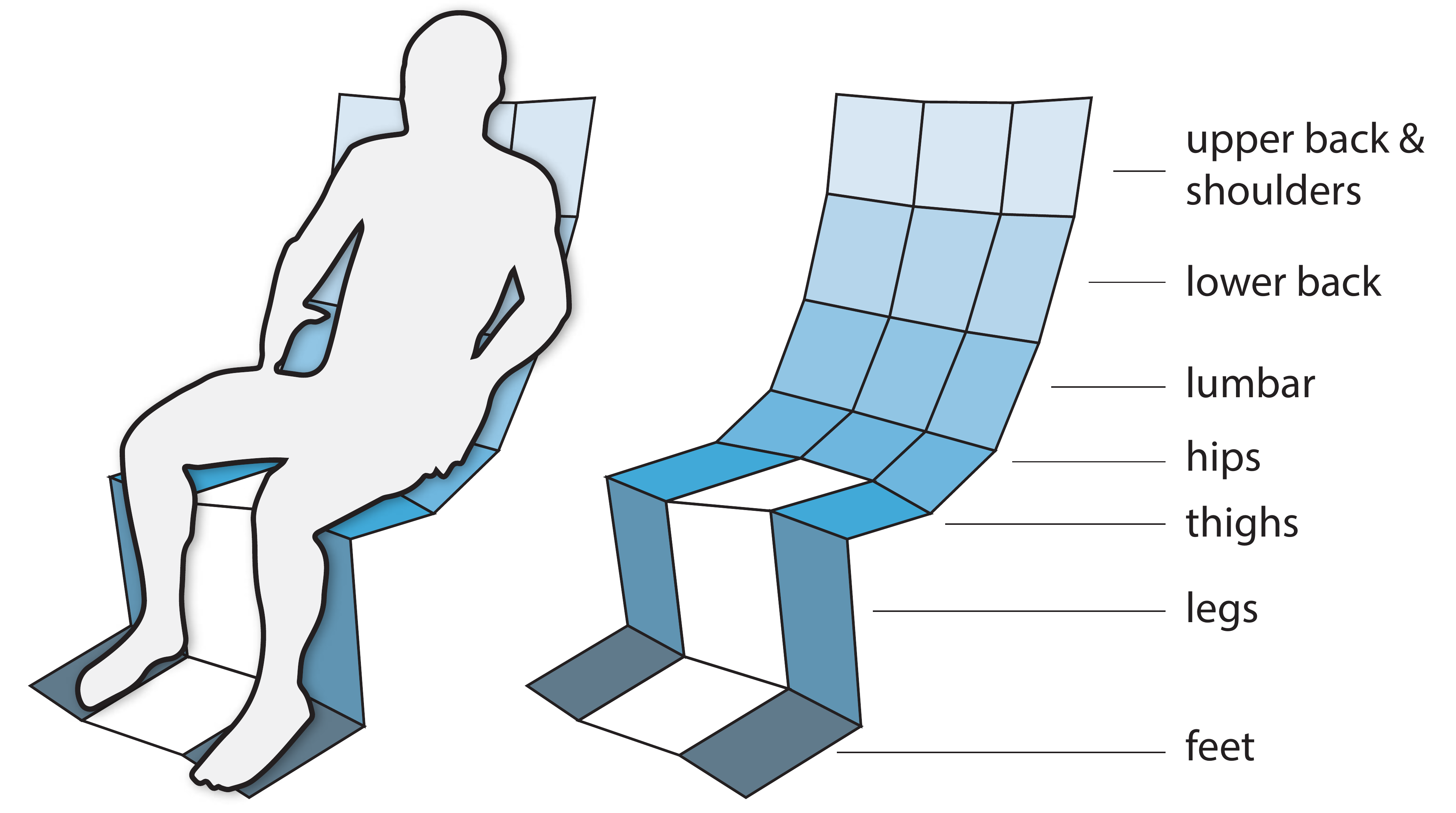}
	\caption{Body part mapping of our template model. The rows of the model are mapped to individual body parts. Within a row, the segment in each column is mapped to a subset of the corresponding body vertices. The leg segments are mapped independently to the corresponding body parts.}
	\label{fig:am_mapping}
\end{figure}

Figure~\ref{fig:am_mapping} shows the assignment of the template faces to each body part. The legs are each mapped to an individual column of $3$ faces (foot, shank and thigh), while each part of the upper body (hips, lumbar, lower back, upper back) is mapped to a row of $3$ faces.
A person's arms are supported by additional faces which are added in a later stage in the algorithm.
%In the advanced template model, the headrest face was omitted. This decision was made based on the fact that the importance weights in the head and neck area of the body was generally very low for most sitting poses in the chosen input domain. Therefore it was concluded that supporting the head was insignificant for fulfilling the chosen comfort requirements.

Since each pose is determined by a $66$ parameter vector, the space of possible poses is vast. This makes it impossible to support all possible poses using a template model with a predefined topology. Problems arise when the projections of the supported body segments onto the ground plane intersect. For a pose to be supported without special treatment, we therefore require that the shortest line between any supported vertex and the ground plane does not intersect the body geometry. Since this requirement significantly limits the space of valid poses, we detect and handle a number of special cases: crossed legs, upper body leaning forward and arms positioned above the body (see Section \ref{sec:specialcase}).

\subsection{Template Fitting}

The process of fitting the model is as follows: 
For each face in the grid, a plane is fitted to the shape of the respective body parts.
The fitting algorithm utilizes the geometry of a human body mesh transformed into specific sitting pose as well as its computed importance map, indicating which vertices are most important to support to reach optimal comfort.
%With better accuracy and expressiveness of the advanced model comes the downside of greater risk for errors. 
As our model consists of 21 free floating planes, hierarchical constraints are introduced to the fitting process to prevent error cases and maintain the general structure of the model.

\paragraph{Mesh Generation}
The mesh generation stage consists of four steps. In the first step, we create the two 3-face columns that support each individual leg. These individual leg supports are then connected by another column of 3 faces in the second step. In the third step, the middle 4-face column of the upper body support is generated. Finally, we complete the mesh generation stage by creating the two outer 4-face columns supporting the upper body.

\begin{figure}[t]
	\centering
	\includegraphics[width=0.48\textwidth]{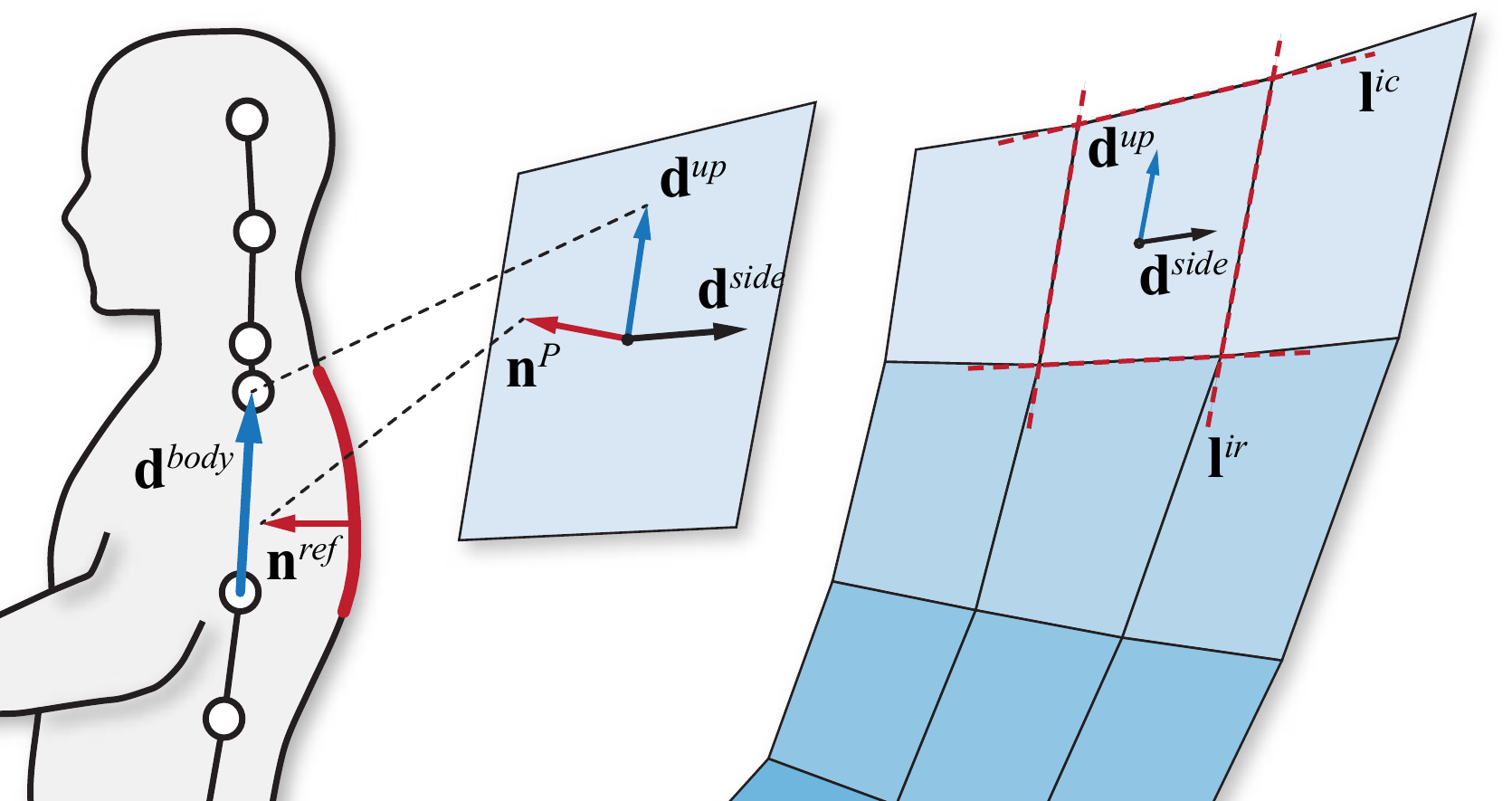}
	\caption{Plane fitting and mesh generation process. Left: Plane fitted to the body segment marked in red, local frame on the plane is given by the normal $\textbf{n}^{P}$,  projection of skeleton segment $\textbf{d}^{body}$ onto the plane $\textbf{d}^{up}$ and the orthogonal vector $\textbf{d}^{side}$. Right: Template mesh during the third step of the mesh generation process with row and column intersection lines $\textbf{l}^{ir}$ and $\textbf{l}^{ic}$ indicated.}
	\label{fig:plane_fitting}
\end{figure}

In the first and third step, we use the RANSAC \cite{fischler1981random} algorithm to find the plane that best supports a given body segment while also satisfying the structural constraints of the model hierarchy. A candidate plane is defined by randomly choosing $3$ vertices of the body segment with probabilities based on their importance. If the candidate plane intersects with the vertices of an adjacent body segment, it is discarded outright. Otherwise we define a local coordinate system on the candidate plane using the unit vectors $\textbf{n}^{P}$, $\textbf{d}^{up}$ and $\textbf{d}^{side}$ (Fig. \ref{fig:plane_fitting}, left). $\textbf{n}^{P}$ is simply the normal vector of the candidate plane. $\textbf{d}^{up}$ is constructed by taking the direction $\textbf{d}^{body}$ of the skeleton bone corresponding to the given body segment and projecting it onto the plane. Finally, we have $\textbf{d}^{side} = \textbf{d}^{up} \times \textbf{n}^{P}$. Additionally, we consider the direction $\textbf{d}^{ir}$ of the line of intersection $\textbf{l}^{ir}$ between the candidate plane and the fitted plane of the previous row of the mesh template.

To evaluate the quality of the candidate plane, we introduce two penalties. The penalty
\begin{equation}
p^r = 1 - \min_{}\left(\frac{4}{\pi} \cdot \left | \atantwo\left( \left\| \mathbf{n}^P \times \mathbf{n}^{ref} \right\|, \left\langle \mathbf{n}^P,  \mathbf{n}^{ref} \right\rangle \right) \right| , 1 \right)
\label{eq:penalty_rv}
\end{equation}
penalizes planes with a normal vector $\textbf{n}^{P}$ deviating from the reference vector $\textbf{n}^{ref}$ which is computed by applying PCA on the vertices of the given body segment. Furthermore, for the upper body, the penalty
\begin{equation}\label{eq:quality}
p^d = 1 - \frac{1}{{m^\alpha}} \min \left( \left| \arccos \left( \frac{ \left\langle \mathbf{d}^{side}, \mathbf{d}^{ir}  \right\rangle}{\| \mathbf{d}^{side}\| \| \mathbf{d}^{ir}\|} \right) \right| , m^{\alpha} \right)  
\end{equation}
ensures that the direction $\mathbf{d}^{ir}$ of the line of intersection between adjacent rows of the template conforms to the body geometry. %This is given by the vector $\mathbf{d}^{side}$ being the vector that is orthogonal to both the normal vector $\textbf{n}^{P}$ and the body up-direction $\mathbf{d}^{up}$ computed from the skeleton. 
Planes with an intersection direction that deviates from $\mathbf{d}^{side}$ by more than a chosen value $m^{\alpha}$ are penalized (cf. Fig. \ref{fig:plane_fitting}, right).

The total quality of a plane is then given by
\begin{equation}
w_P =  \sum_{\mathbf{v} \in \mathbf{V}_P}  w(\mathbf{v})  (1-\lambda^{r} + p^{r}\lambda^{r} )  (1-\lambda^{d} + p^{d}\lambda^{d} )
\end{equation}
with $\mathbf{V}_P$ being the set of body segment vertices within a set distance of the plane, $w(\mathbf{v})$ being the importance of vertex $\mathbf{v}$ based on the pressure value, and the weights $\lambda^r$ and $\lambda^d$ to set the influence of each penalty term.

To create the actual mesh geometry, we first estimate the width of the faces for the already fitted planes by taking a line with direction $\mathbf{d}^{up}$ and offsetting it in the positive and negative direction of $\mathbf{d}^{side}$ by half (or in the case of the upper body less then half) the width of the body segment to obtain the column intersection lines $\mathbf{l}^{ic}$ (cf. Fig. \ref{fig:plane_fitting}, right). By intersecting the row and column intersection lines $\mathbf{l}^{ir}$ and $\mathbf{l}^{ic}$, we obtain estimates for the corner vertices of the current face. The final coordinates of the corner vertices are obtained by averaging the positions of the estimated vertices of adjacent faces.

In the final step of the mesh generation process, the outer column faces of the template are determined again via mesh fitting while utilizing the inner column segments as hard constraints, ie., the two vertices incident to both the inner face and outer face of the row are fixed, so only one additional vertex is necessary to construct a plane.  
%For the third vertex that is necessary to construct the plane, 
We iterate over all relevant vertices of the body segment to find the plane with the best support, using Eq. \ref{eq:quality} as a quality measure. However, if the angle between the new row intersection line and the previous row intersection line is too large, the face in the previous row could degenerate into a triangle. In such cases we reject the plane. 

Once a suitable plane has been found, the two inner vertices are then shifted along the corresponding intersection lines by a set distance to create the remaining two vertices of the outer face.
The resulting geometry is a connected 3x7 grid of non-planar quadrilateral faces fitted to the given body shape.

\paragraph{Special Case Handling}
~\label{sec:specialcase}
The proposed algorithm is capable of providing suitable solutions for basic sitting poses.
However, certain orientations of body parts in sitting poses can cause errors and require additional measures.
%The framework must be able to detect these special cases and adjust the surface fitting process accordingly.
In this stage, we identify two primary cases that require special attention:
Poses where the person is leaning forward as well as poses where the person's legs are in a crossed position.

In the first case, the back cannot be actively supported by a chair's backrest. This is easily detected by evaluating the vertex weights on the corresponding body parts. If the back does not need any support, no backrest is created.

\begin{figure}[t]
	\centering
	\includegraphics[width=0.45\textwidth]{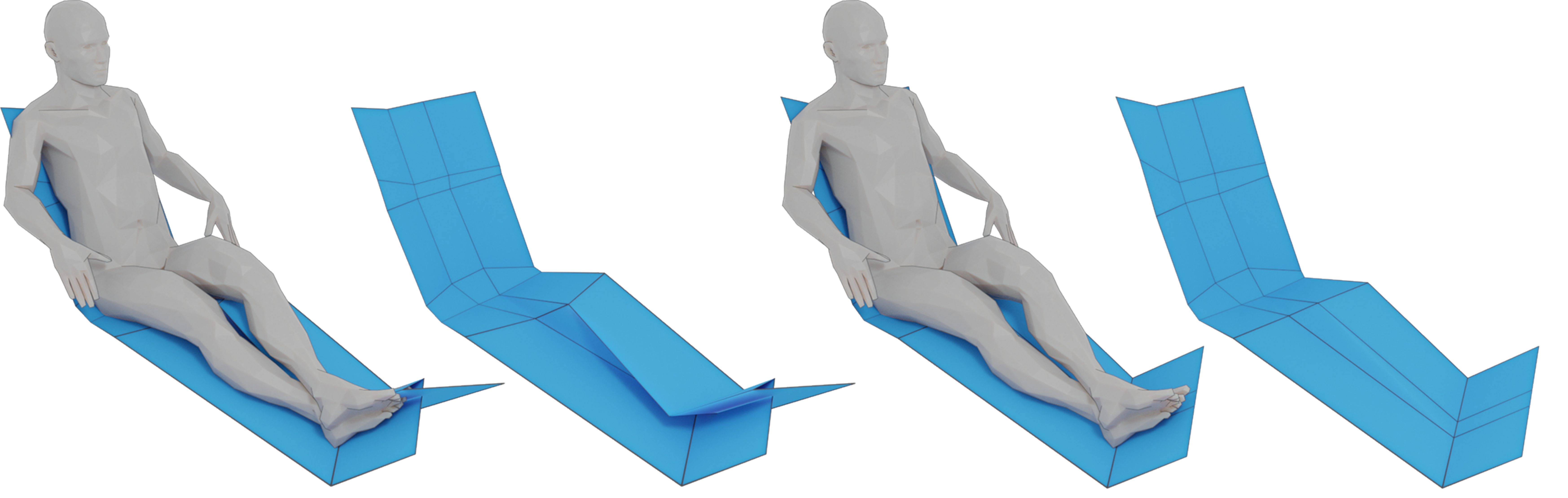}
	\setlength{\belowcaptionskip}{-11pt}
	\caption{Error handling on a pose where one foot rests on top of the other. Left: surface model generated without error handling showing intersections. Right: corrected surface mesh.}
	\label{fig:crossedlegs}
\end{figure}

To detect crossed legs, we evaluate the distance between the computed planes for the outer columns of the respective rows. When the distance is under a defined minimal value, we assume that it is not possible to support both legs individually and instead fit a single plane for the combined vertices of both legs. Figure~\ref{fig:crossedlegs} shows an example of a pose where one foot rests on top of the other, so the initial surface mesh needs to be corrected.

\paragraph{Refinement Stage}

In the final stage, we add armrests to the model if they are required and connect the borders of the model to the ground. We start by constructing the armrests:

First, the algorithm starts by finding optimal planes supporting the person's upper arms and forearms.
For this task, regular mesh fitting is performed on the respective body parts, using PCA and an unconstrained RANSAC variant. We then find the minimal spanning rectangle on the computed plane that contains all relevant vertices that lie within supporting distance of the plane. %This rectangular face represents the ideal armrest regarding comfort and visual quality.

The next step is the integration of the armrest into the mesh grid structure. This is only possible if the armrest does not intersect the body and if it is sufficiently far away from the mesh grid. If the requirements are met, two additional columns are added to the mesh, one containing the armrest itself and another to connect the first column to the geometry.

Finally, the mesh grid is expanded in each direction by two additional rows or columns of quadrilateral faces.
The outermost vertices of the resulting geometry are moved to ground height and arranged to form a rectangle.
In case the surface geometry contains overhanging faces, invalid quadrilateral faces in the outermost columns of the model are possible. To correct these issues, linear optimization is performed on the outer vertices on each side of the model.
This process rearranges the corresponding vertices so that each outer column face is convex.

\begin{figure}[t]
	\centering
	\includegraphics[width=0.40\textwidth]{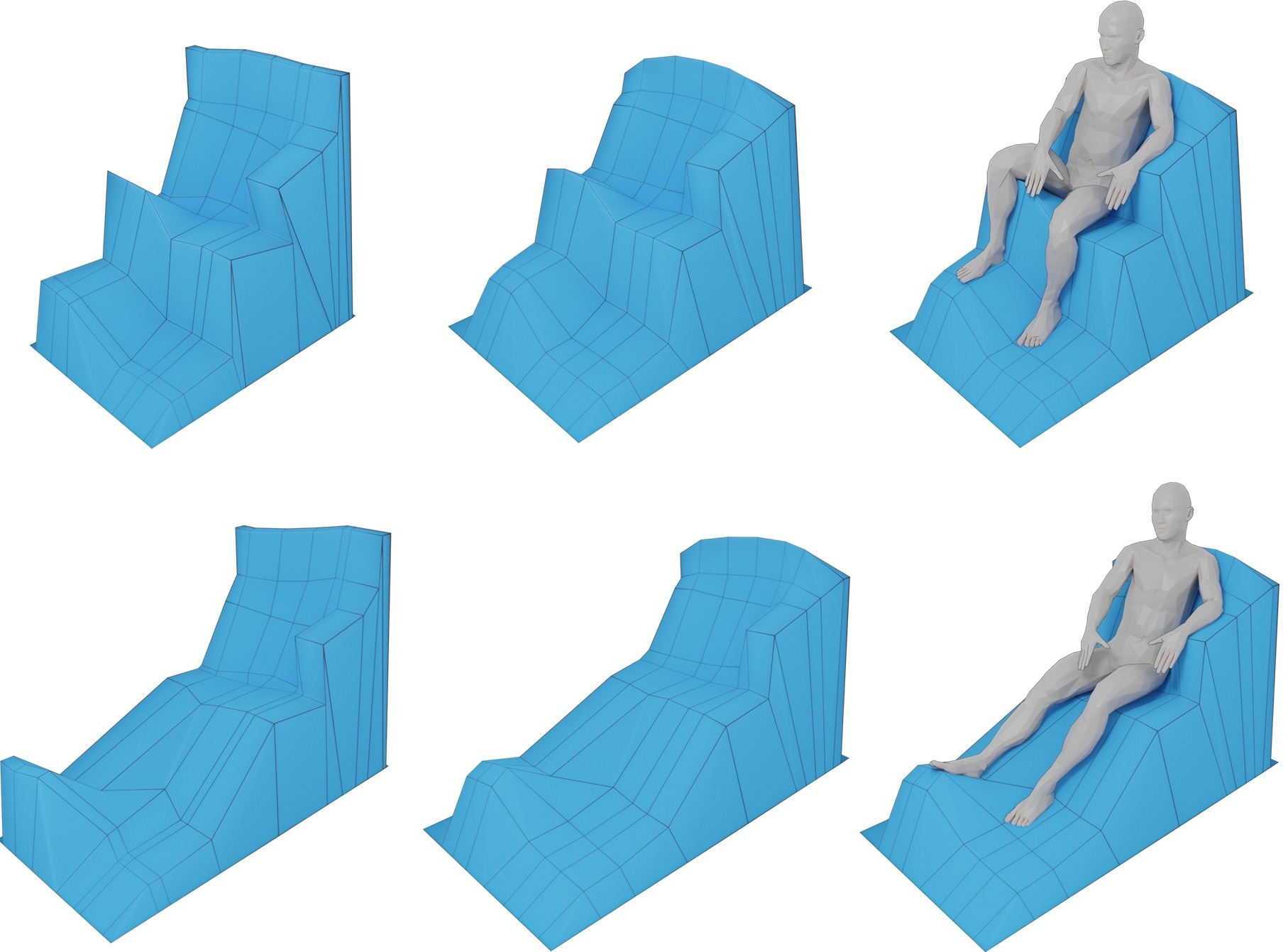}
	\setlength{\belowcaptionskip}{-11pt}
	\caption{Finalized seating surface results after the optimization process. Left: seating surface before optimization. Center/Right: results after optimization %(see Section \ref{chap:optimization})
	.}
	\label{fig:optim_results1}
\end{figure}

The left side of Figure~\ref{fig:optim_results1} shows visual examples for intermediate results generated from the advanced model after the refinement stage.
The added border sections are lacking in visual quality in regards to planarity and regularity.
%Therefore, further processing is required.
Therefore, we apply an additional optimization step in which we aim to smooth the geometry and improve the planarity and regularity of the faces, while keeping the functional aspects of the surface intact. %For detailed information about this optimization process, we refer the reader to the supplementary material. The results of the optimization step are illustrated in Figure~\ref{fig:optim_results1}.

 \begin{figure*}[t]
	\begin{subfigure}[t]{0.48\textwidth}
		\centering
		\includegraphics[width=0.30\textwidth]{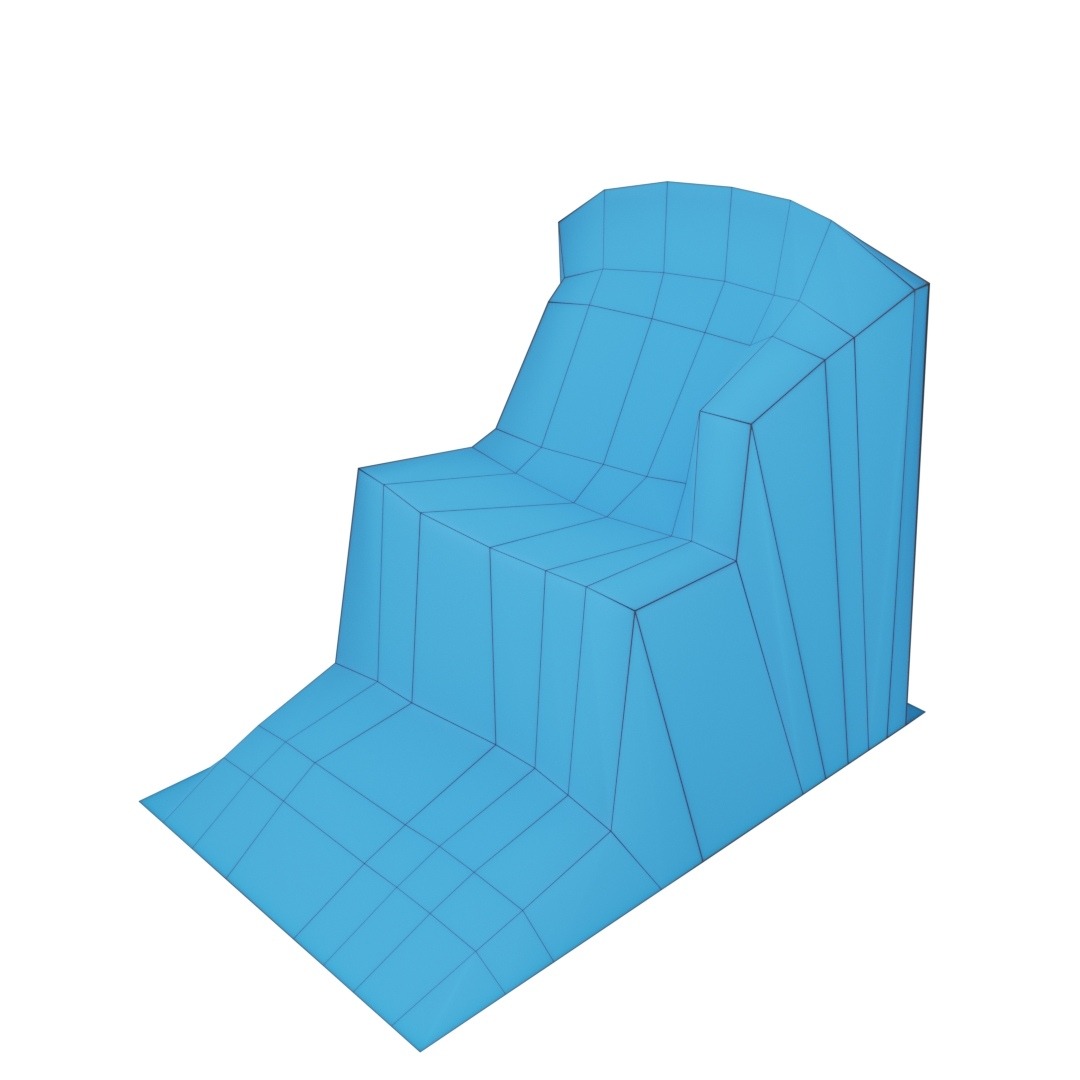}
		\includegraphics[width=0.30\textwidth]{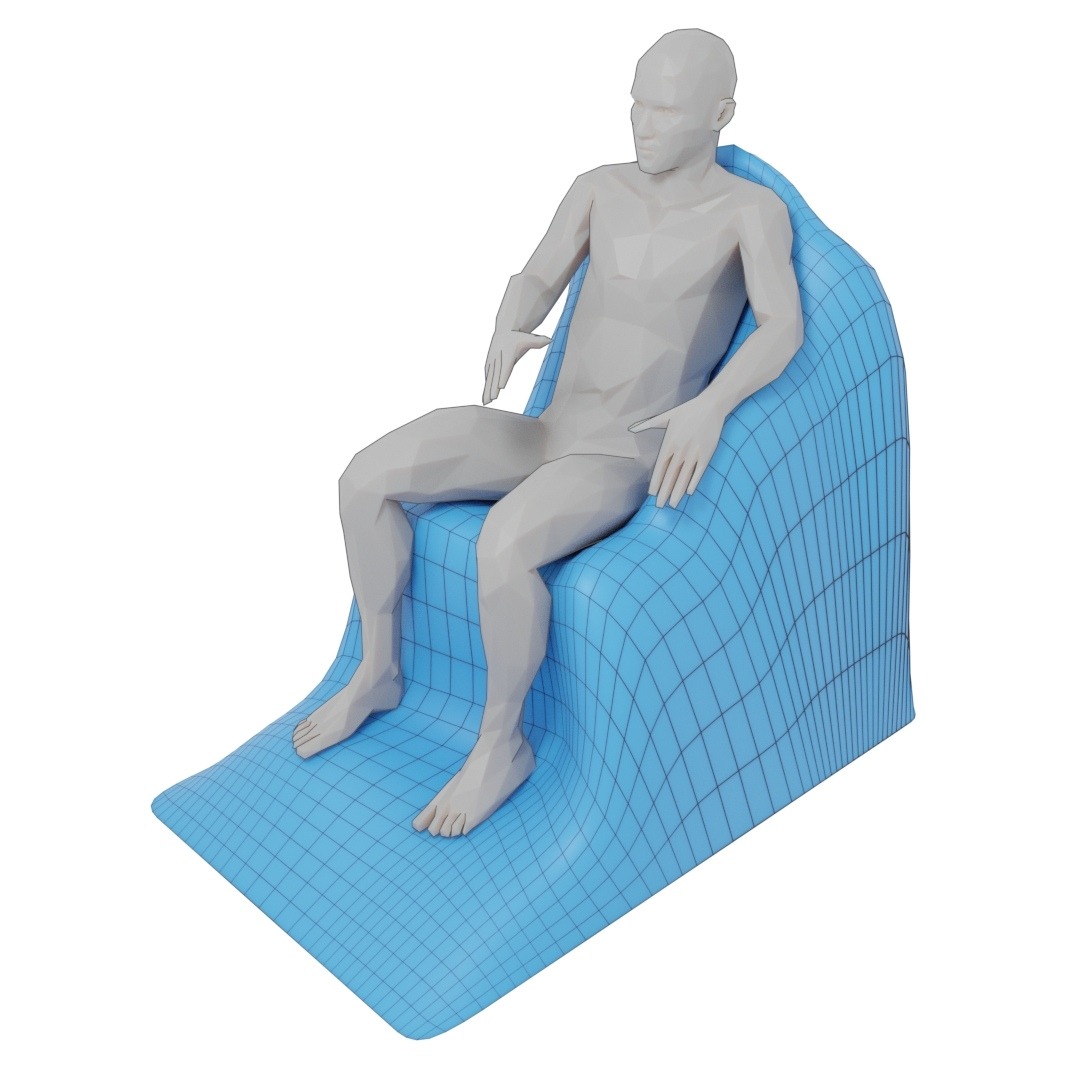}
		\includegraphics[width=0.30\textwidth]{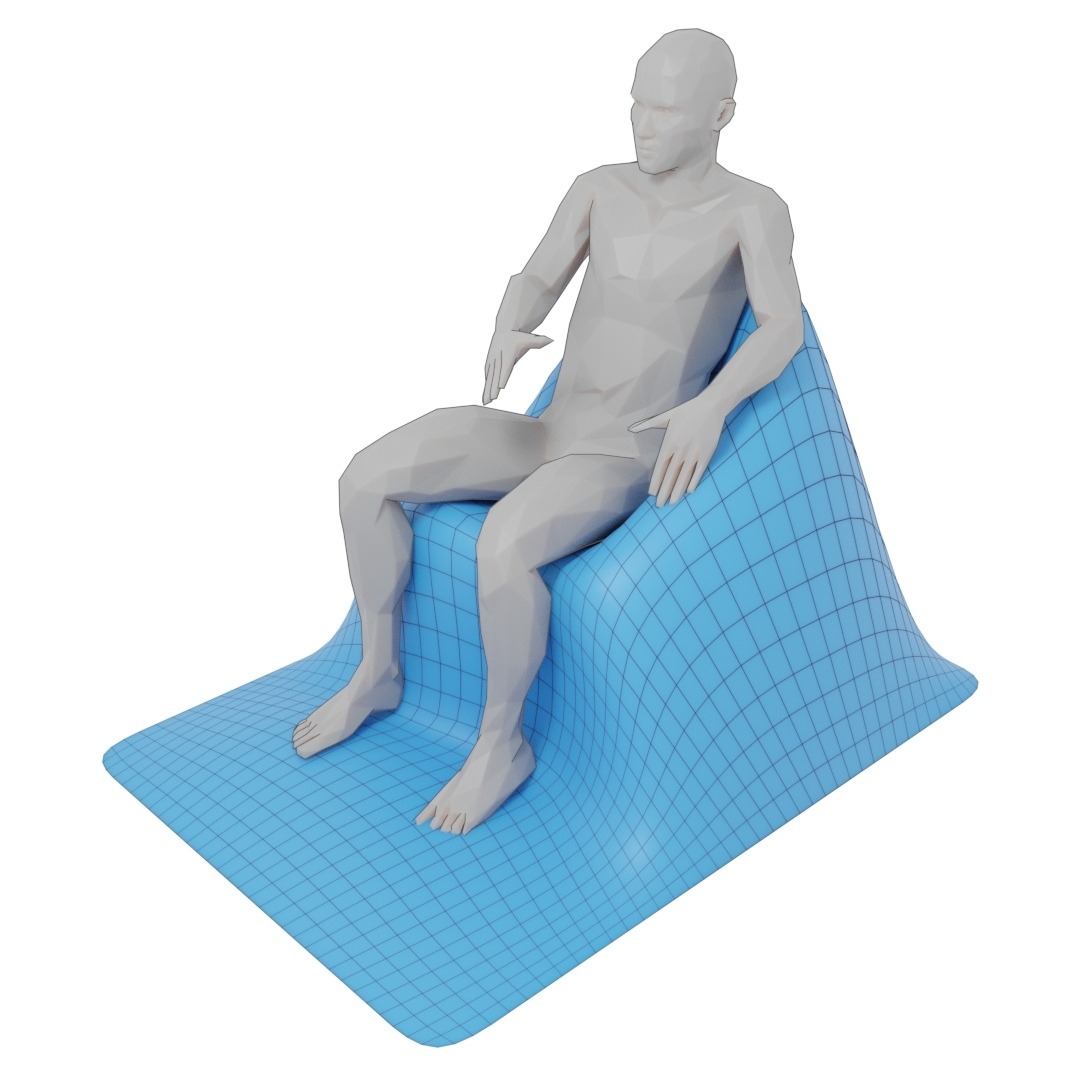}
		\captionsetup{justification=centering}
		\caption[caption]{Pose 1}
	\end{subfigure}\hfill
	%\hspace{0.5cm}
	\begin{subfigure}[t]{0.48\textwidth}
		\centering
		\includegraphics[width=0.30\textwidth]{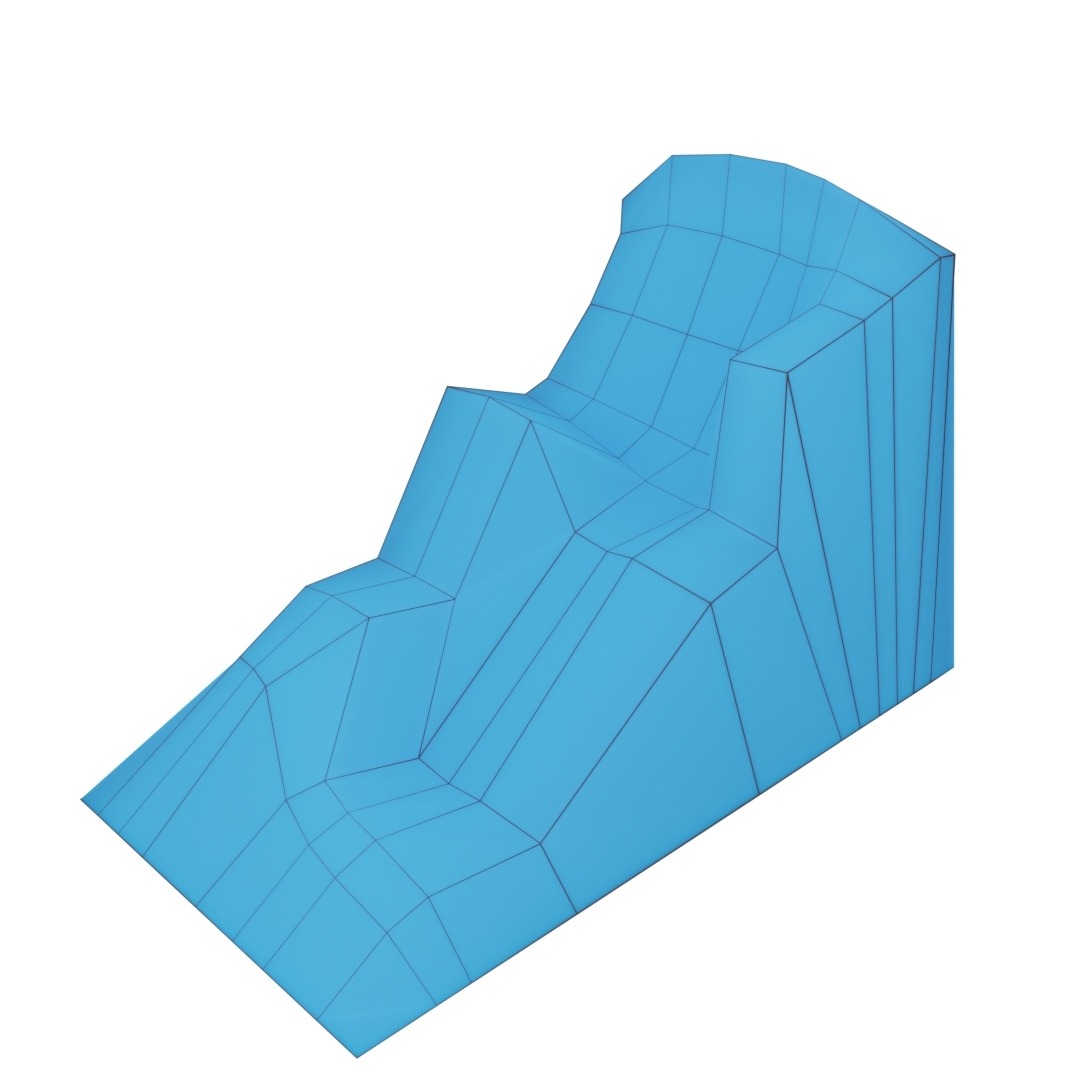}
		\includegraphics[width=0.30\textwidth]{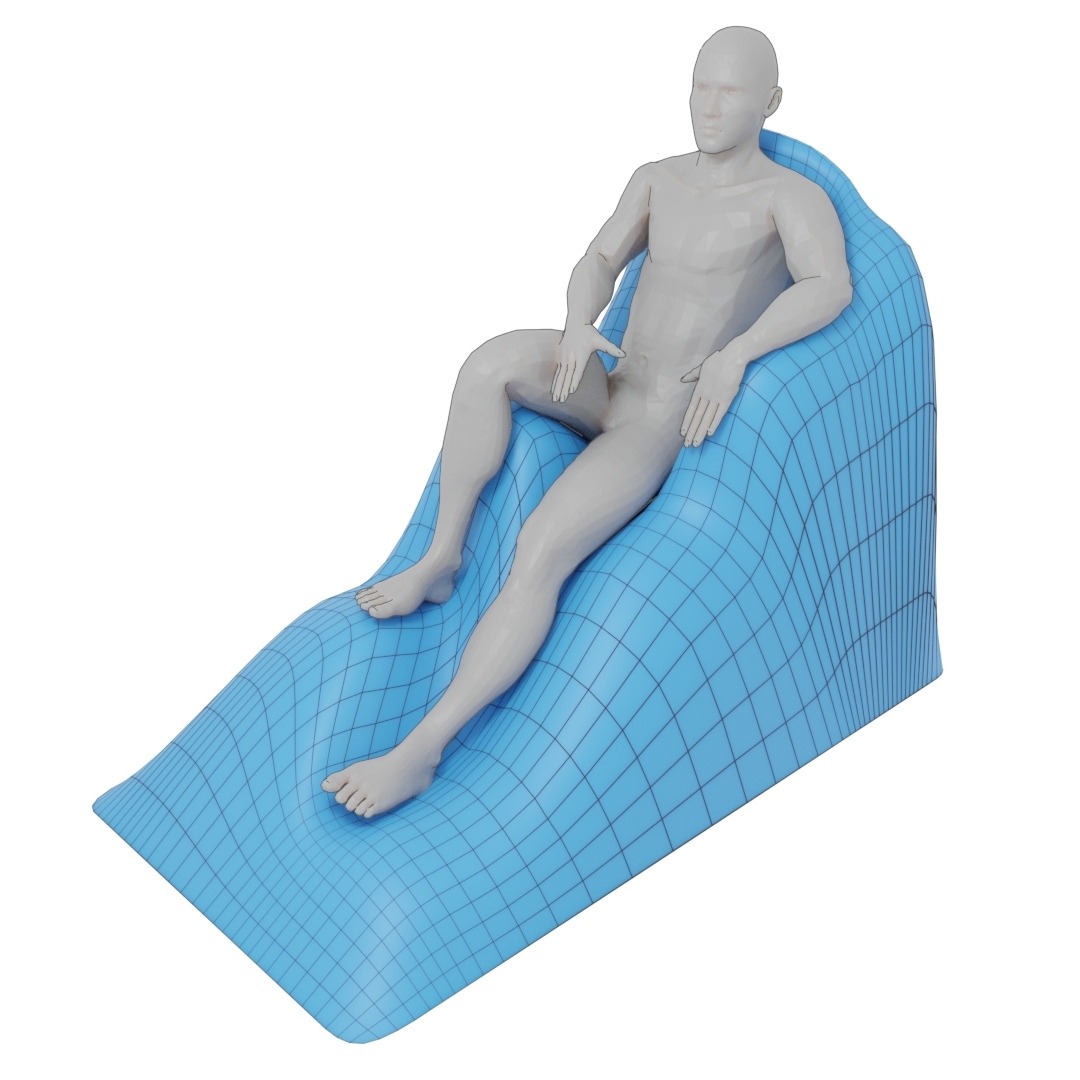}
		\includegraphics[width=0.30\textwidth]{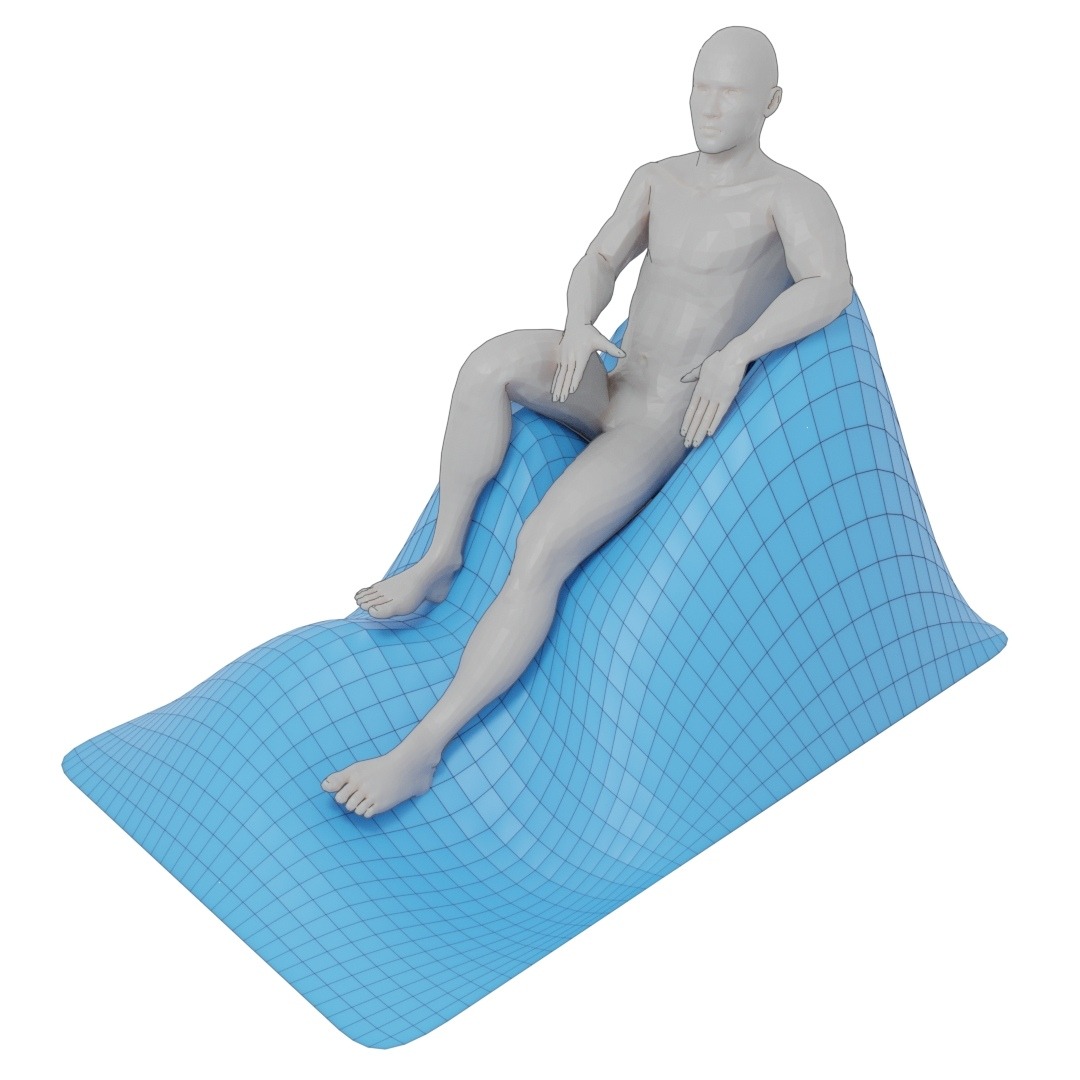}
		\captionsetup{justification=centering}
		\caption[caption]{Pose 2} 
	\end{subfigure}%\hfill
	\par
	\vspace{0.2cm}
	\begin{subfigure}[t]{0.48\textwidth}
		\centering
		\includegraphics[width=0.30\textwidth]{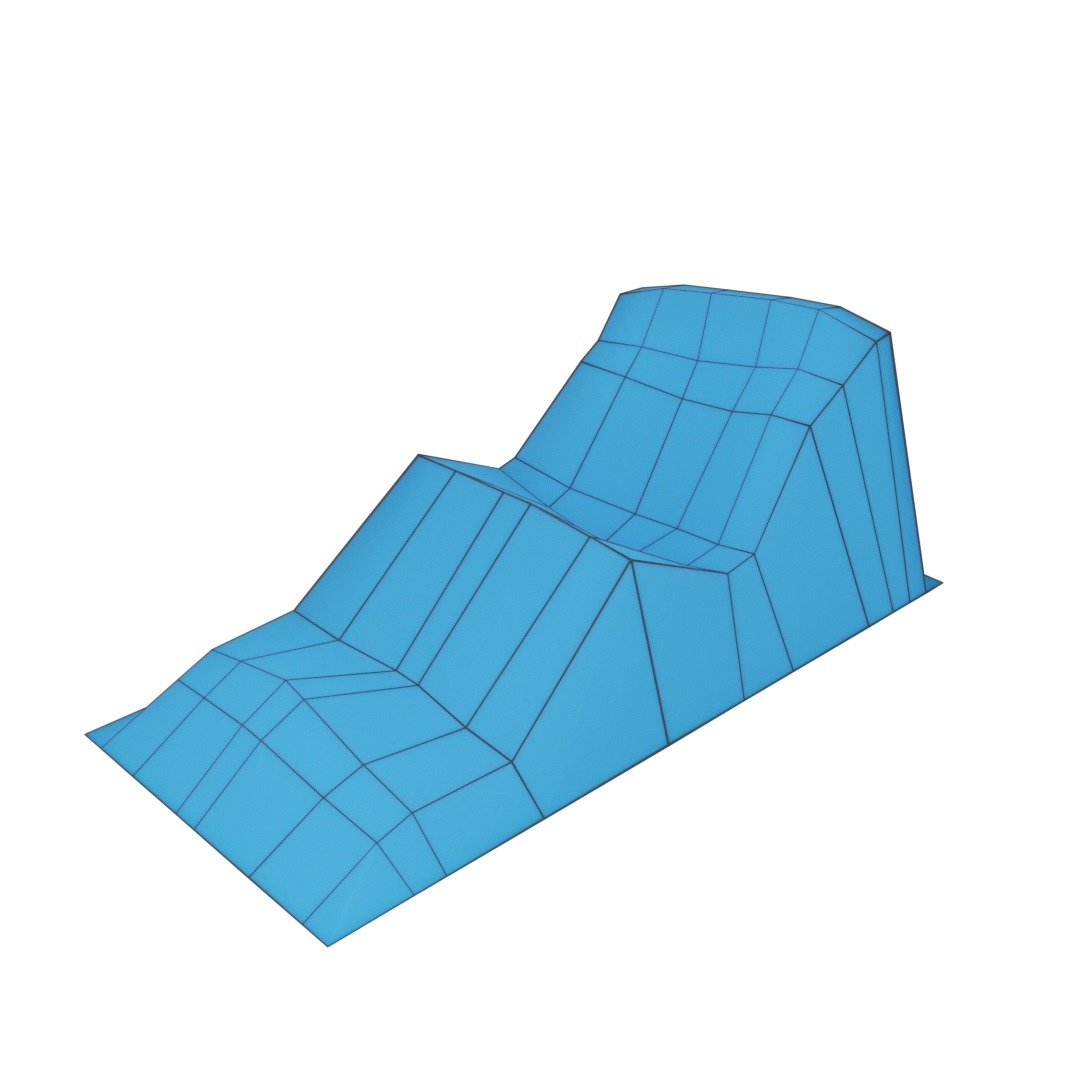}
		\includegraphics[width=0.30\textwidth]{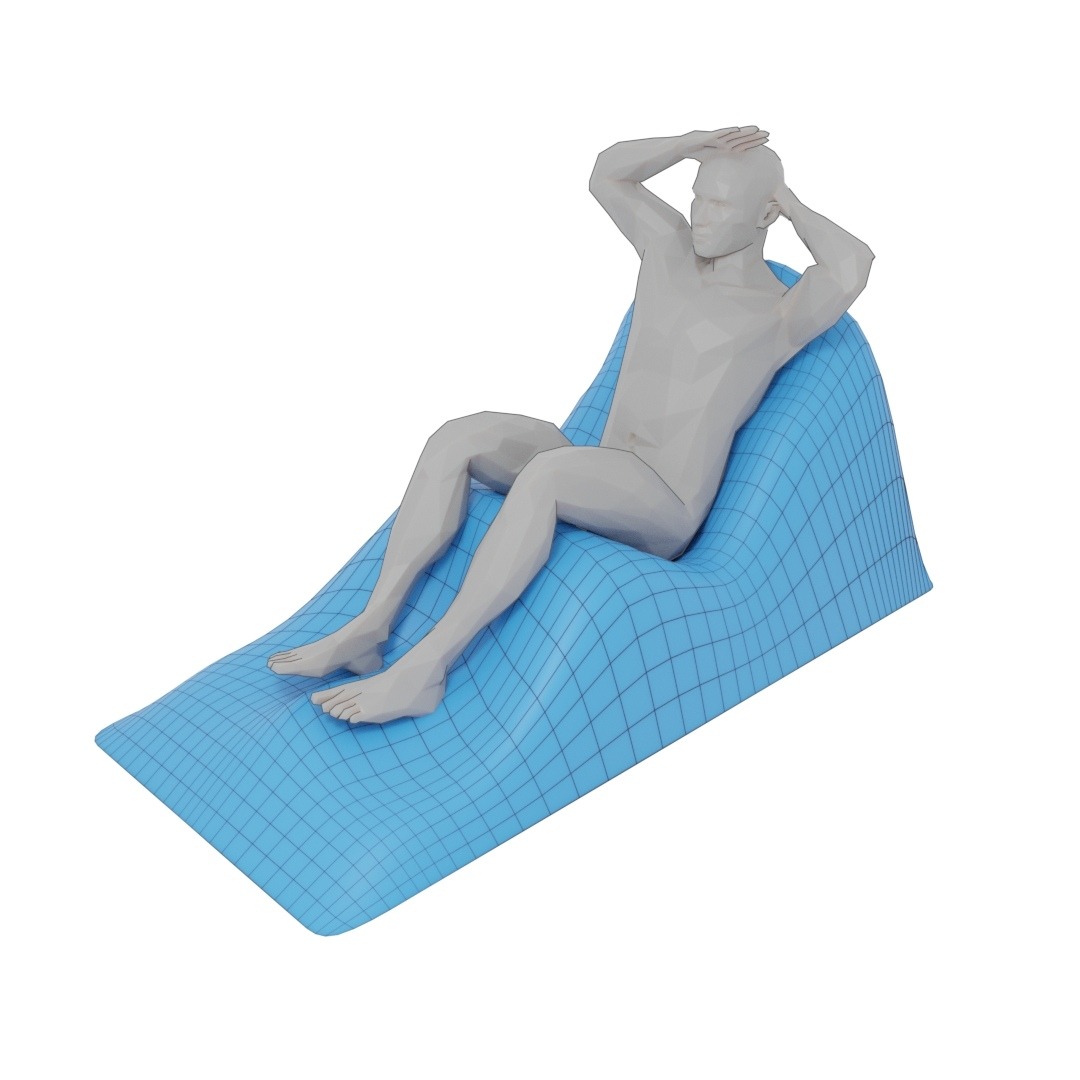}
		\includegraphics[width=0.30\textwidth]{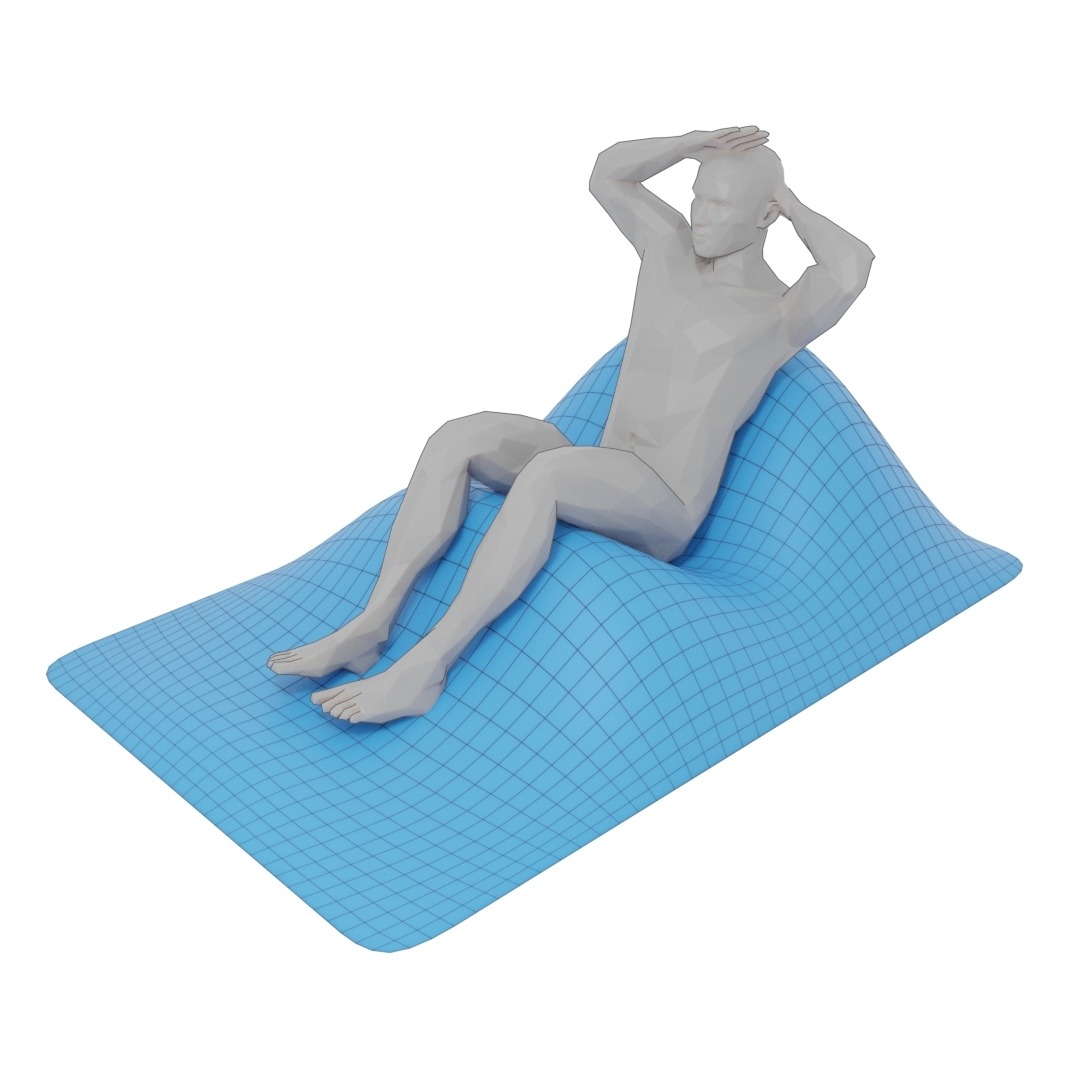}
		\captionsetup{justification=centering}
		\caption[caption]{Pose 3}
	\end{subfigure}\hfill
	\hspace{0.5cm}
	\begin{subfigure}[t]{0.48\textwidth}
		\centering
		\includegraphics[width=0.30\textwidth]{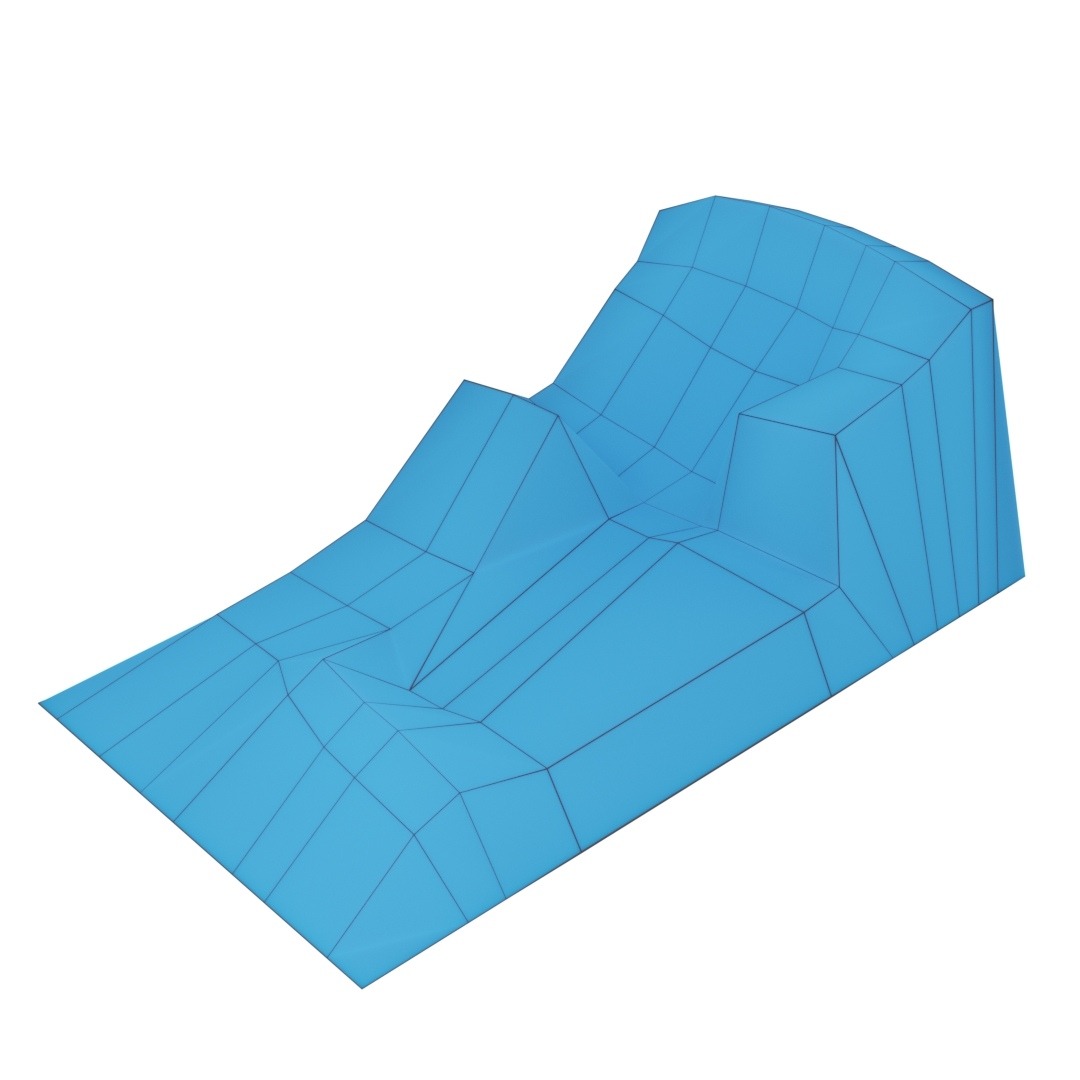}
		\includegraphics[width=0.30\textwidth]{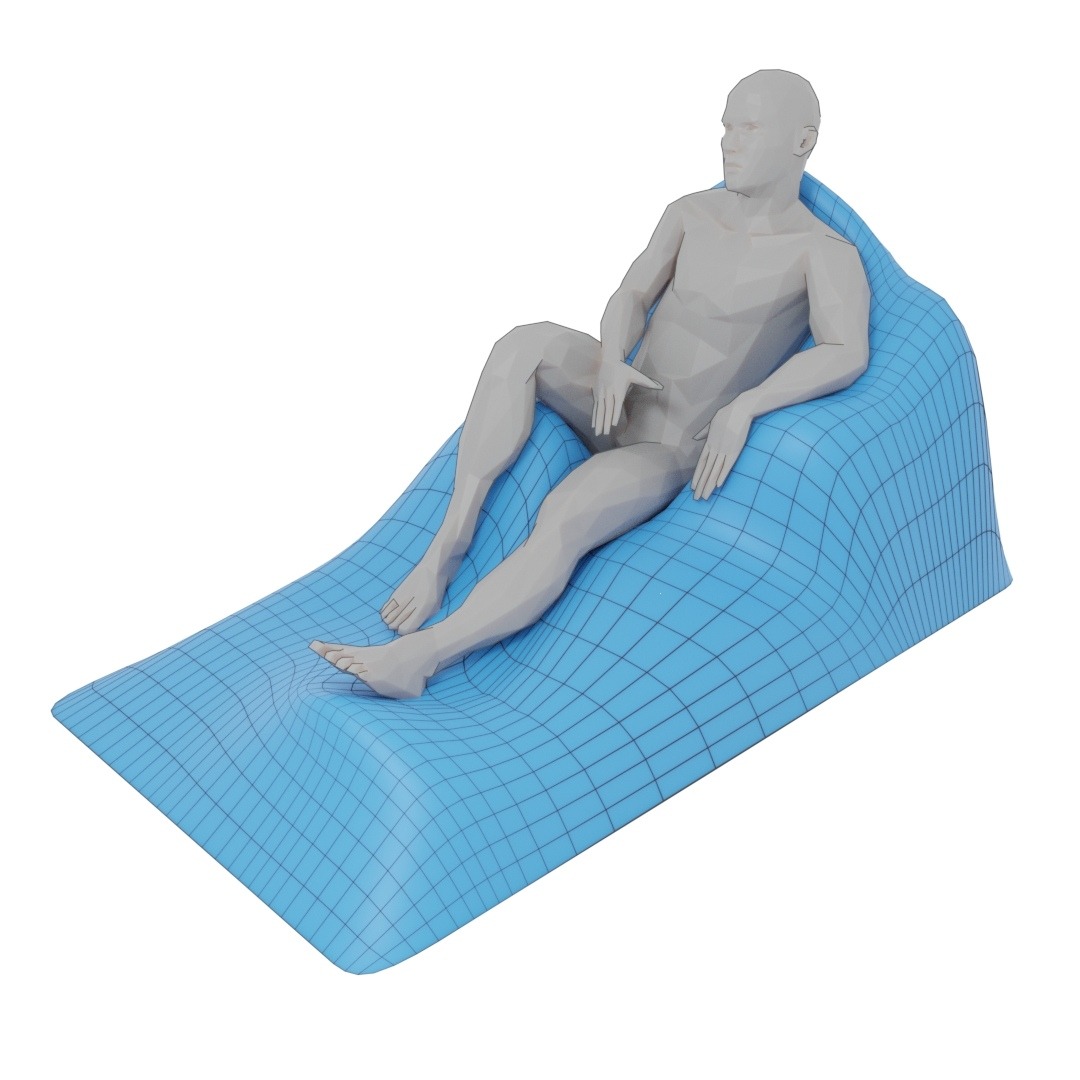}
		\includegraphics[width=0.30\textwidth]{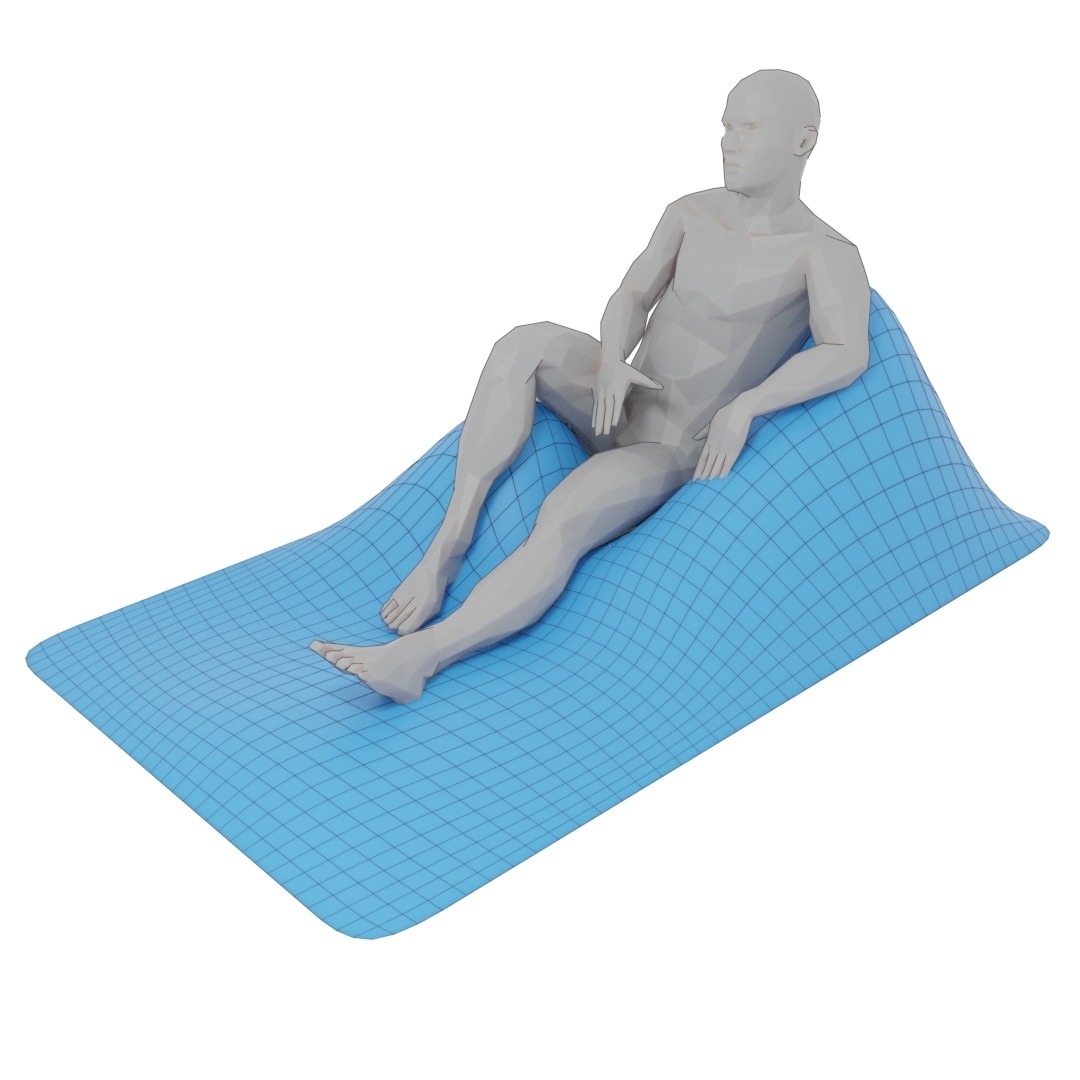}
		\captionsetup{justification=centering}
		\caption[caption]{Pose 4}
	\end{subfigure}%\hfill
	\par
	\vspace{0.2cm}
	\begin{subfigure}[t]{0.48\textwidth}
		\centering
		\includegraphics[width=0.30\textwidth]{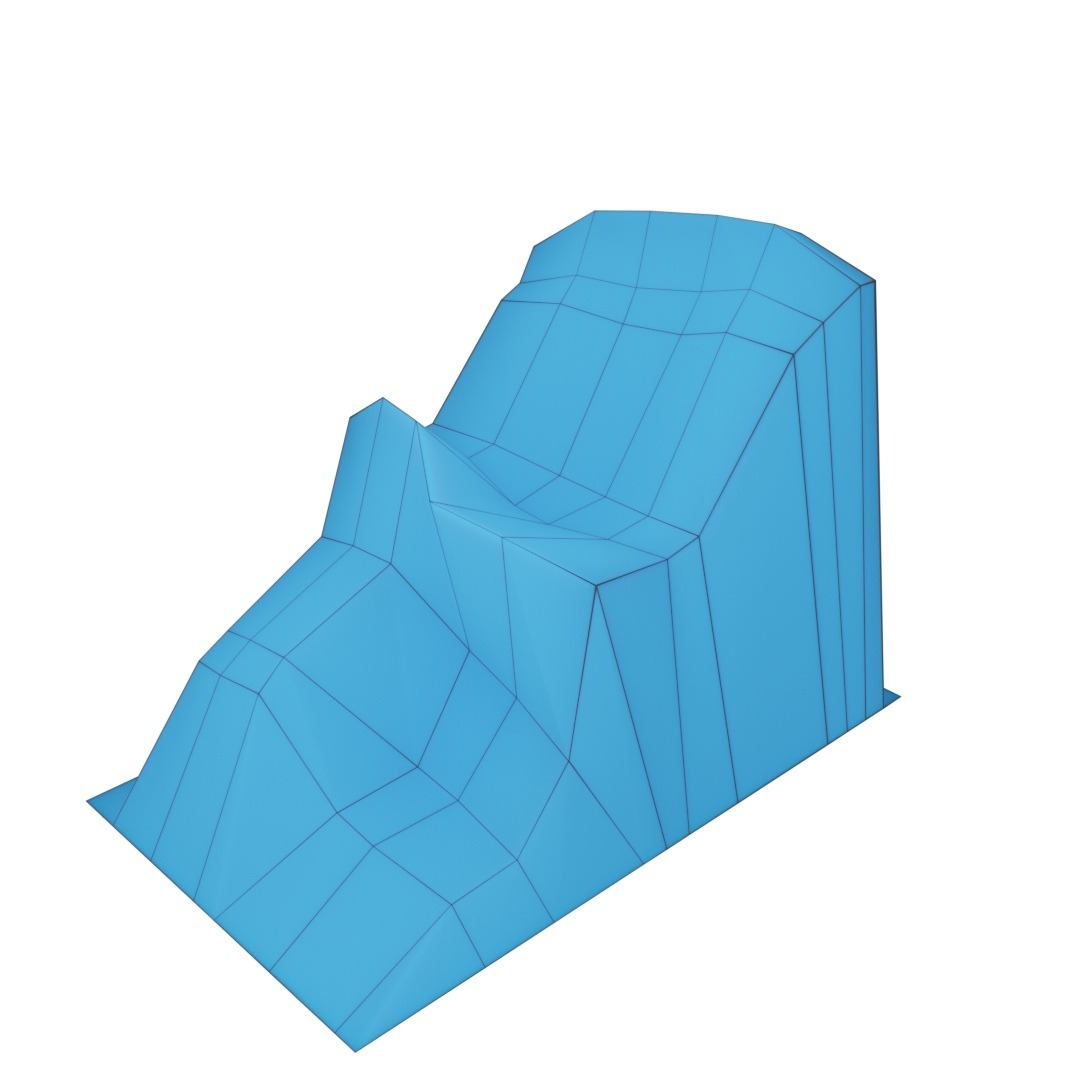}
		\includegraphics[width=0.30\textwidth]{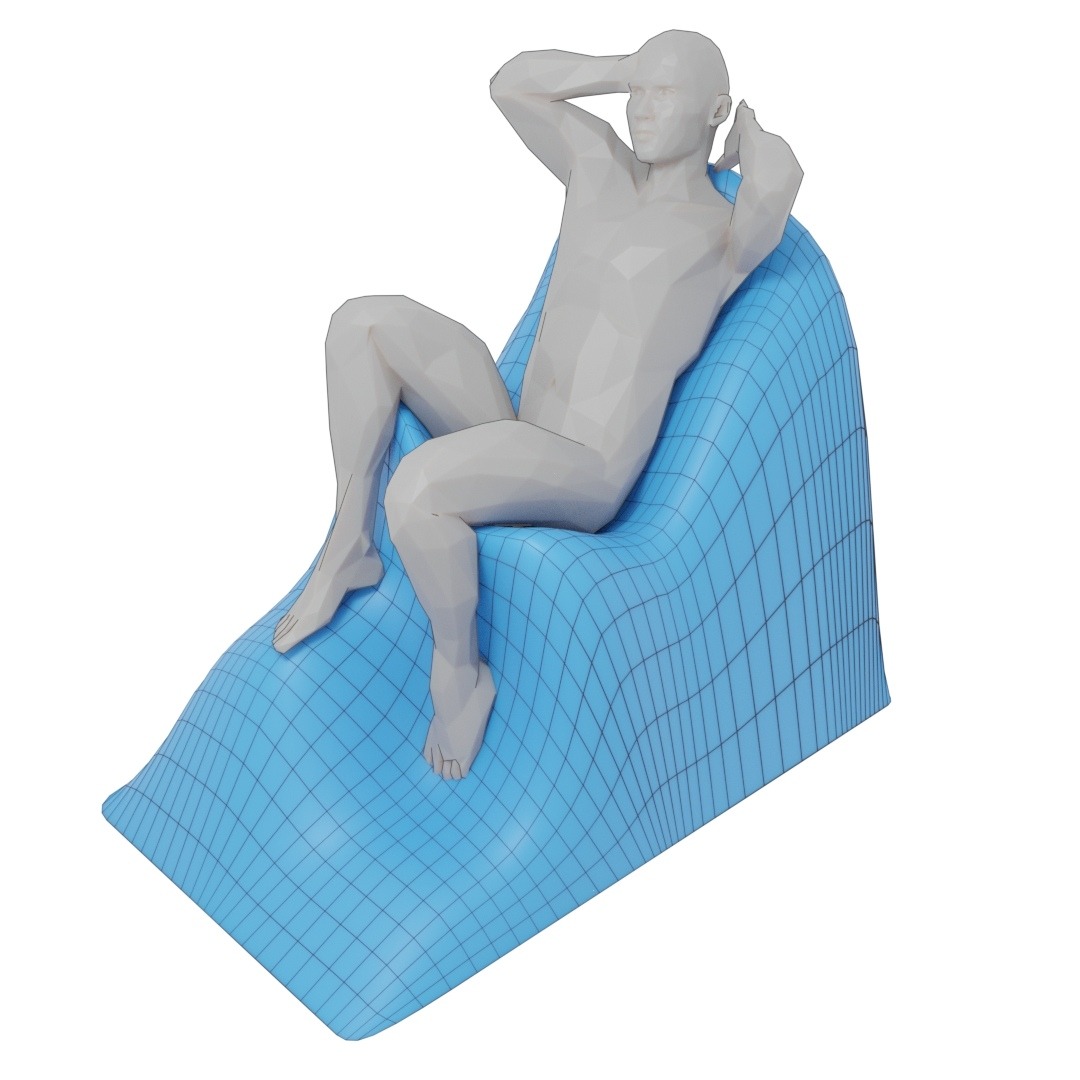}
		\includegraphics[width=0.30\textwidth]{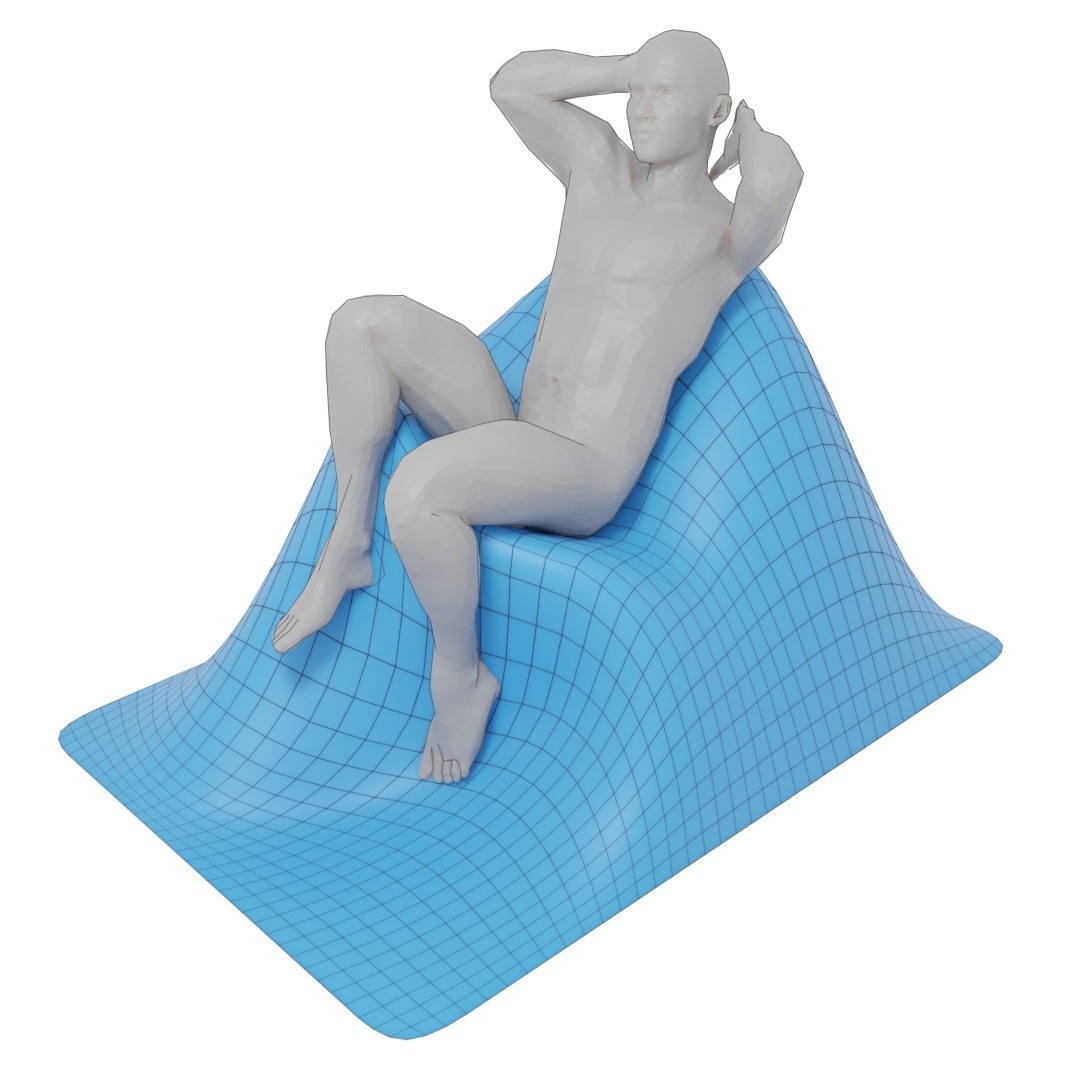}
		\captionsetup{justification=centering}
		\caption[caption]{Pose 5}
	\end{subfigure}\hfill
	%\hspace{0.5cm}
	\begin{subfigure}[t]{0.48\textwidth}
		\centering
		\includegraphics[width=0.30\textwidth]{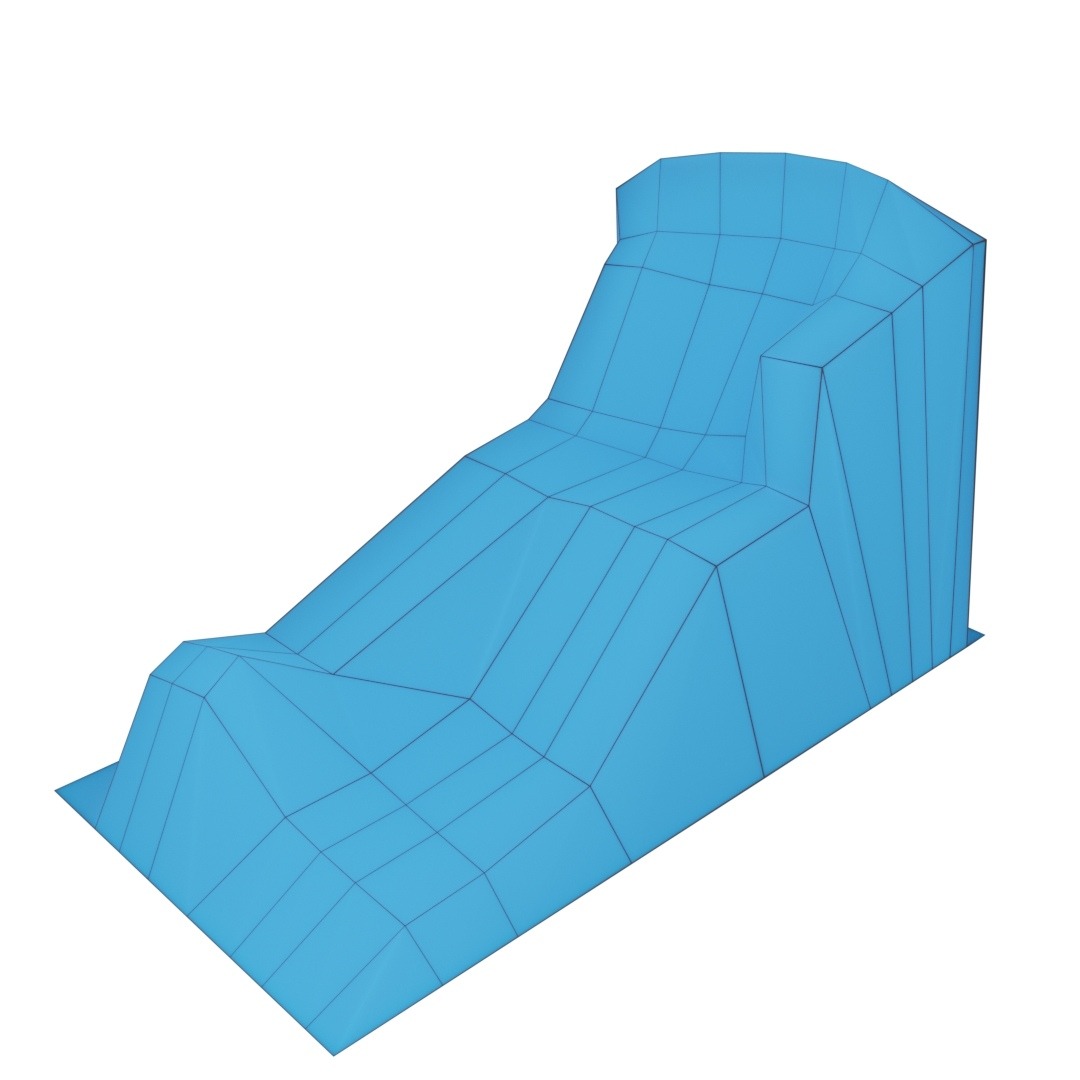}
		\includegraphics[width=0.30\textwidth]{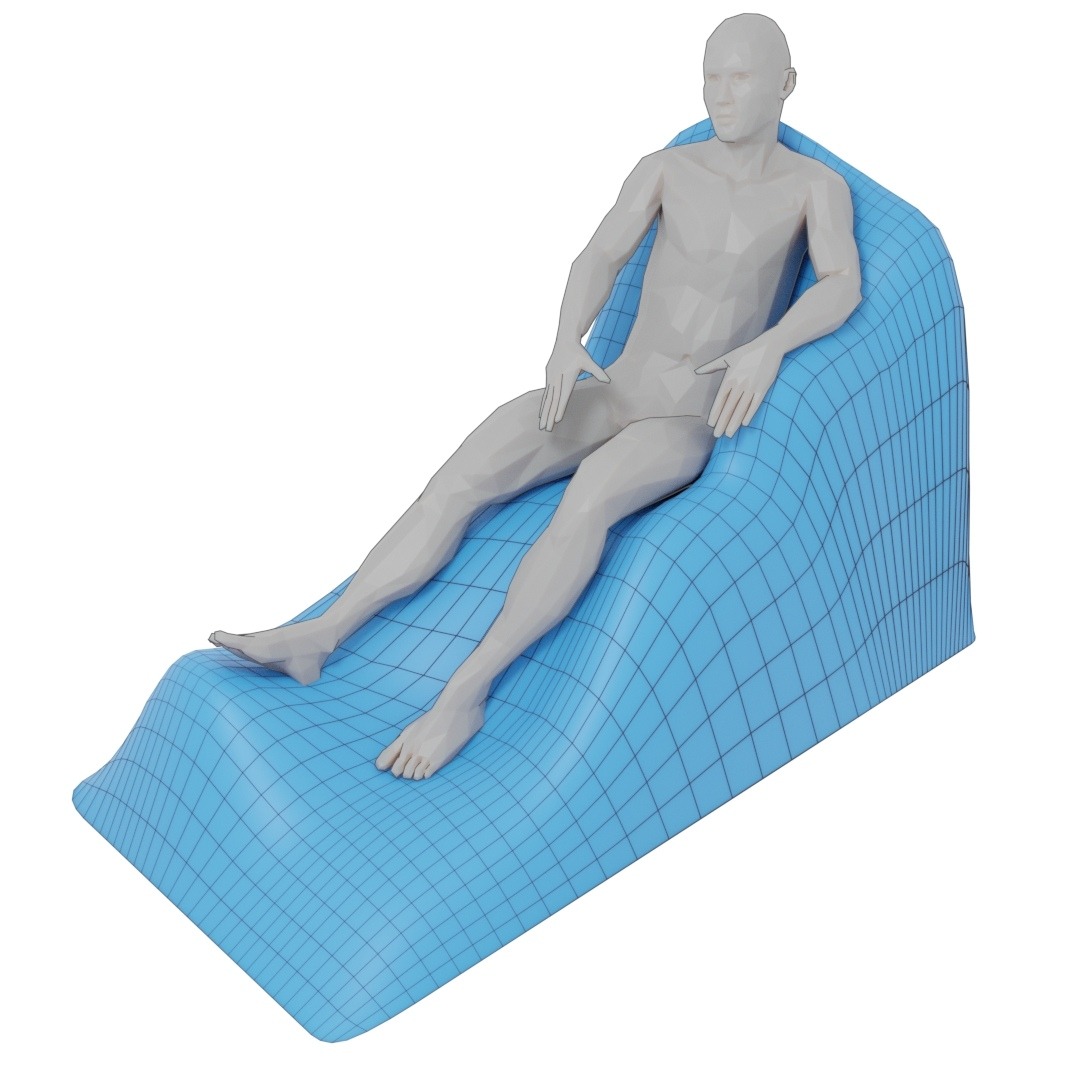}
		\includegraphics[width=0.30\textwidth]{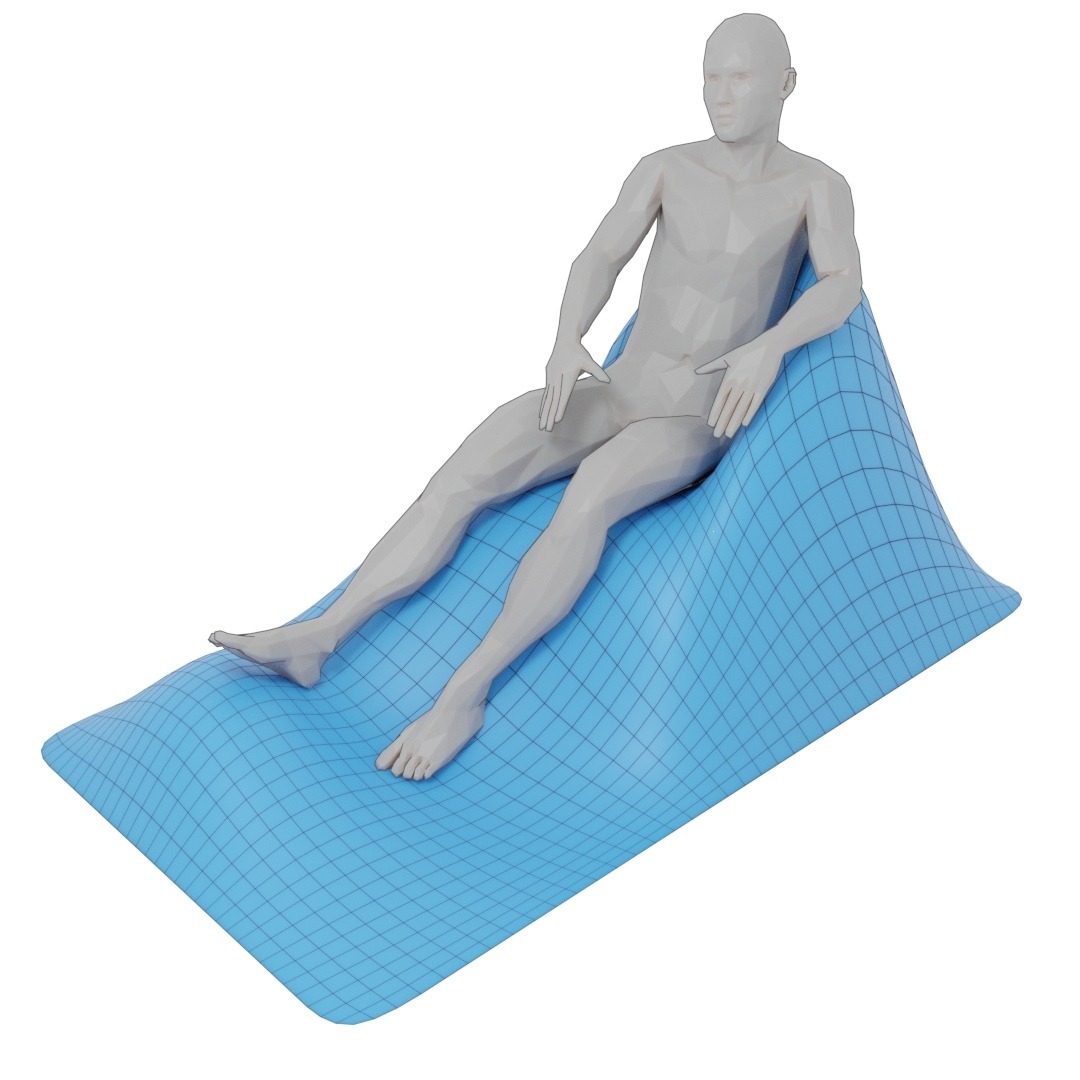}
		\captionsetup{justification=centering}
		\caption[caption]{Pose 6}
	\end{subfigure}%\hfill
	\caption{Results of our method. Left: control mesh generated with our method. Center: fitting algorithm of Leimer \etal~\cite{leimer2018sit} applied to our control mesh. Right: the method of Leimer \etal~applied to a flat patch serving as the control mesh.}
	\label{fig:results_pressure}
\end{figure*}

\subsection{Mesh Optimization}\label{sub:optimization}

While the functional requirements of our furniture model are now satisfied to an adequate degree, the visual mesh quality can still be improved. For this task we apply a non-linear local optimization process.

We formulate this as an energy minimization problem containing two terms. The first is the data term, which is used to preserve the initial configuration as much as possible, since it is the one that best satisfies the functional requirements. The second term is the mesh term, which describes the visual mesh quality regarding the smoothness of the surface as well as the regularity and planarity of its faces. 
The energy function is defined as
\begin{equation}
E = \lambda_S \, (S_L + S_A) +  \lambda_D \, (D_V + D_P) \,,
\end{equation}
where $(S_L + S_A)$ is the mesh term, $(D_V + D_P)$ is the data term and $\lambda_S$ and $\lambda_D$ are global weights balancing the two terms.

An \textit{error metric} based on the discrete Laplacian $S_L$ is computed as the sum of squared distances between the vertex positions and the average position of their neighboring vertices:
\begin{equation}
S_L = \sum_{i=1}^{n^V}\left\|  \mathbf{v}_i - \frac{\sum\limits_{j \in N_1(i)} \mathbf{v}_j w_j}{\sum\limits_{j \in N_1(i)} w_j} \right\|^2 \lambda_S^l
\label{eq:ST_lapl}
\end{equation}
with $N_1(i)$ being the 1-ring neighborhood of $\mathbf{v}_i$, $w_j$ being the importance weight for $\mathbf{v}_j$ and $\lambda_S^l$ being a global weight for the Laplacian error metric term.

The \textit{angle based smoothing} term $S_A$ is defined as %consists of two separate error metrics as described in Section~\ref{sec:qualitymeasures}. The angle based error metrics are accumulated over the faces of the surface model.
\begin{equation}
S_{A} = \sum_{j=1}^{n^F} \left(\left( \sum_{i \in F_j}\alpha_i - 2\pi\right)^2  w^1_j  \lambda_S^{A_1} + \left(\sum_{i \in F_j}(\alpha_i - \pi) ^2 \right )  w^2_j \lambda_S^{A_2}\right)
\label{eq:ST_angle}
\end{equation}
with $n^F$ being the total number of faces, $\alpha_i$ being the $i$th interior angle of the face $F_j$, $w^1_j$ and $w^2_j$ being term-specific importance weights for each face, and $\lambda_S^{A_1}$ and $\lambda_S^{A_2}$ being global weights. The first part of the term penalizes non-regular faces, while the second part aims to maximize each interior angle.

The \textit{vertex distance term} $D_V$ is computed from the sum of squared distances between the vertex positions of the current configuration and their corresponding original positions:
\begin{equation}
D_V = \lambda_D^V \sum_{i=1}^{n^v}  \left \| \mathbf{v}_i - \tilde{\mathbf{v}}_i \right \|^2  w_i
\end{equation}
with $\tilde{\mathbf{v}}_i$ being the original position of vertex $\mathbf{v}_i$, $w_i$ being the importance weight of $\mathbf{v}_i$, and $\lambda_D^V$ being a global weight for the term.

Finally, the \textit{plane distance term} $D_P$ utilizes the supporting planes that were computed in the mesh fitting stage of the algorithm. Each face in the current configuration is compared to its supporting plane by computing the distance to the plane for each corner vertex: %The same vertex importance weights are applied to each plane distance value. The plane distance error metric is the sum of plane distances weighted by a global scaling factor.
\begin{equation}
D_P = \lambda_D^P \, \sum_{j=1}^{n^F}\left( \sum_{i \in F_j} \, w_i \, \left \langle \mathbf{v}_i -\mathbf{c}^P_j, \mathbf{n}^P_j \right \rangle^2 \right) ,
\end{equation}
where $\mathbf{c}^P_j$ and $\mathbf{n}^P_j$ are the center position and surface normal of the supporting plane for face $F_j$, $w_i$ is the importance weight of vertex $\mathbf{v}_i$, and $\lambda_D^P$ is a global weight for the term.

The vertex weights are chosen such that the data term is given more importance for vertices belonging to faces that support a large area of the body, while other vertices can be moved more freely to improve the mesh term. We furthermore add $2$ kinds of hard constraints: first, we need to constrain the position of the border vertices to stay on the edges of the rectangular base, and second, we define a minimal edge length between vertices to prevent degeneration of the geometry.

To improve the performance of solving the optimization problem, we furthermore compute the analytical gradient of the objective function. A detailed description of the gradient can be found in the supplementary material. To solve the problem, we use \matlab{}'s \texttt{fmincon} function. A comparison of results from before and after the optimization can be seen in Figure \ref{fig:optim_results1}.

\section{Results and Discussion}\label{sec:results}

\begin{figure}[t!]
	\centering
	\iflowres
	\includegraphics[width=.22\textwidth]{./renders/replacements/lores_f1_simple}
	\includegraphics[width=.22\textwidth]{./renders/replacements/lores_f30_simple}
	\else
	\includegraphics[width=.22\textwidth]{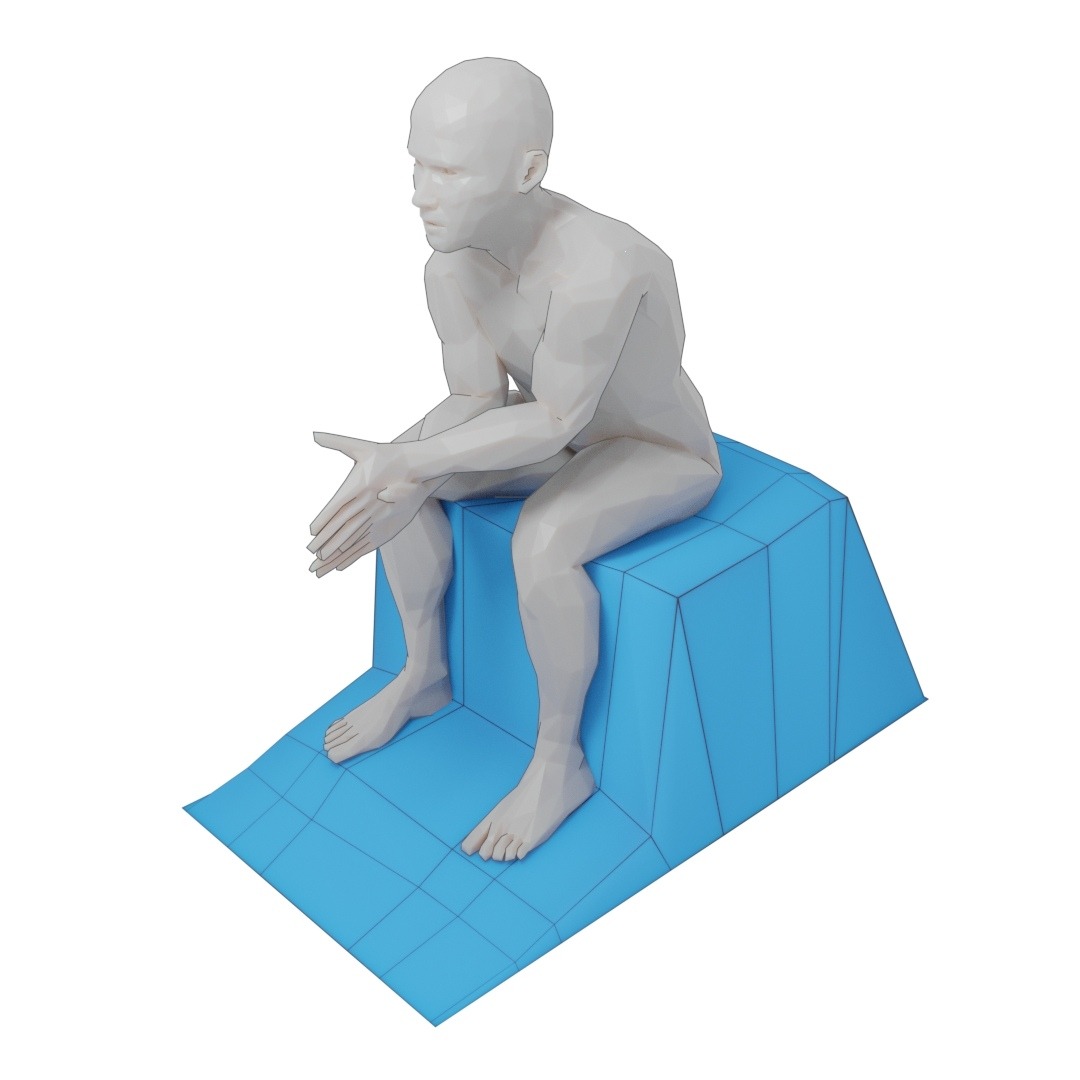}
	\includegraphics[width=.22\textwidth]{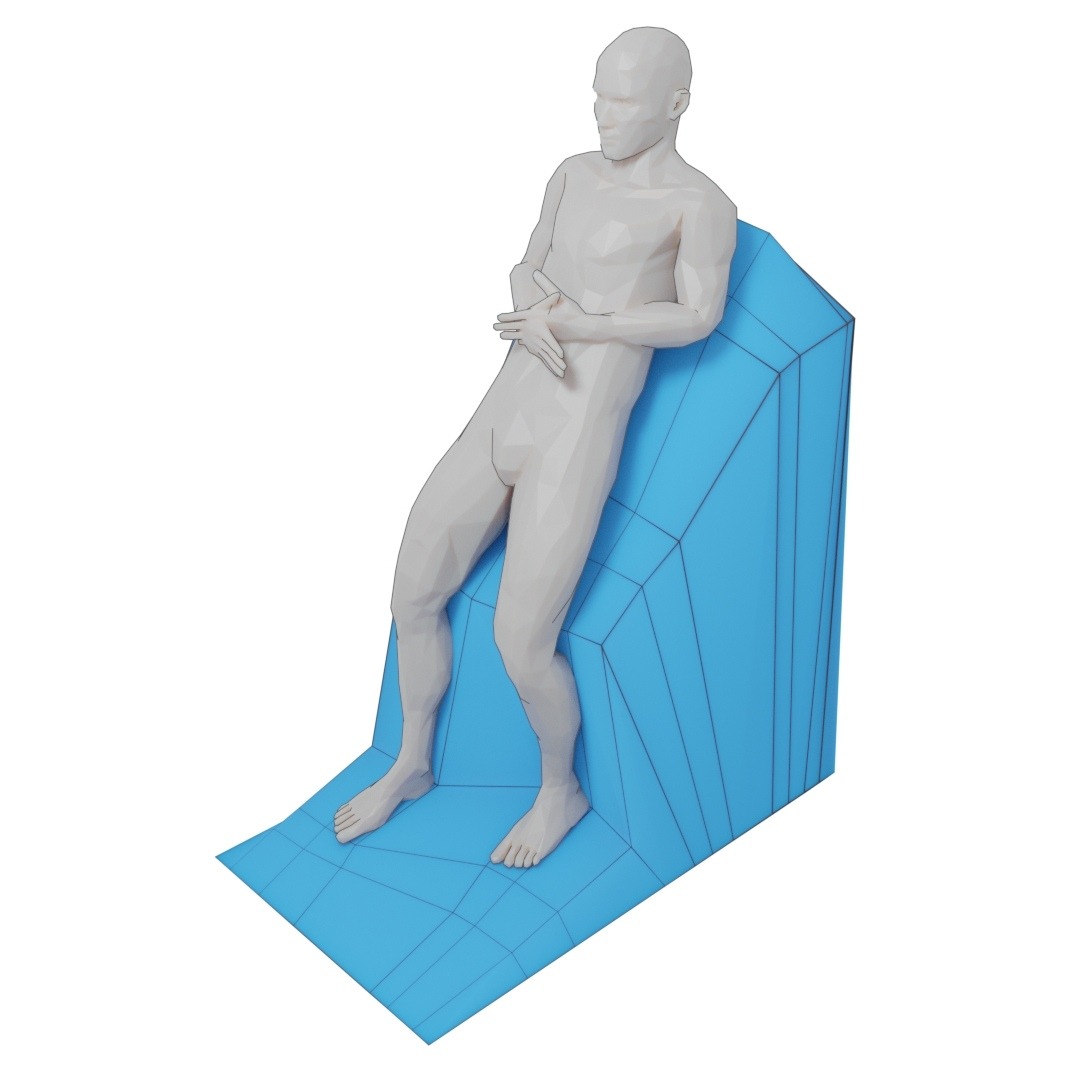}
	\fi
	\setlength{\belowcaptionskip}{-4pt}
	\caption{Forward-leaning poses and leaning back while standing are also supported by our method.}
	\label{fig:results_special}
\end{figure}

We apply our surface generation algorithm to a number of different poses to create a variety of body-supporting surfaces. We furthermore apply the surface fitting algorithm of Leimer \etal~\cite{leimer2018sit} using our generated surfaces as the input for the control mesh and compare the results to surfaces created using a flat patch as the control mesh, as is usually the case in their work. 

\begin{figure*}[t!]
	\begin{subfigure}[t]{0.96\textwidth}
		\centering
		\includegraphics[width=0.16\textwidth,trim=1cm 0.5cm 1cm 0.2cm,clip]{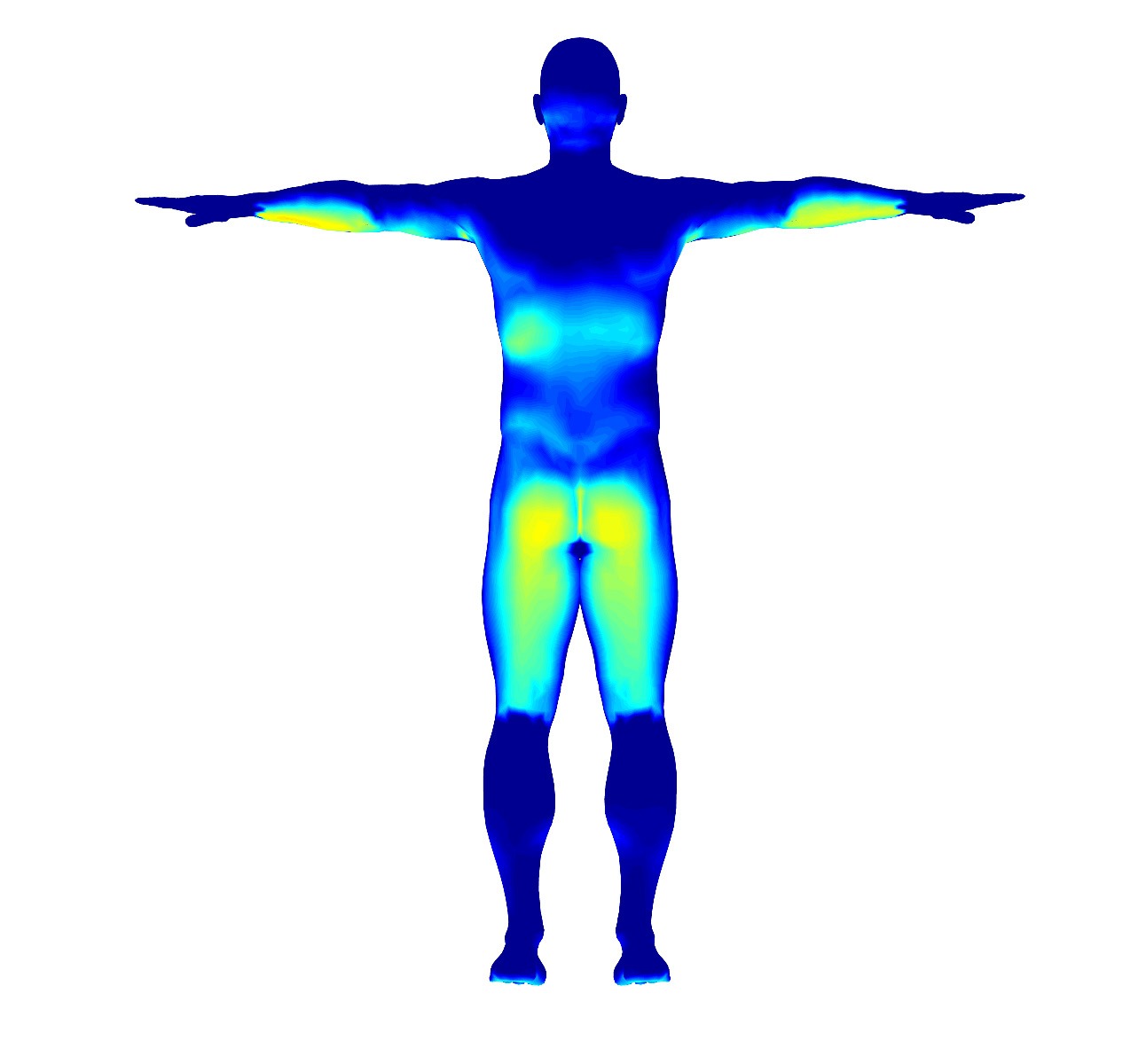}
		\scalebox{-1}[1]{\includegraphics[width=0.16\textwidth,trim=1cm 0.5cm 1cm 0.2cm,clip]{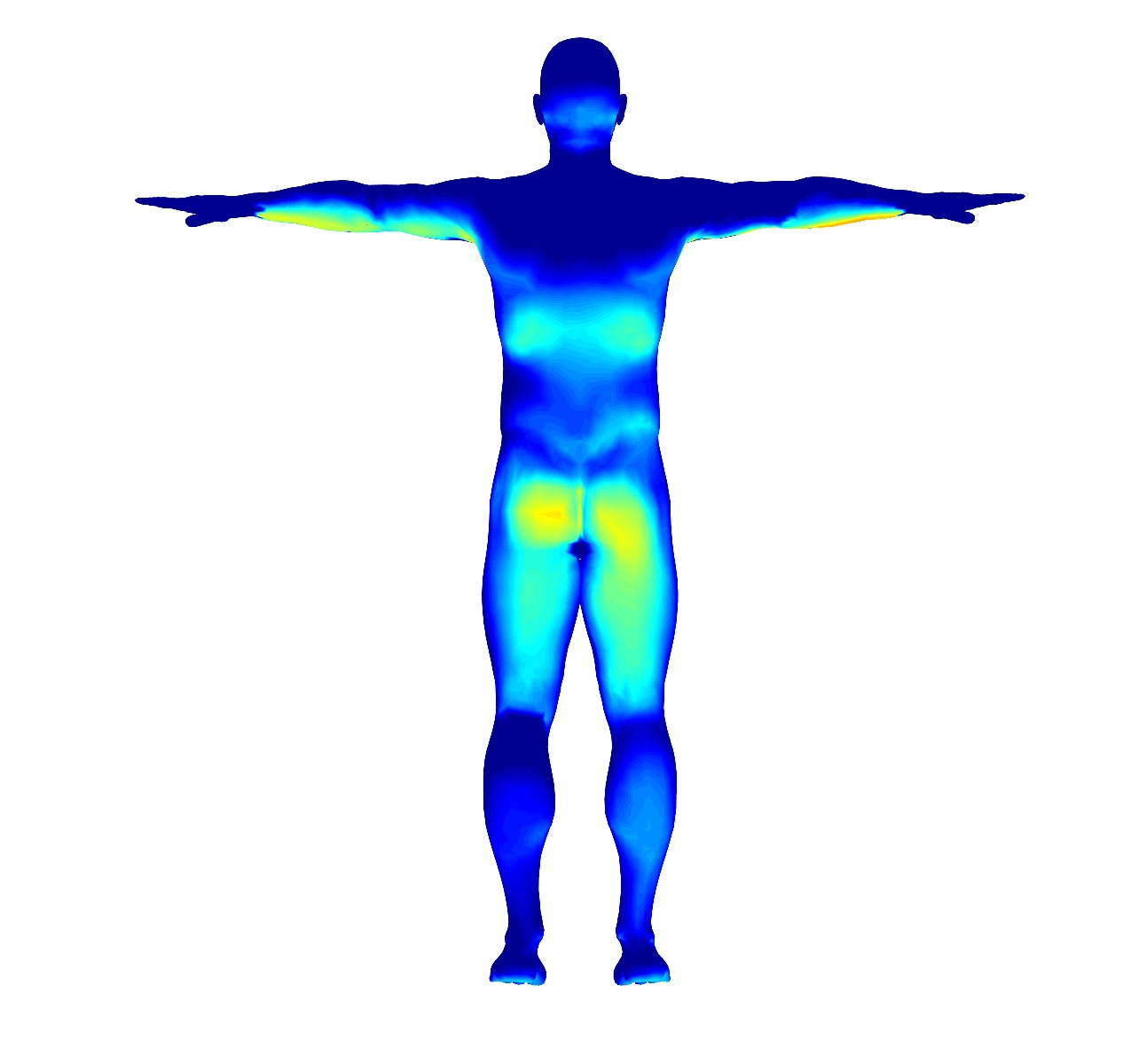}}
		\includegraphics[width=0.16\textwidth,trim=1cm 0.5cm 1cm 0.2cm,clip]{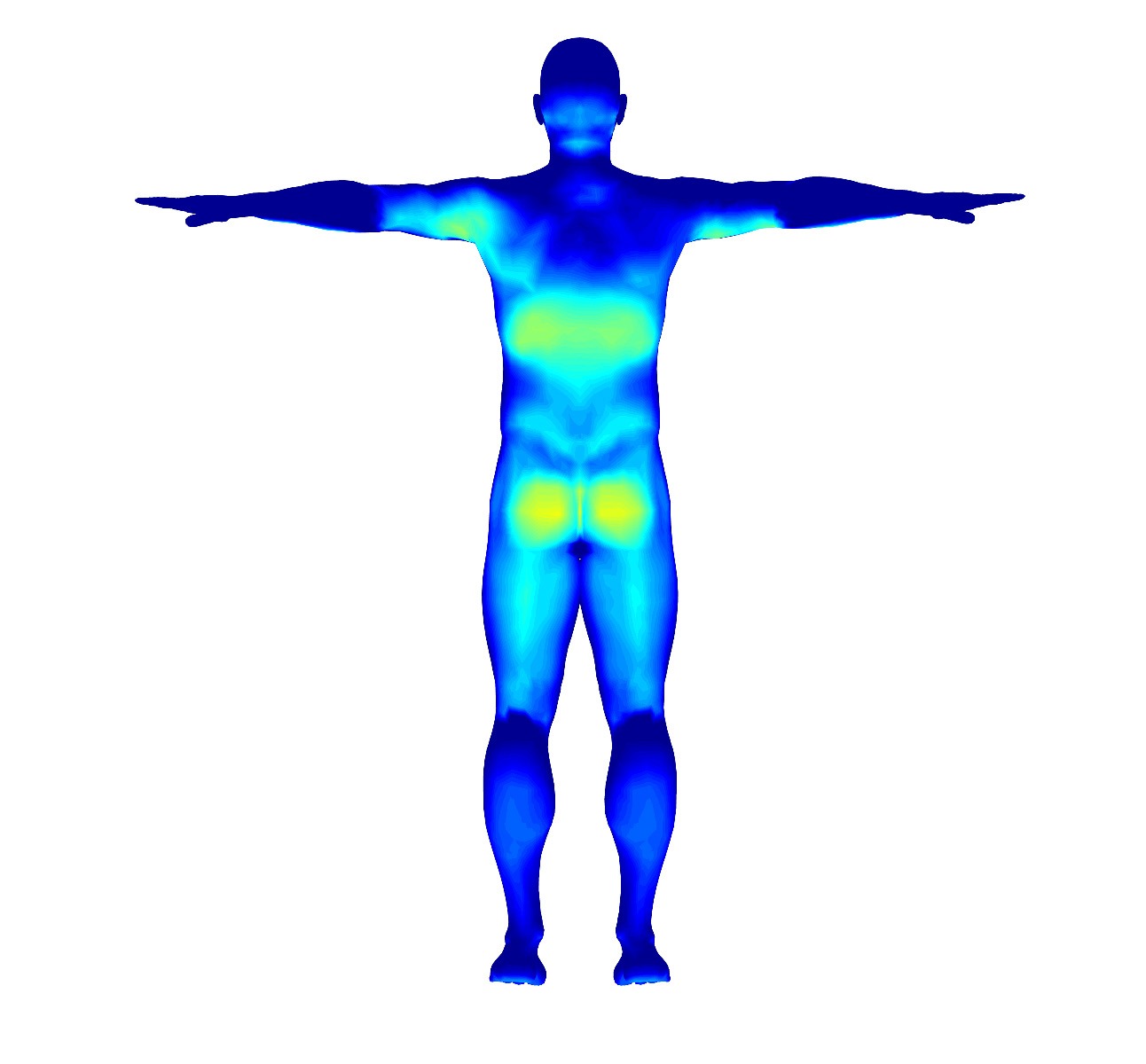}
		\scalebox{-1}[1]{\includegraphics[width=0.16\textwidth,trim=1cm 0.5cm 1cm 0.2cm,clip]{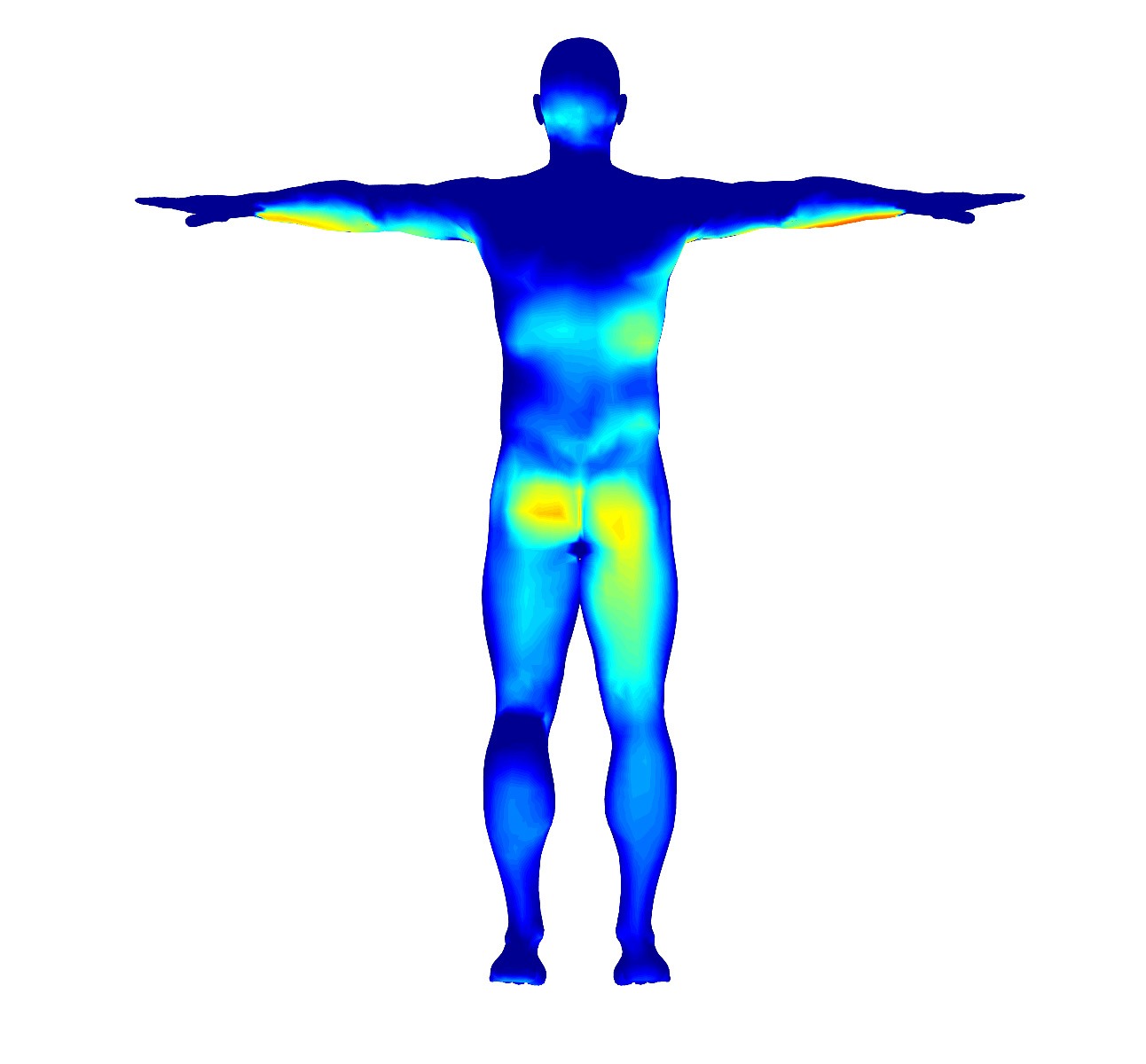}}
		\includegraphics[width=0.16\textwidth,trim=1cm 0.5cm 1cm 0.2cm,clip]{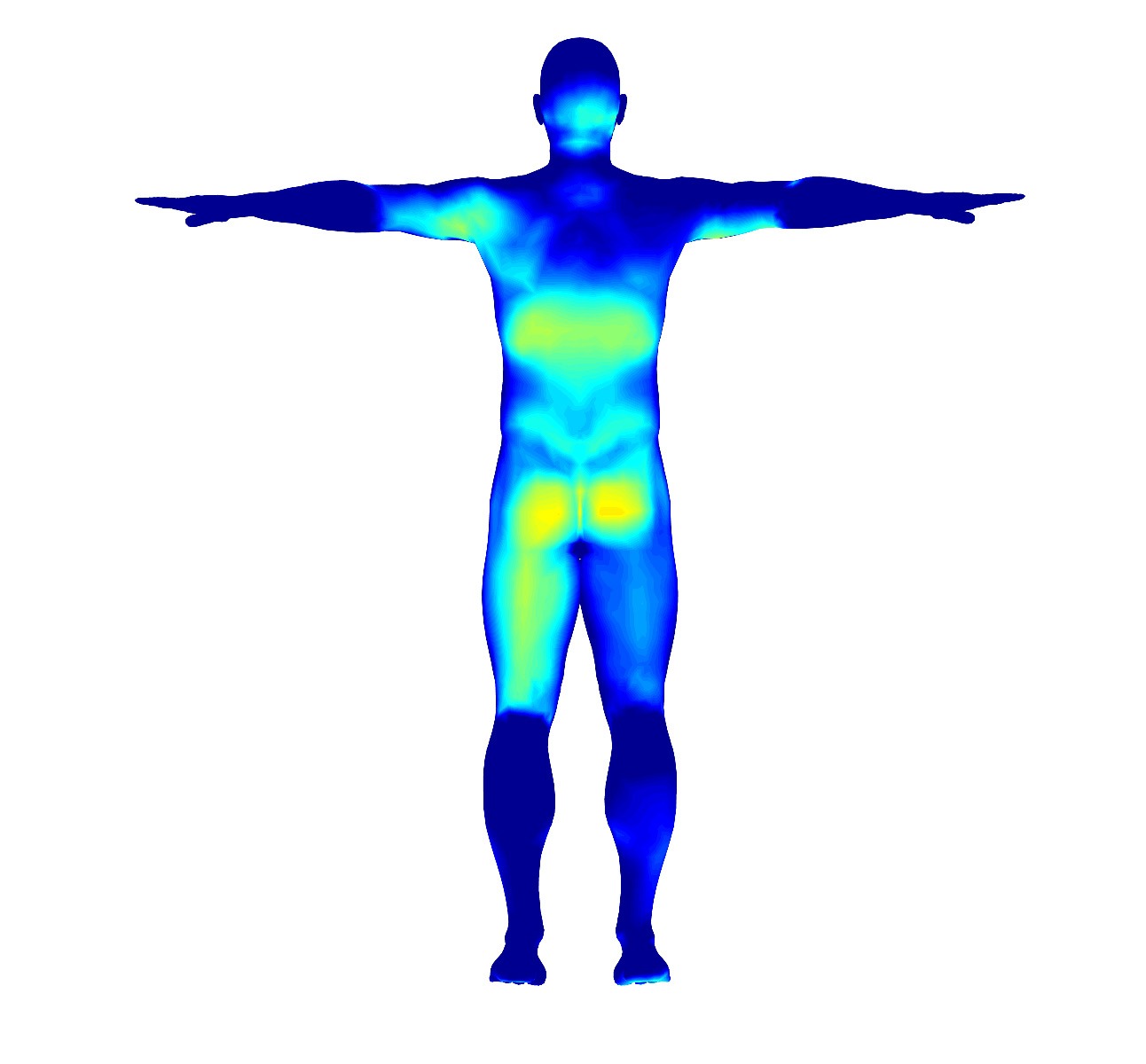}
		\includegraphics[width=0.16\textwidth,trim=1cm 0.5cm 1cm 0.2cm,clip]{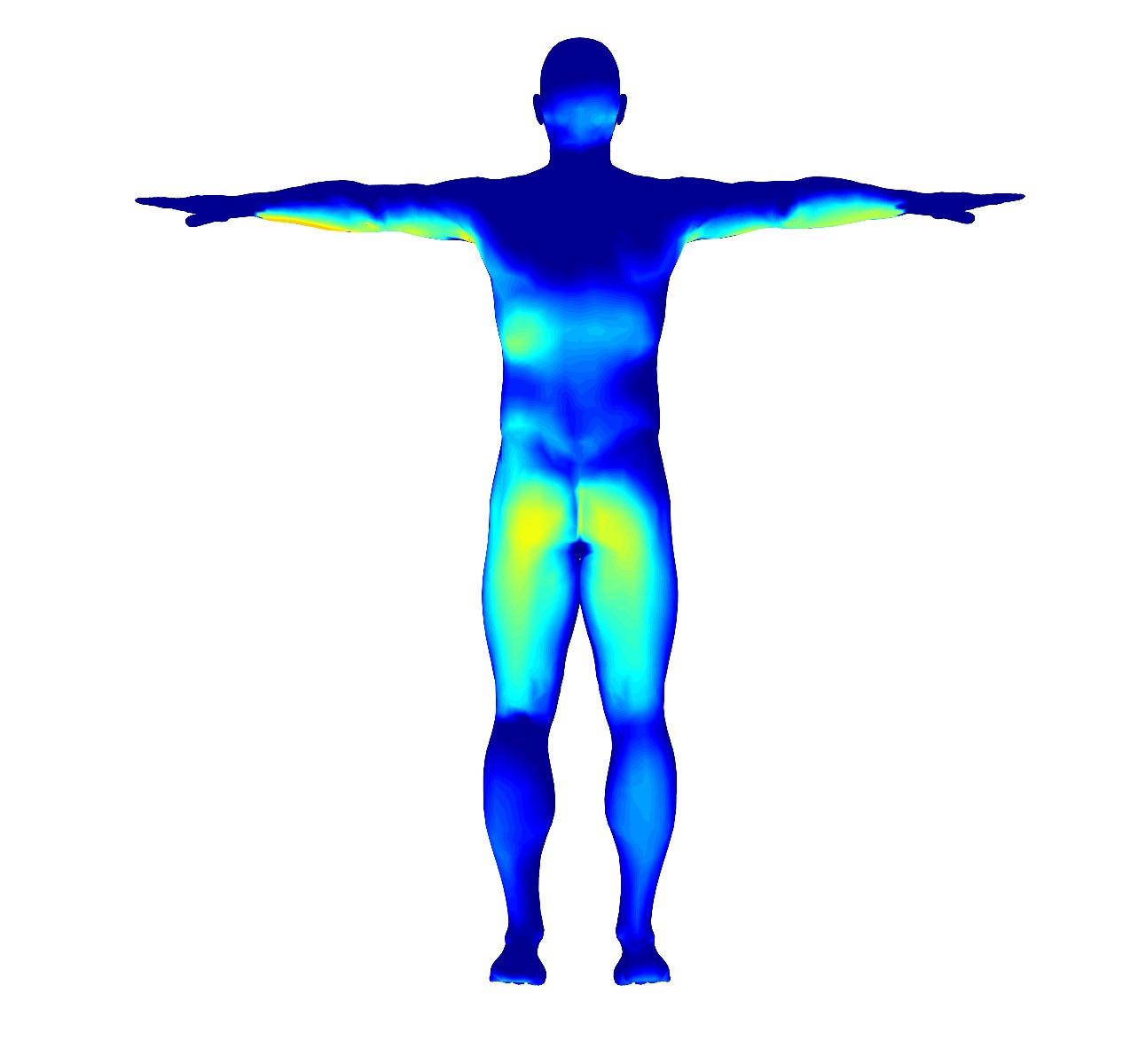}
		\captionsetup{justification=centering}
		\caption[caption]{Input contact pressure computed by assuming that the input body is perfectly supported. This optimal pressure serves as an importance map in the mesh generation stage.}
	\end{subfigure}
	\begin{subfigure}[t]{0.96\textwidth}
		\centering
		\includegraphics[width=0.16\textwidth,trim=1cm 0.5cm 1cm 0.2cm,clip]{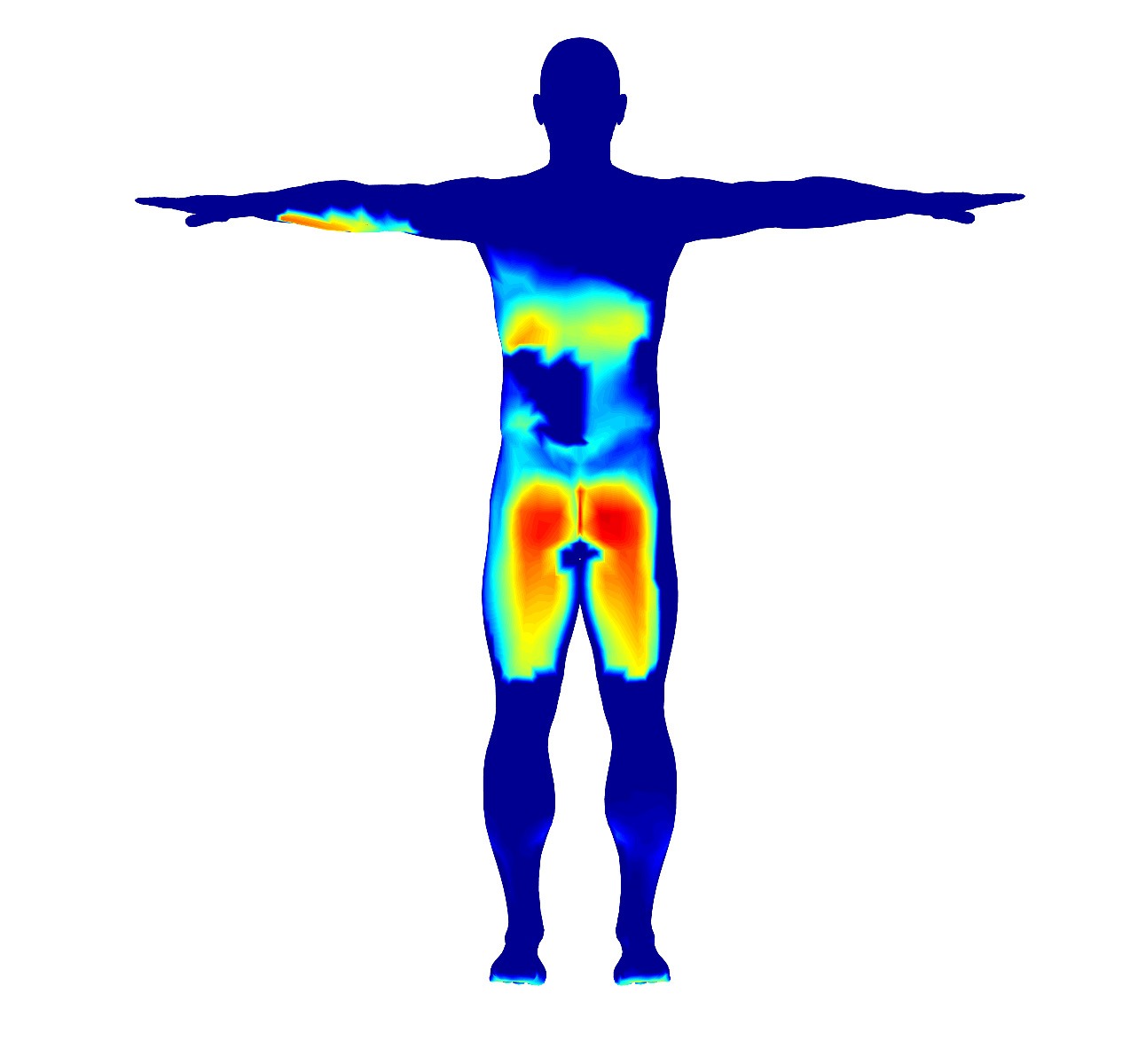}
		\scalebox{-1}[1]{\includegraphics[width=0.16\textwidth,trim=1cm 0.5cm 1cm 0.2cm,clip]{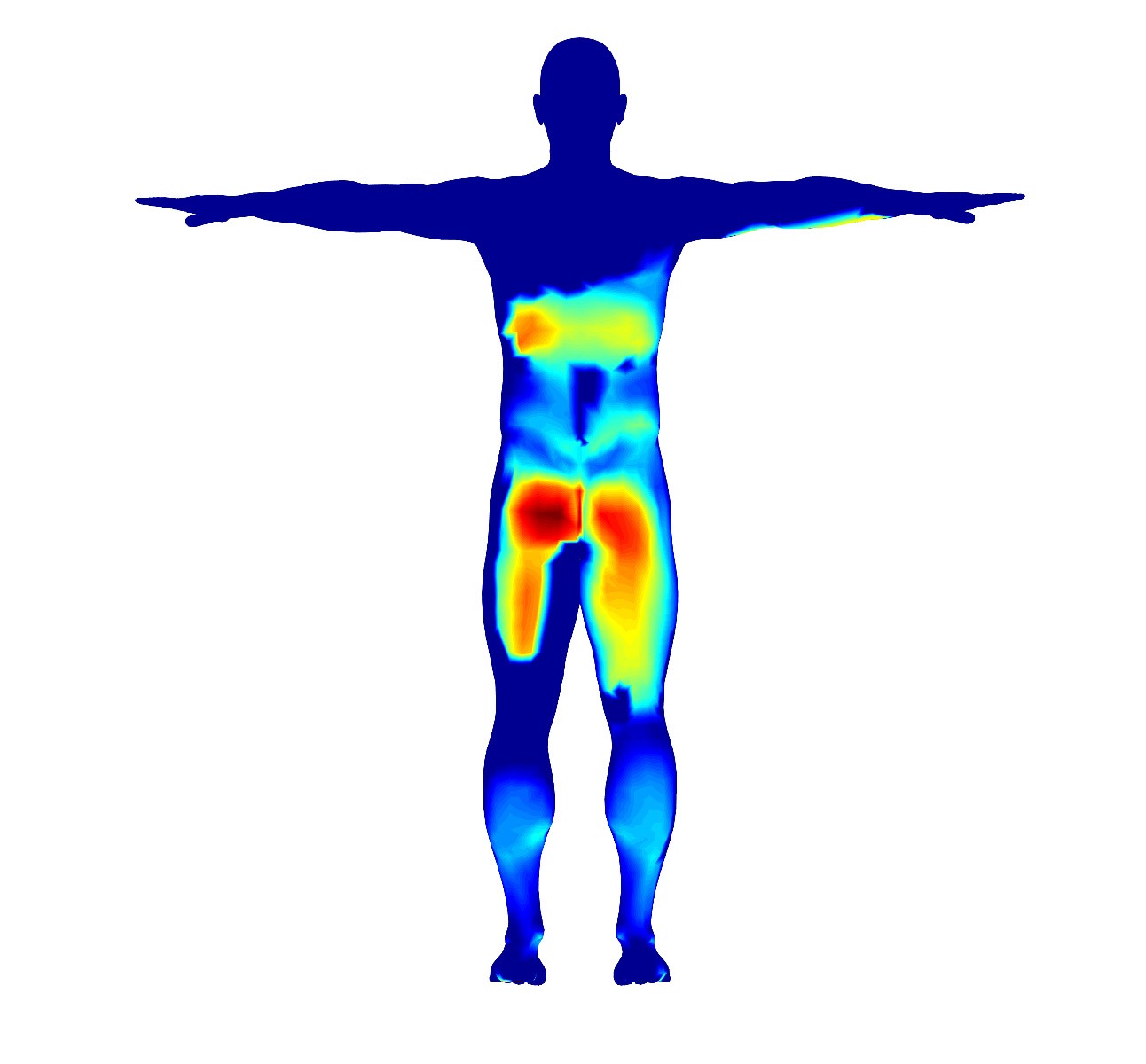}}
		\includegraphics[width=0.16\textwidth,trim=1cm 0.5cm 1cm 0.2cm,clip]{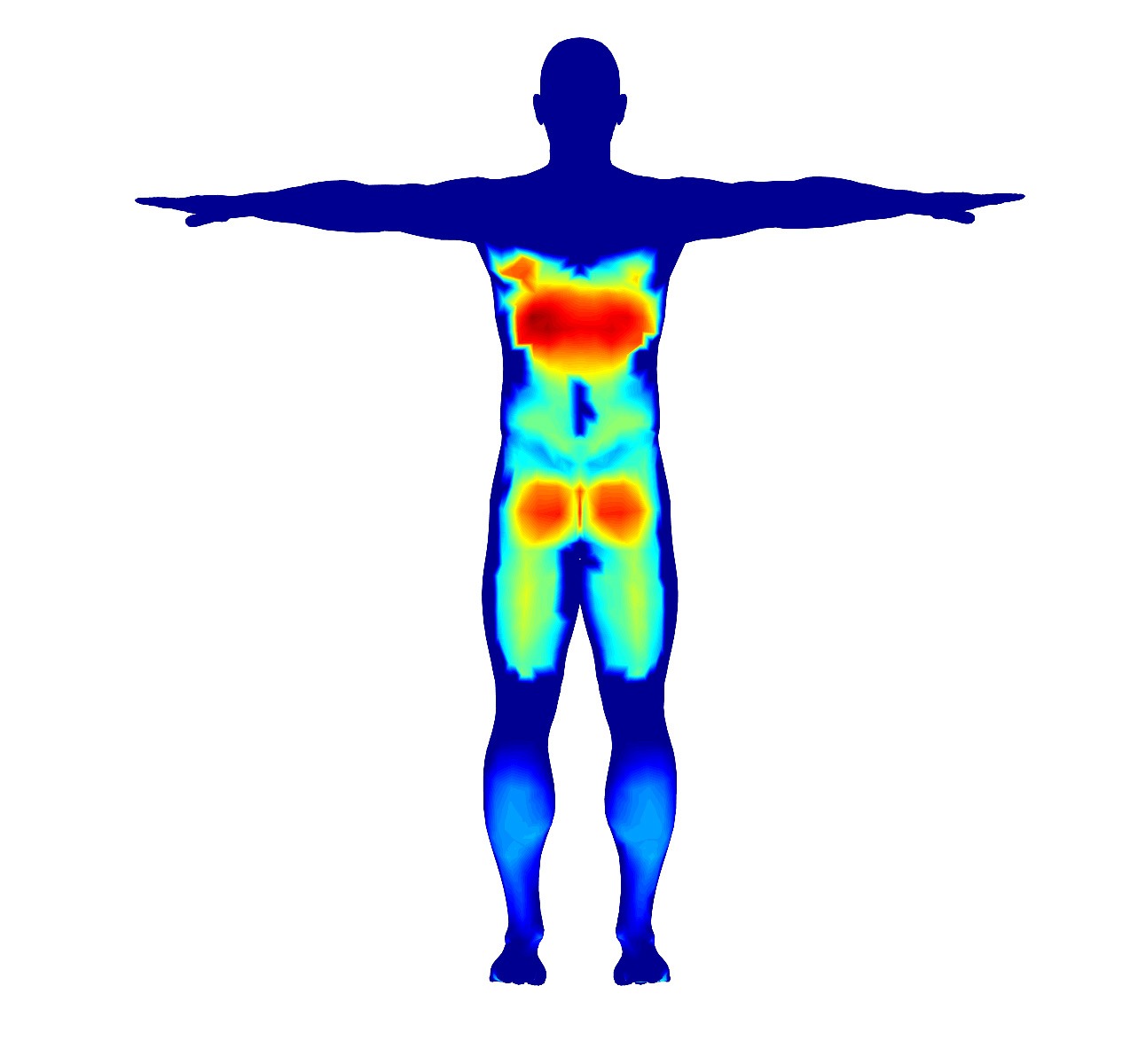}
		\scalebox{-1}[1]{\includegraphics[width=0.16\textwidth,trim=1cm 0.5cm 1cm 0.2cm,clip]{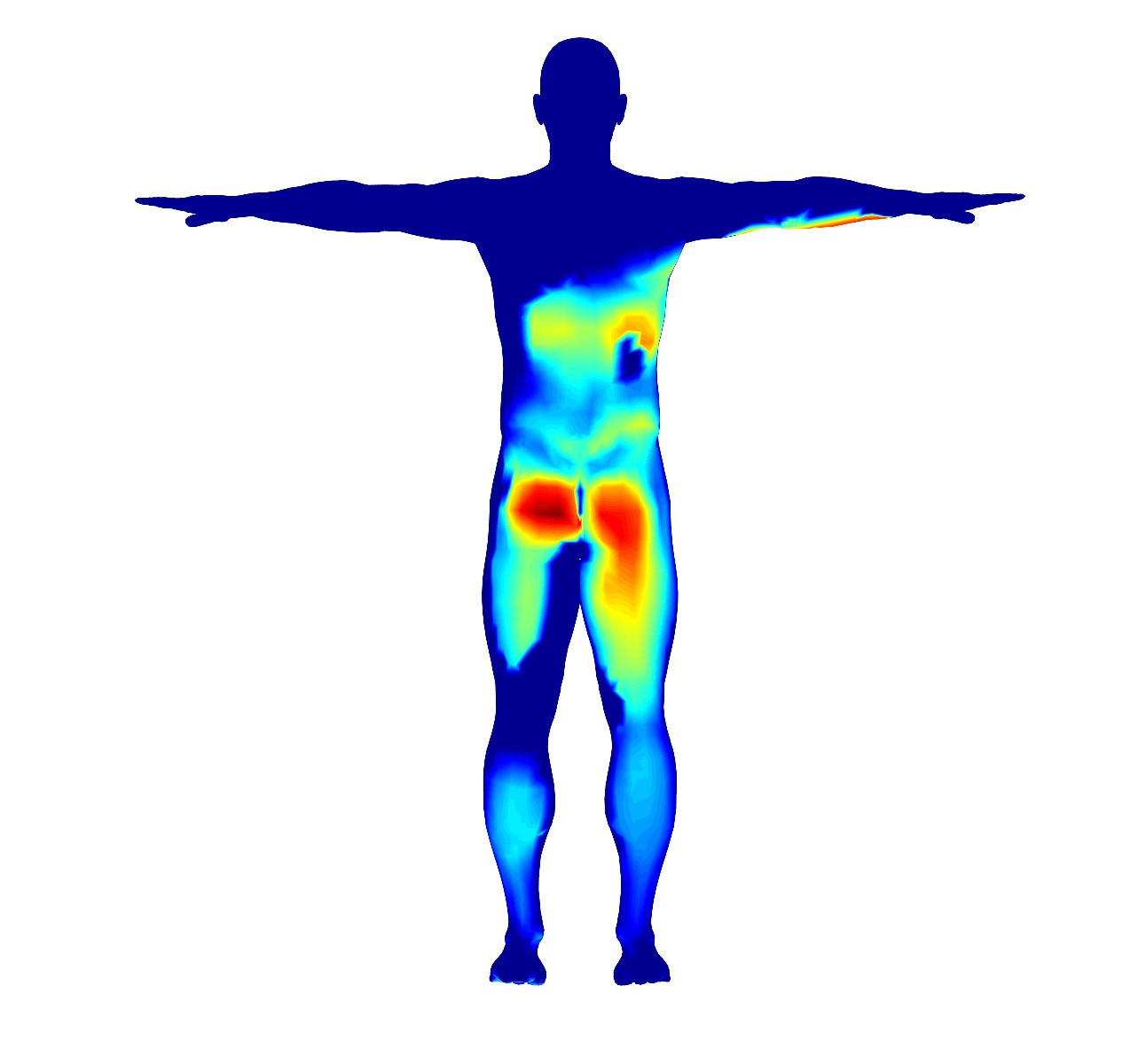}}
		\includegraphics[width=0.16\textwidth,trim=1cm 0.5cm 1cm 0.2cm,clip]{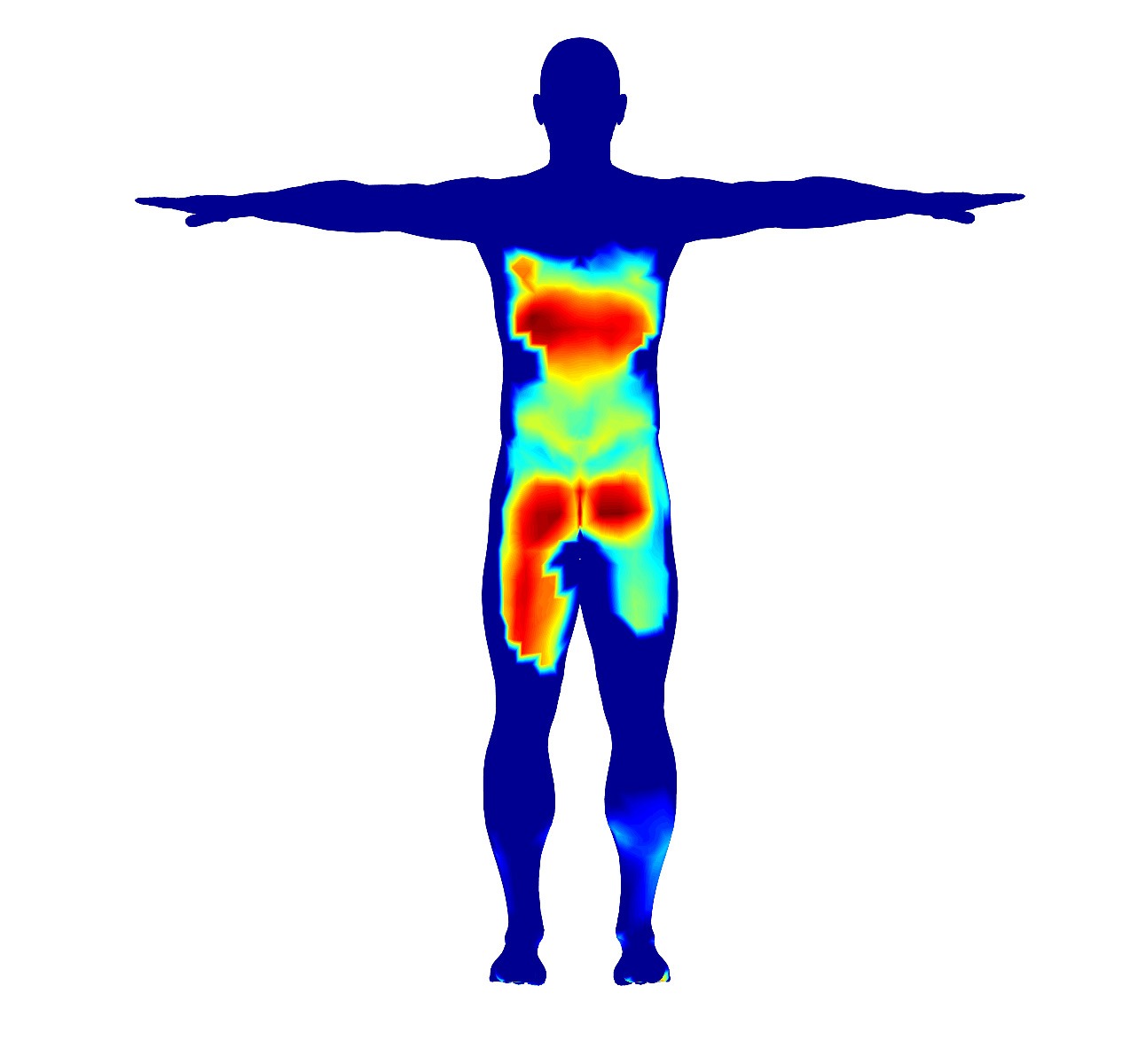}
		\includegraphics[width=0.16\textwidth,trim=1cm 0.5cm 1cm 0.2cm,clip]{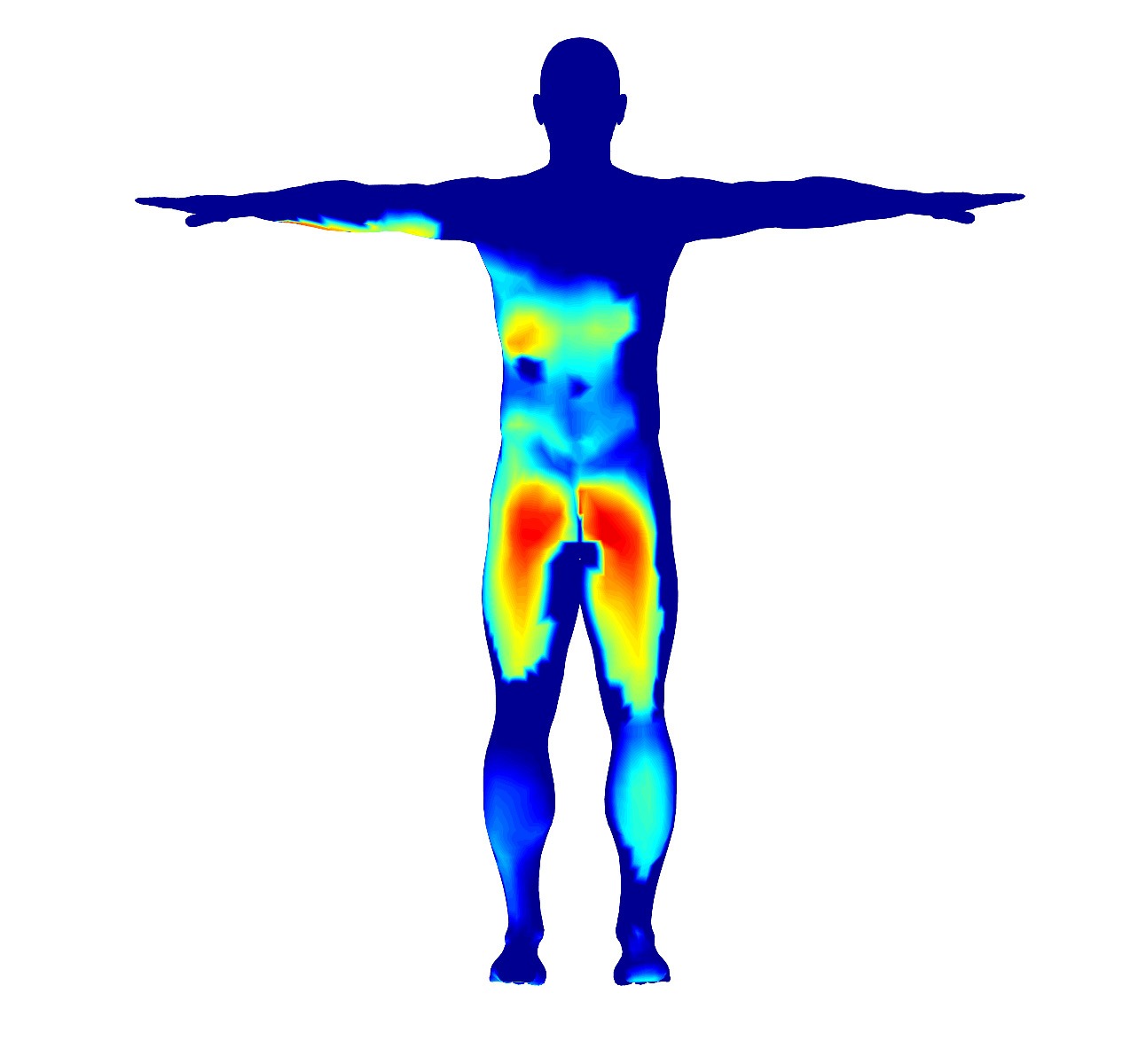}
			\includegraphics[width=0.16\textwidth,trim=1.5cm 1.5cm 4cm 1cm,clip]{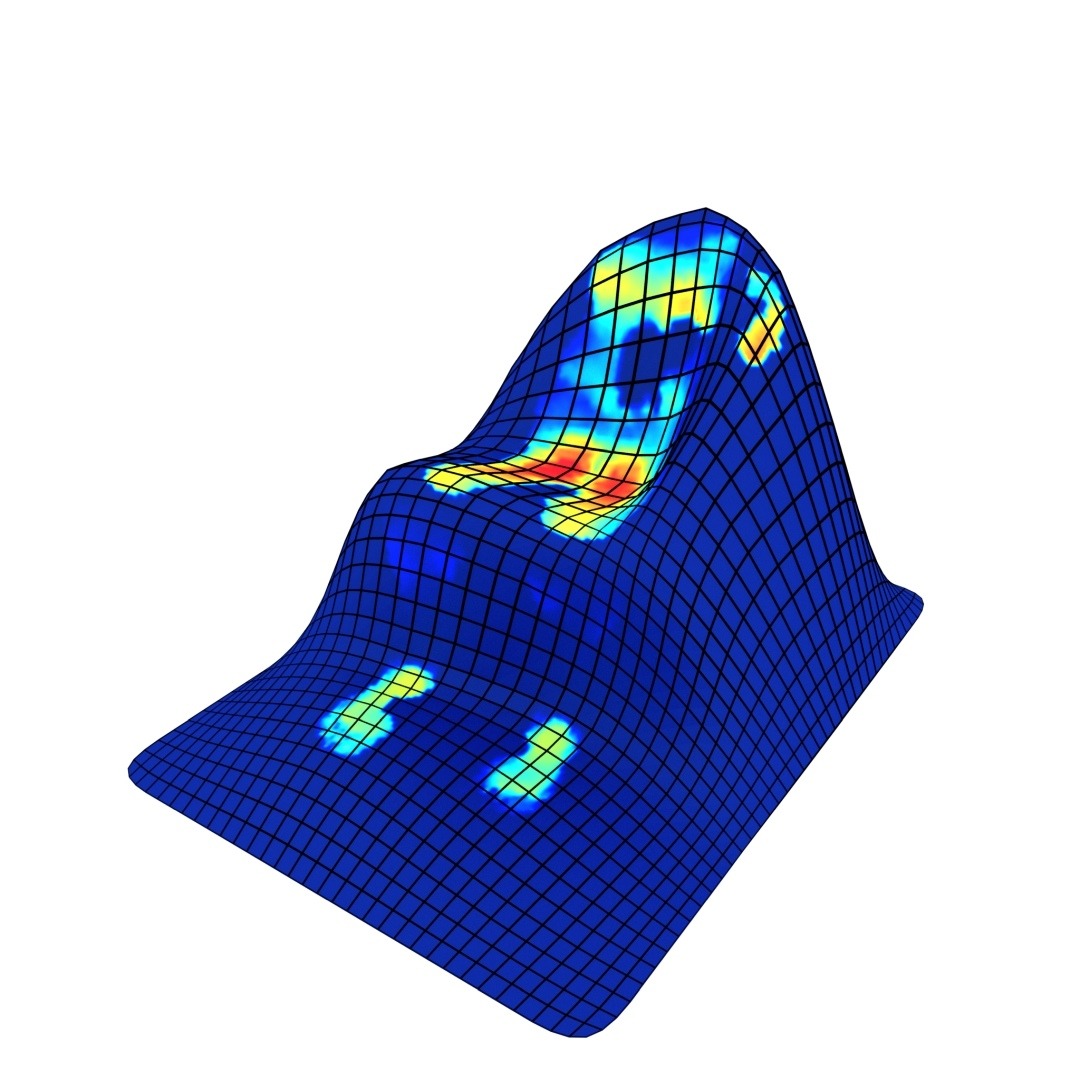}
			\includegraphics[width=0.16\textwidth,trim=1.5cm 1.5cm 4cm 1cm,clip]{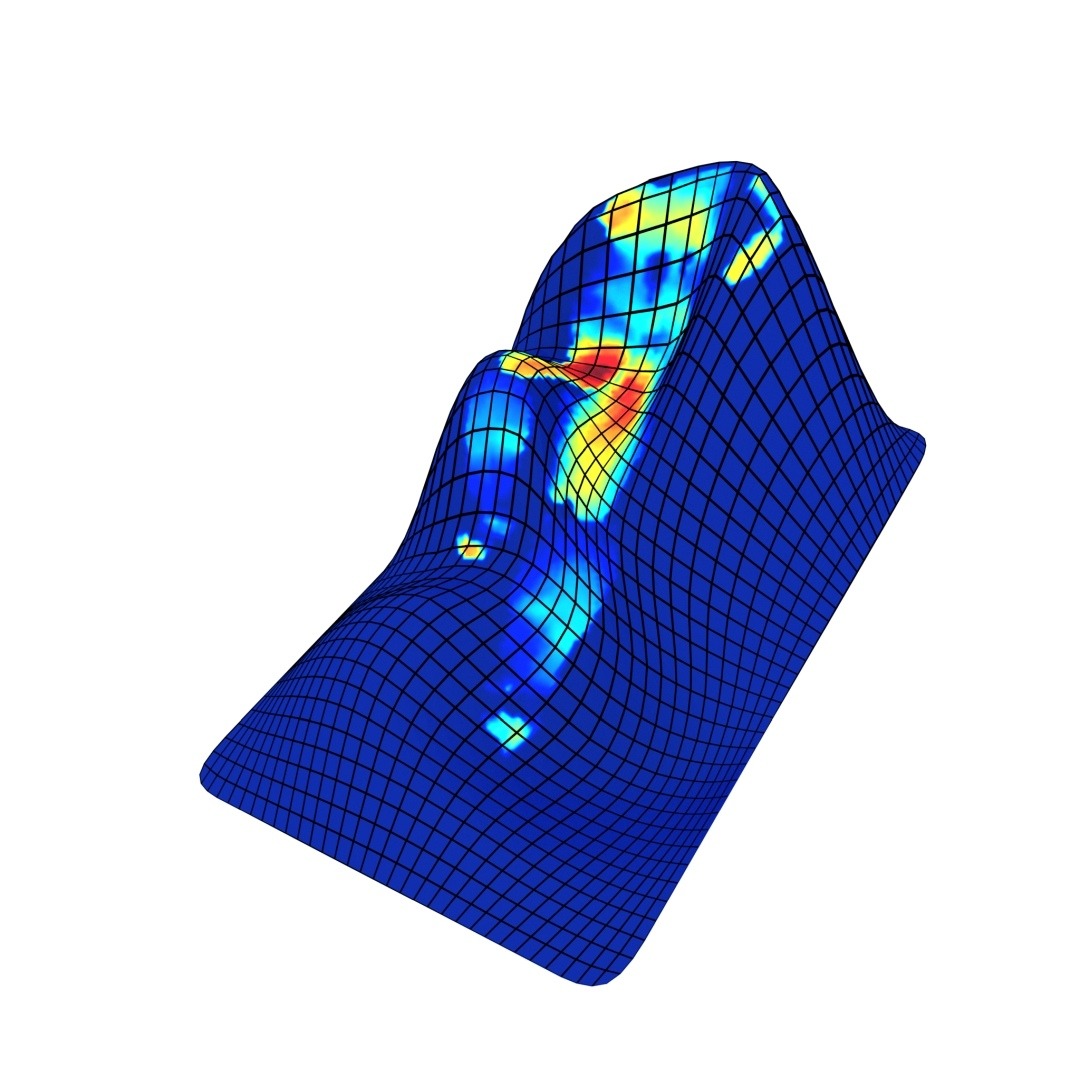}
			\includegraphics[width=0.16\textwidth,trim=1.5cm 1.5cm 4cm 1cm,clip]{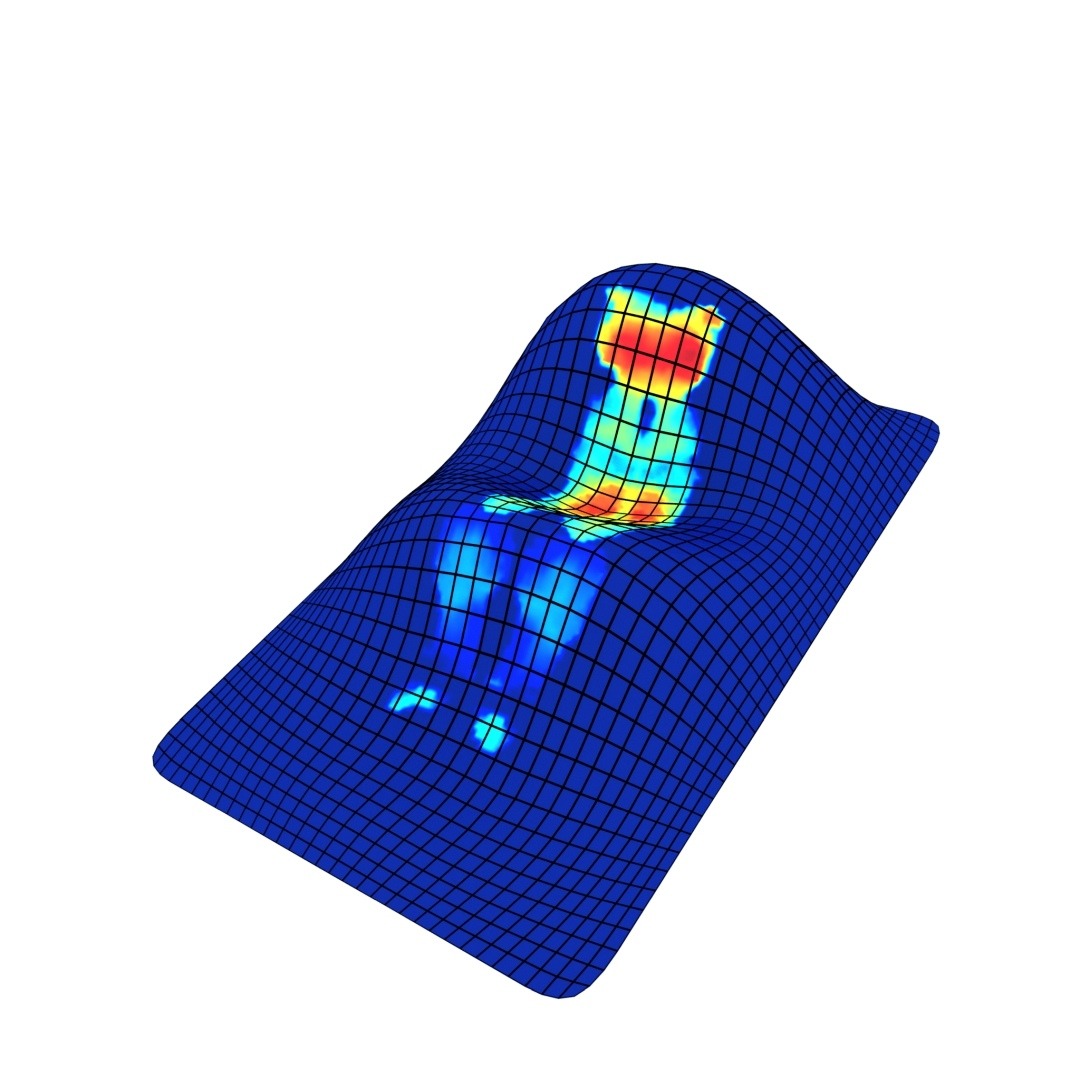}
			\includegraphics[width=0.16\textwidth,trim=1.5cm 1.5cm 4cm 1cm,clip]{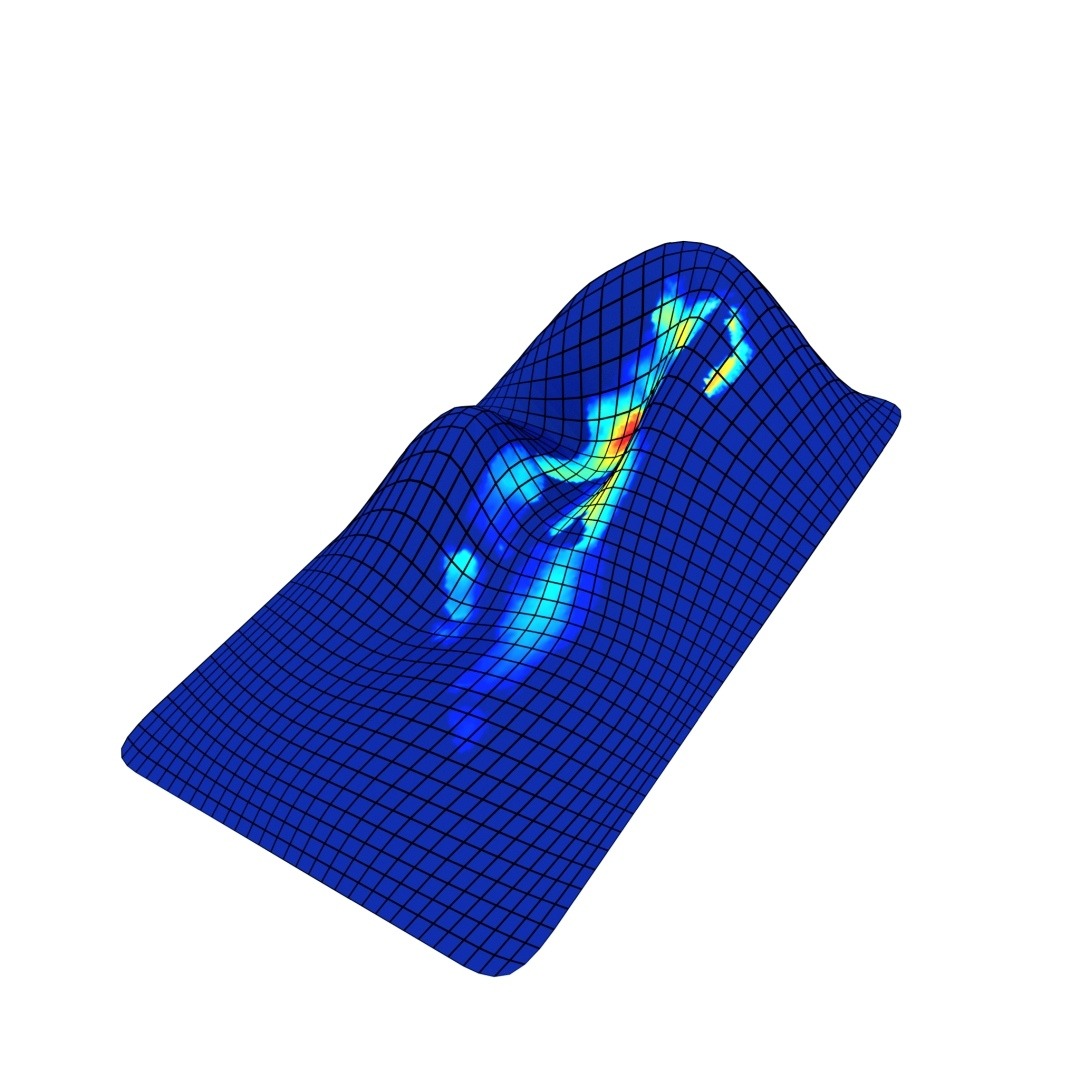}
			\includegraphics[width=0.16\textwidth,trim=1.5cm 1.5cm 4cm 1cm,clip]{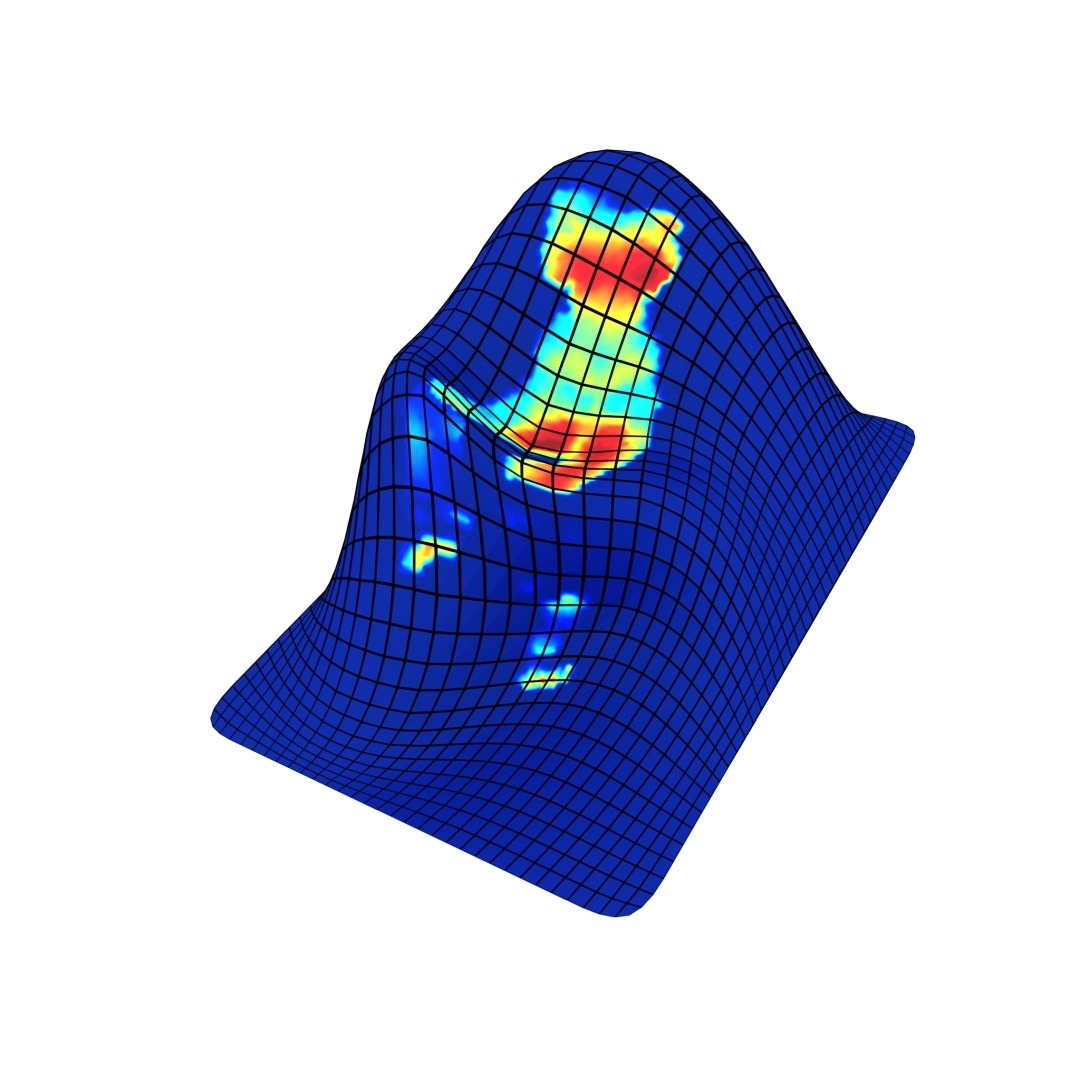}
			\includegraphics[width=0.16\textwidth,trim=1.5cm 1.5cm 4cm 1cm,clip]{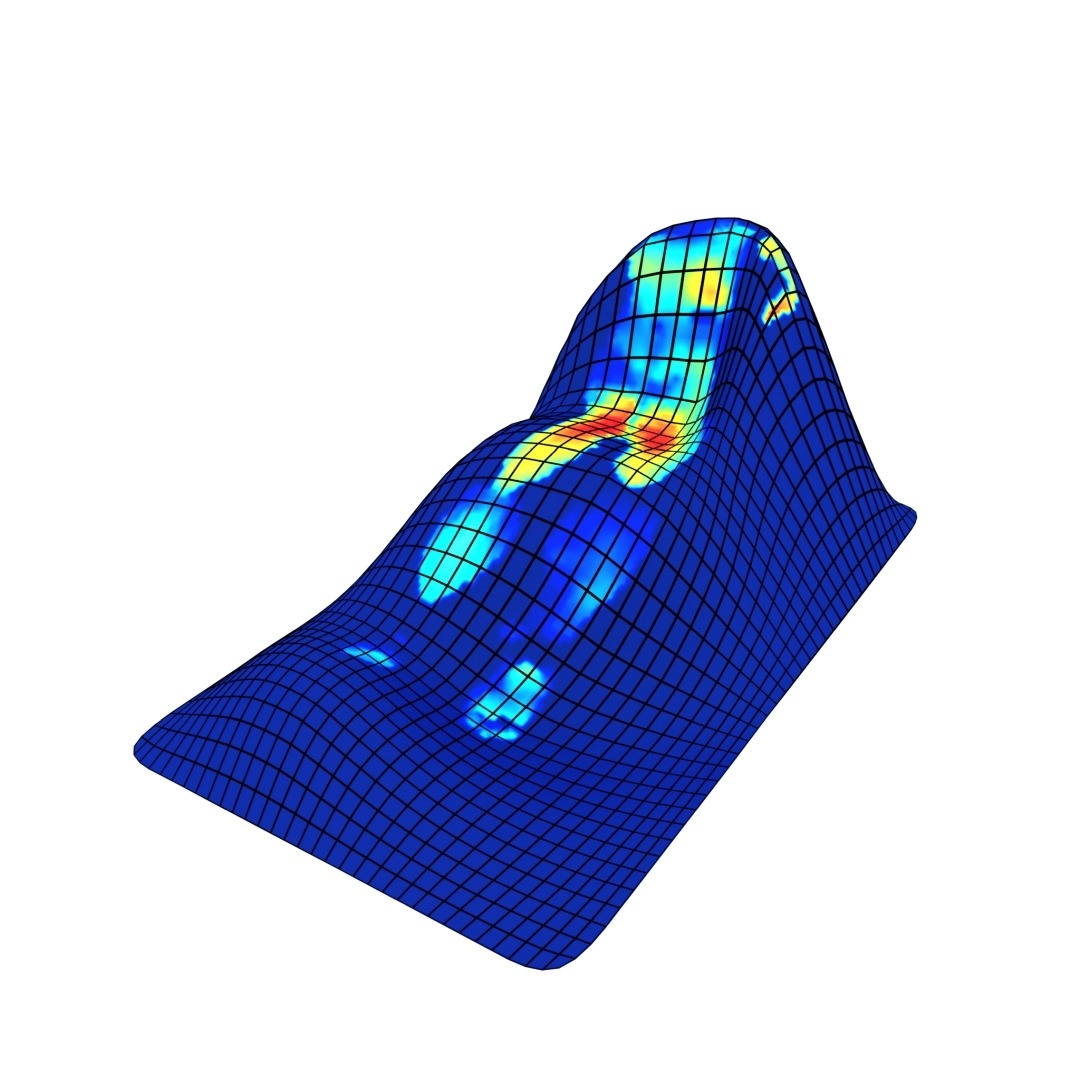}
		\captionsetup{justification=centering}
		\caption[caption]{Contact pressure mapped onto surfaces generated by applying the fitting algorithm of Leimer \etal~\cite{leimer2018sit} to a flat patch serving the control mesh. Please note the high pressure peaks and holes in the pressure distribution where no contact to the surface exists.}
	\end{subfigure}
	\begin{subfigure}[t]{0.96\textwidth}
		\centering
		\includegraphics[width=0.16\textwidth,trim=1cm 0.5cm 1cm 0.2cm,clip]{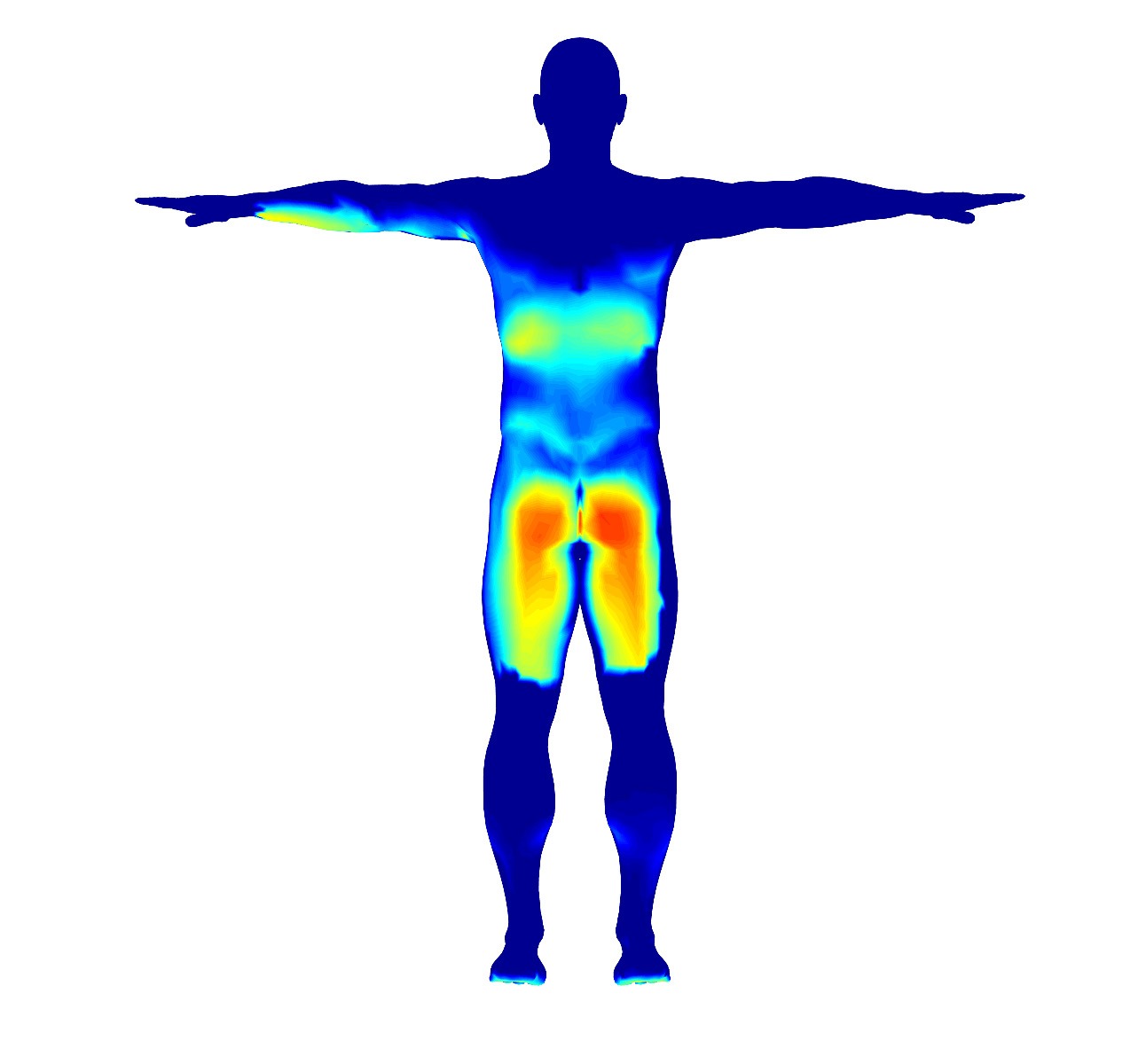}
		\scalebox{-1}[1]{\includegraphics[width=0.16\textwidth,trim=1cm 0.5cm 1cm 0.2cm,clip]{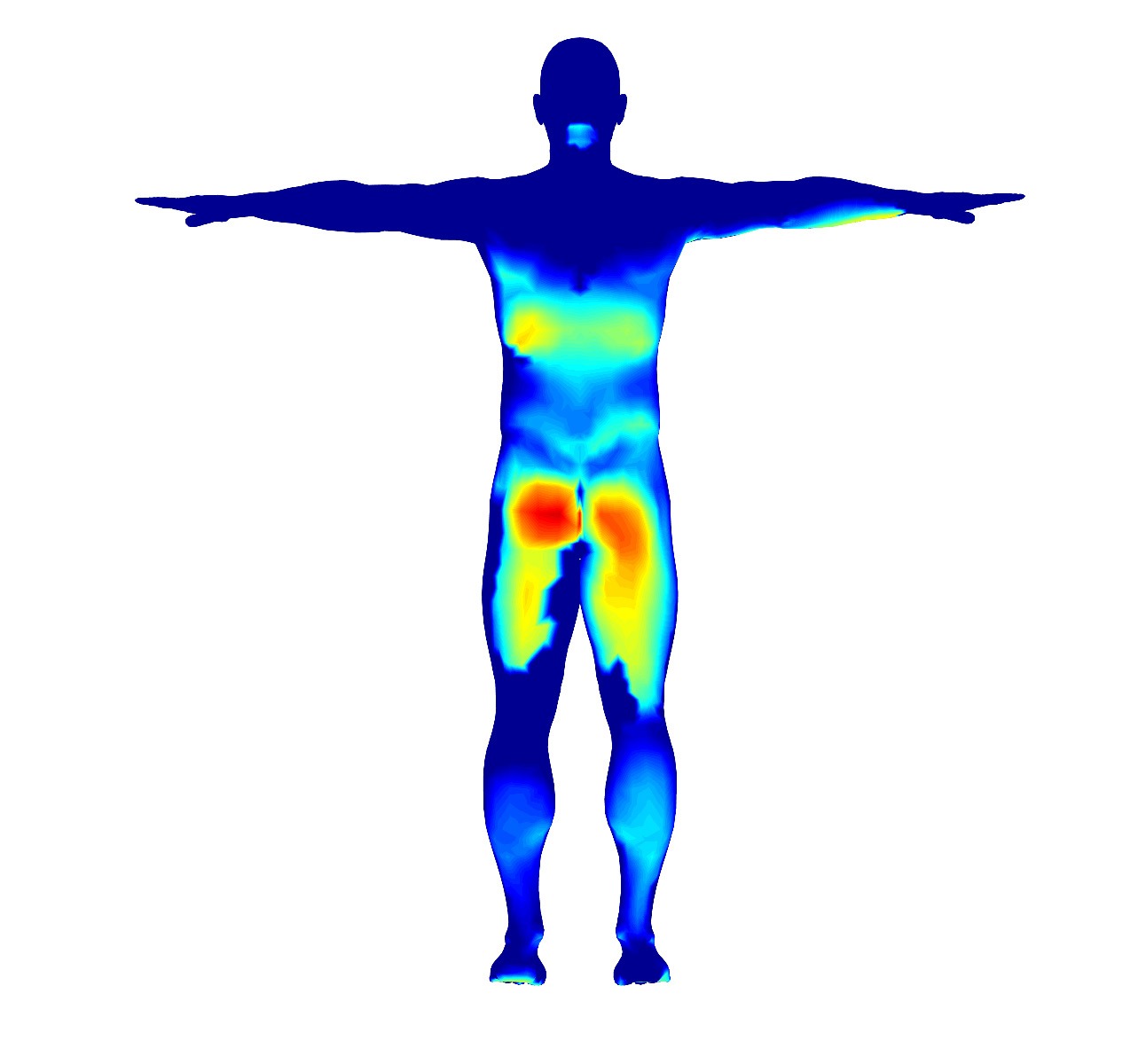}}
		\includegraphics[width=0.16\textwidth,trim=1cm 0.5cm 1cm 0.2cm,clip]{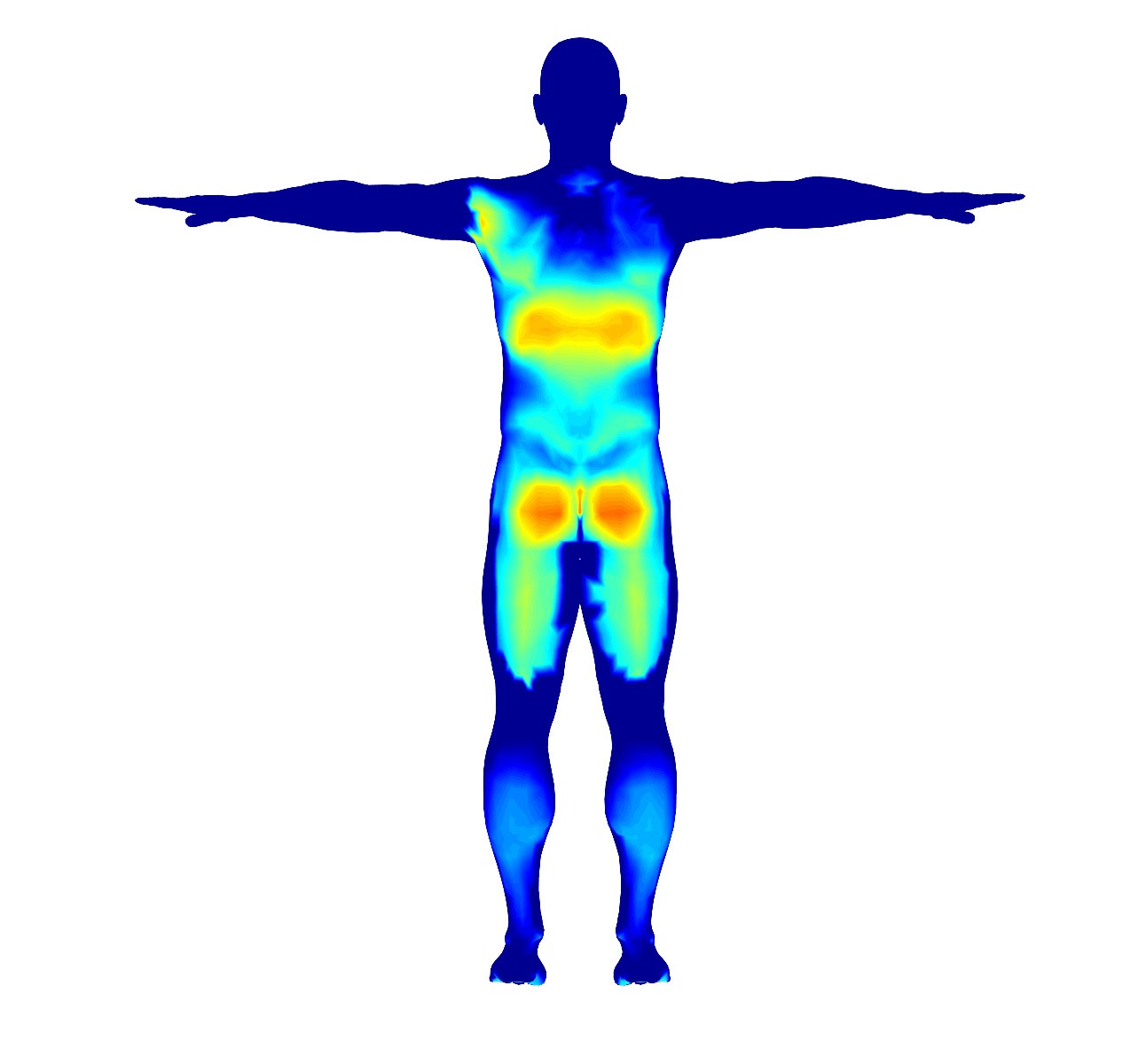}
		\scalebox{-1}[1]{\includegraphics[width=0.16\textwidth,trim=1cm 0.5cm 1cm 0.2cm,clip]{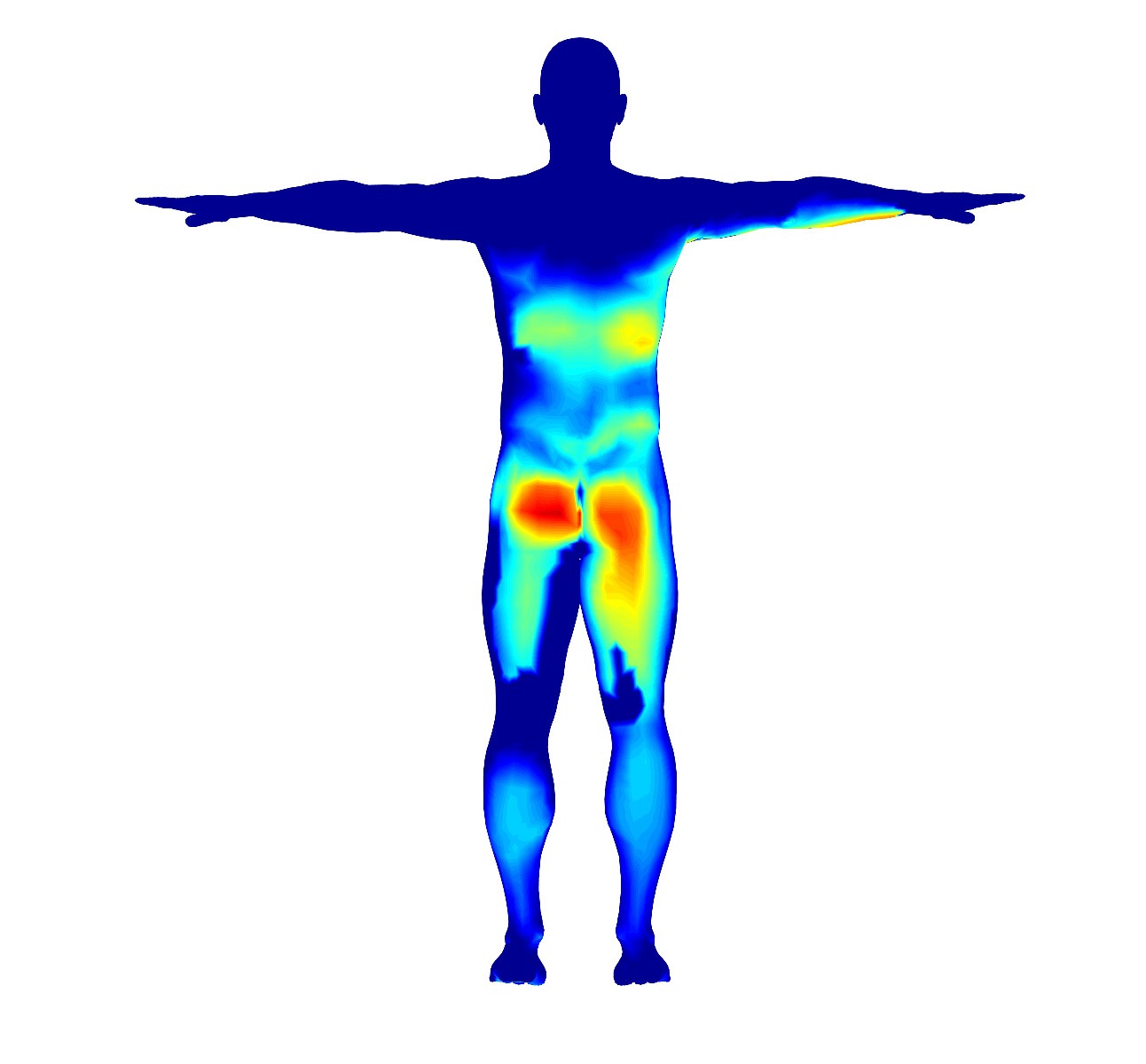}}
		\includegraphics[width=0.16\textwidth,trim=1cm 0.5cm 1cm 0.2cm,clip]{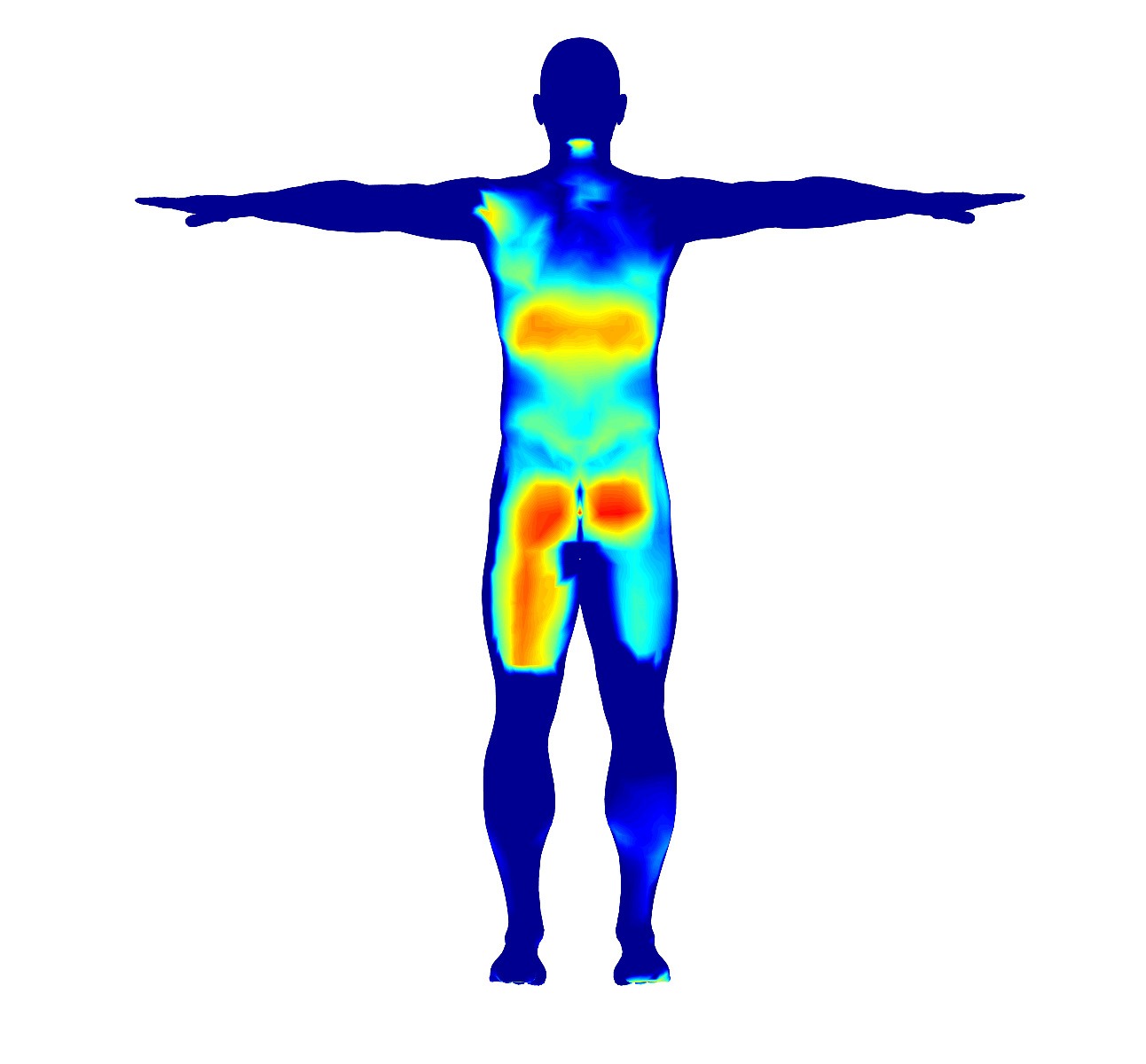}
		\includegraphics[width=0.16\textwidth,trim=1cm 0.5cm 1cm 0.2cm,clip]{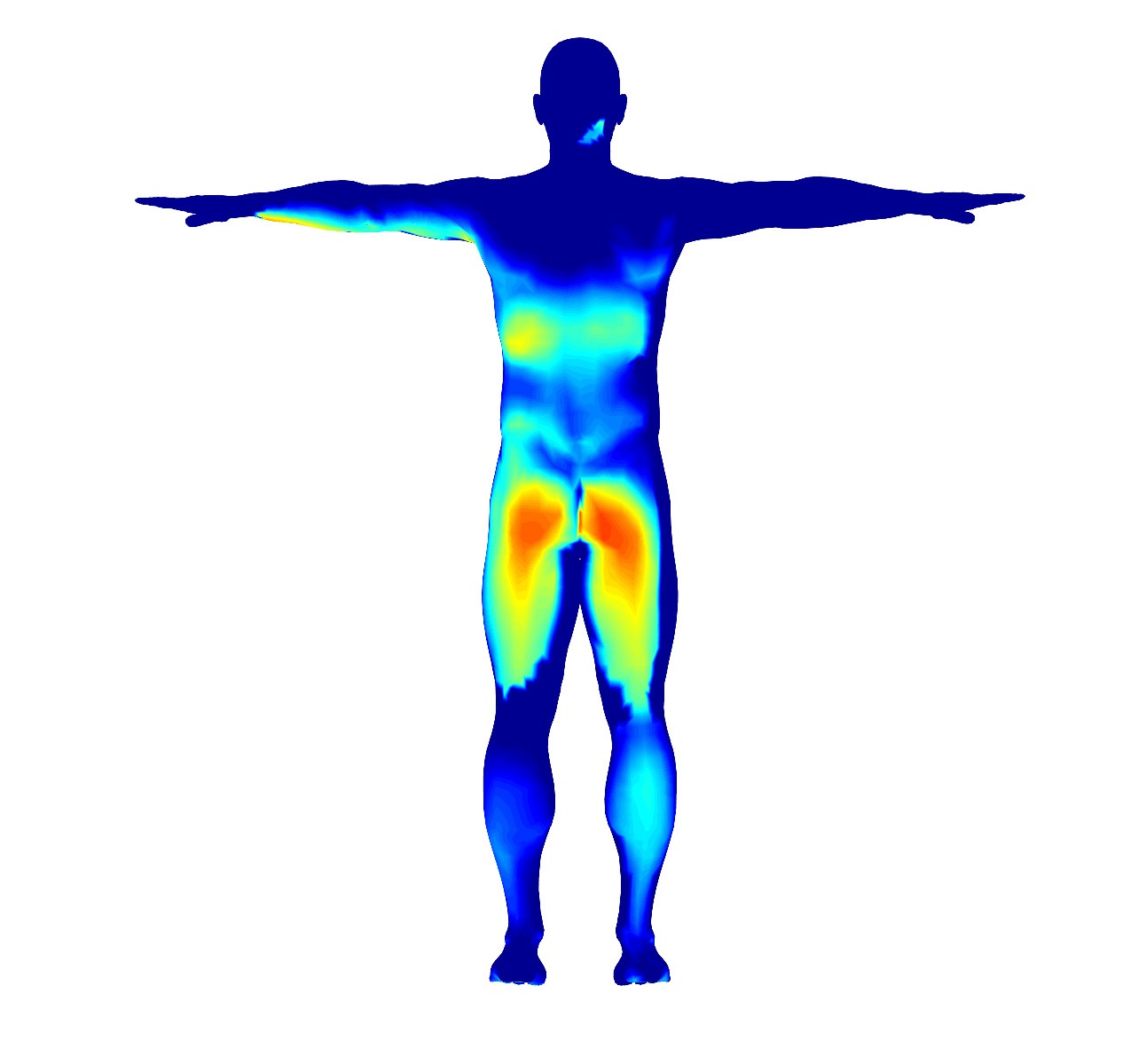}	
			\includegraphics[width=0.16\textwidth,trim=1.5cm 1.5cm 4cm 1cm,clip]{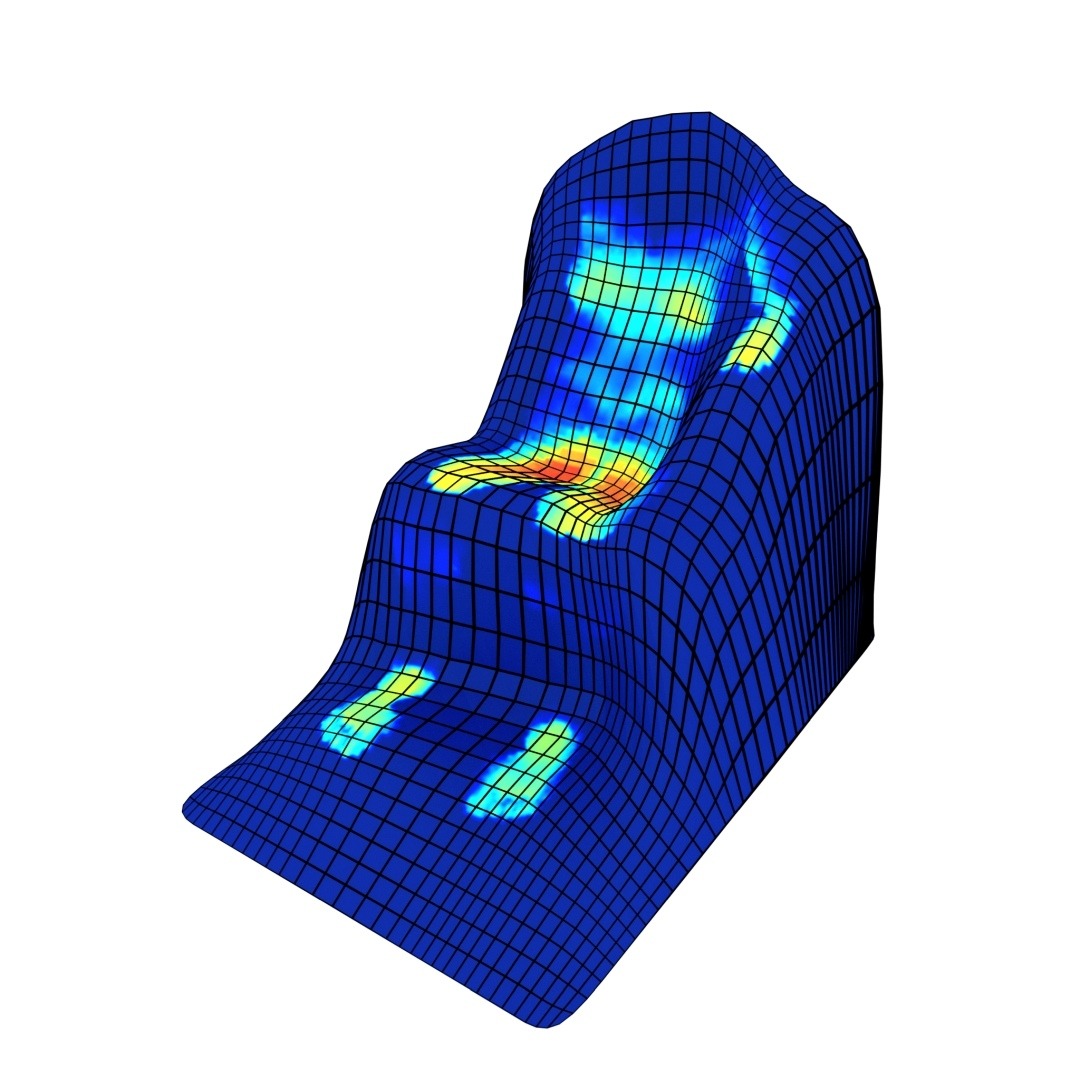}
			\includegraphics[width=0.16\textwidth,trim=1.5cm 1.5cm 4cm 1cm,clip]{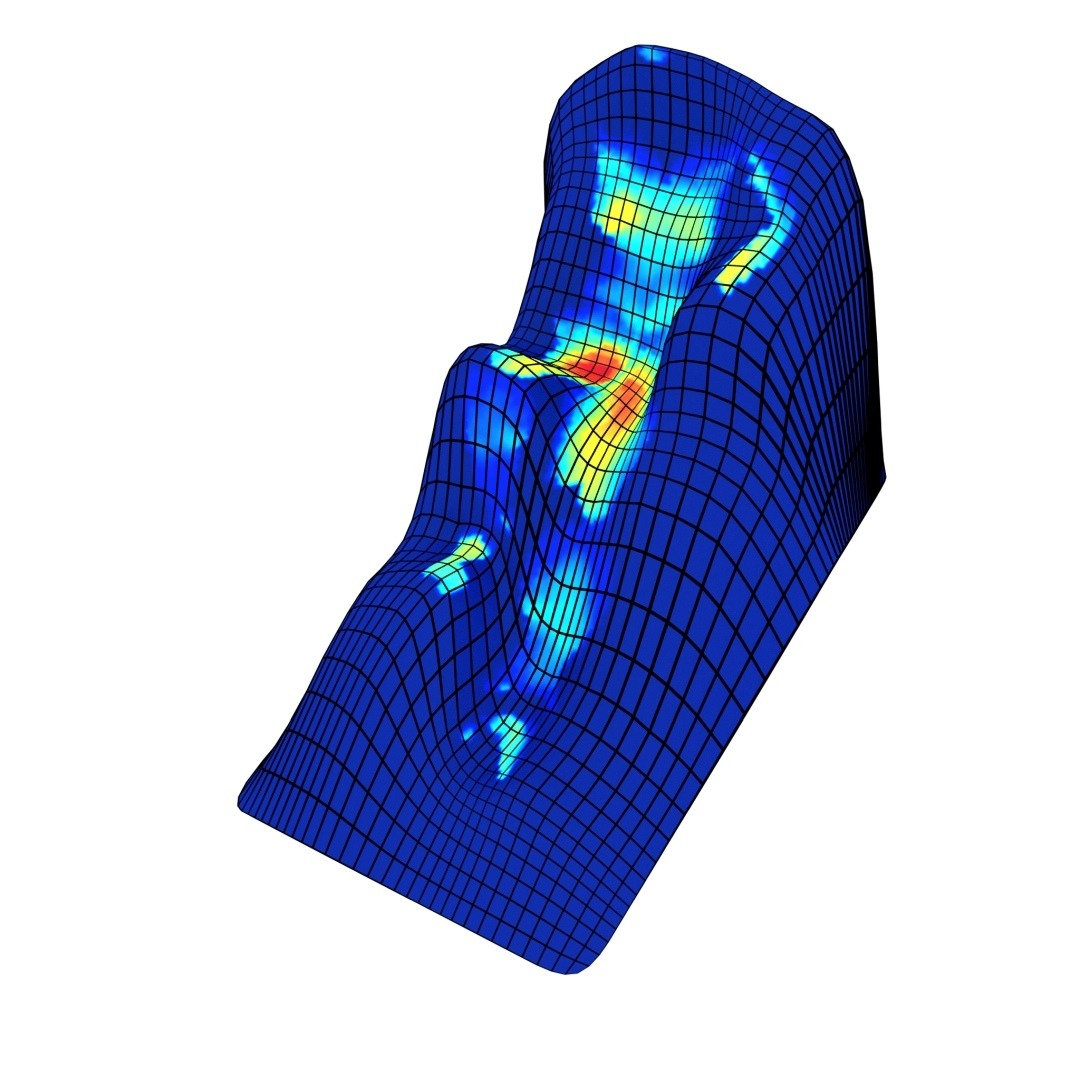}
			\includegraphics[width=0.16\textwidth,trim=1.5cm 1.5cm 4cm 1cm,clip]{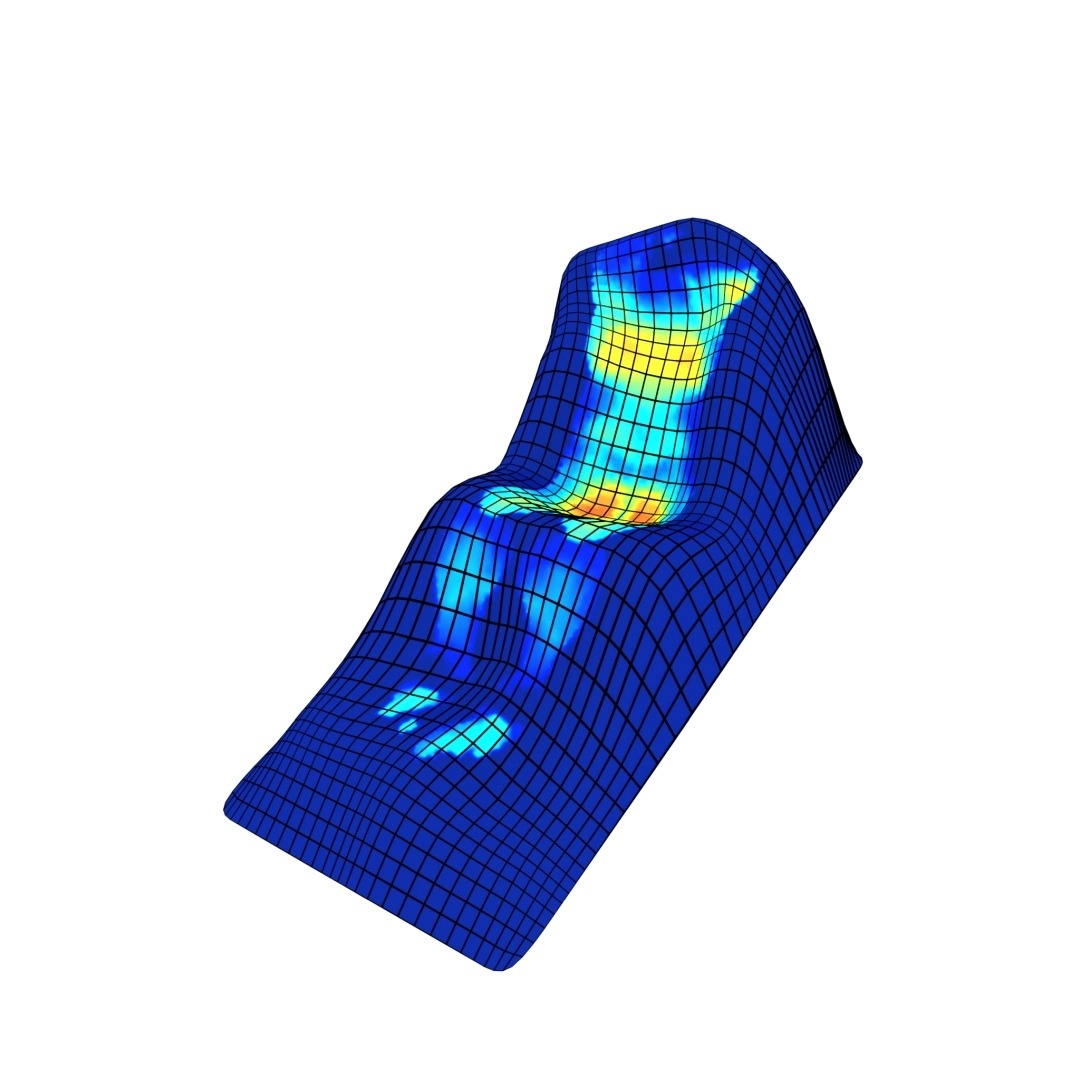}
			\includegraphics[width=0.16\textwidth,trim=1.5cm 1.5cm 4cm 1cm,clip]{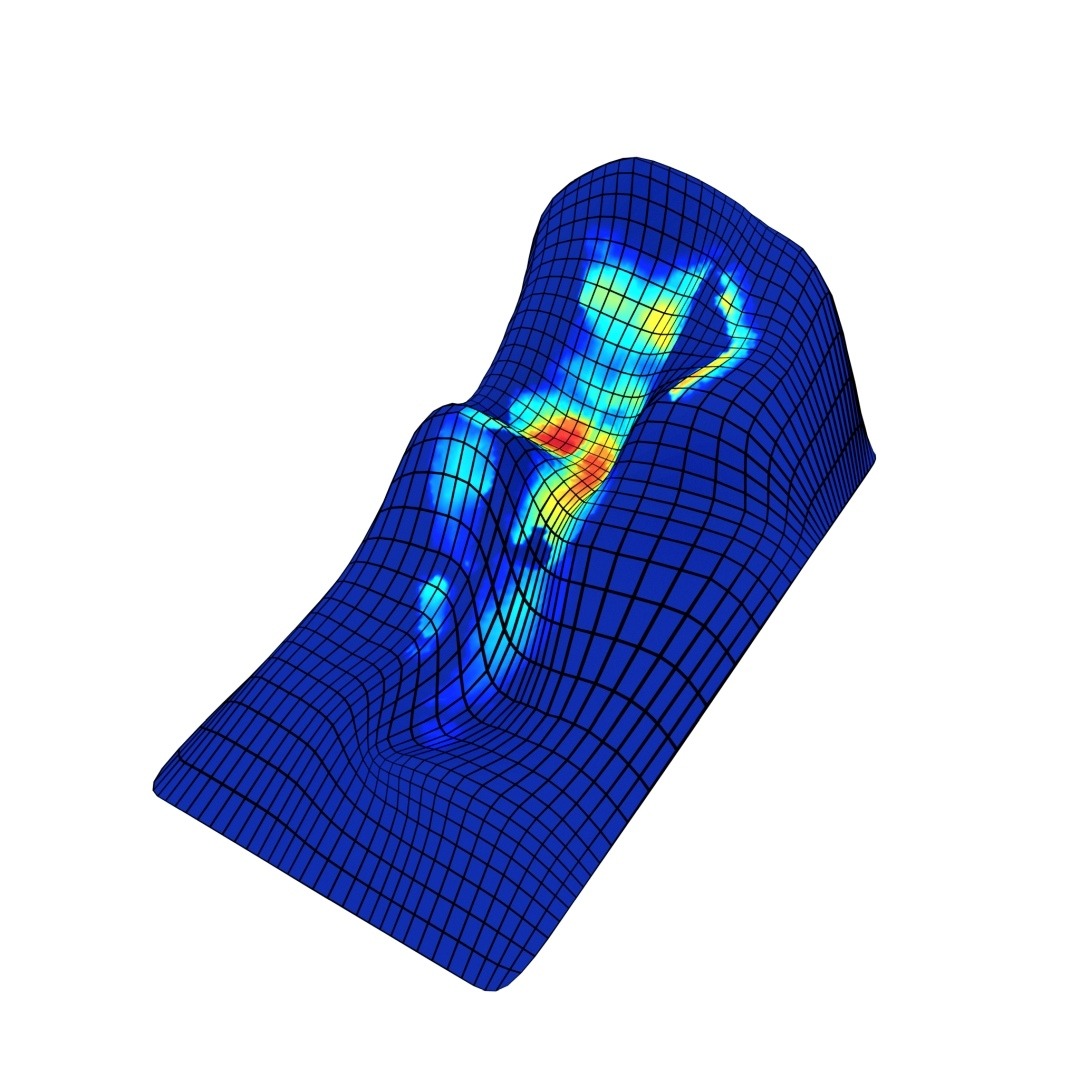}
			\includegraphics[width=0.16\textwidth,trim=1.5cm 1.5cm 4cm 1cm,clip]{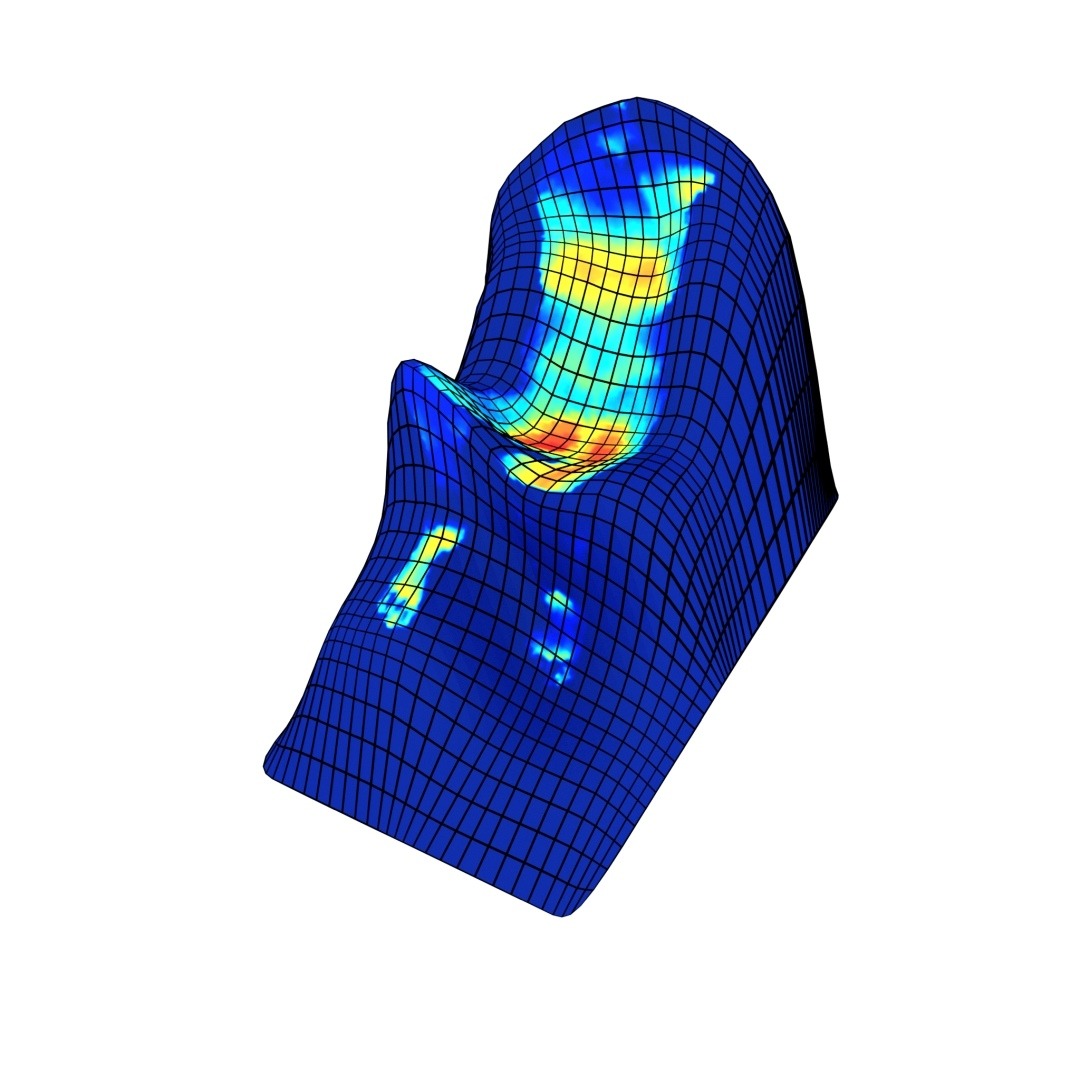}
			\includegraphics[width=0.16\textwidth,trim=1.5cm 1.5cm 4cm 1cm,clip]{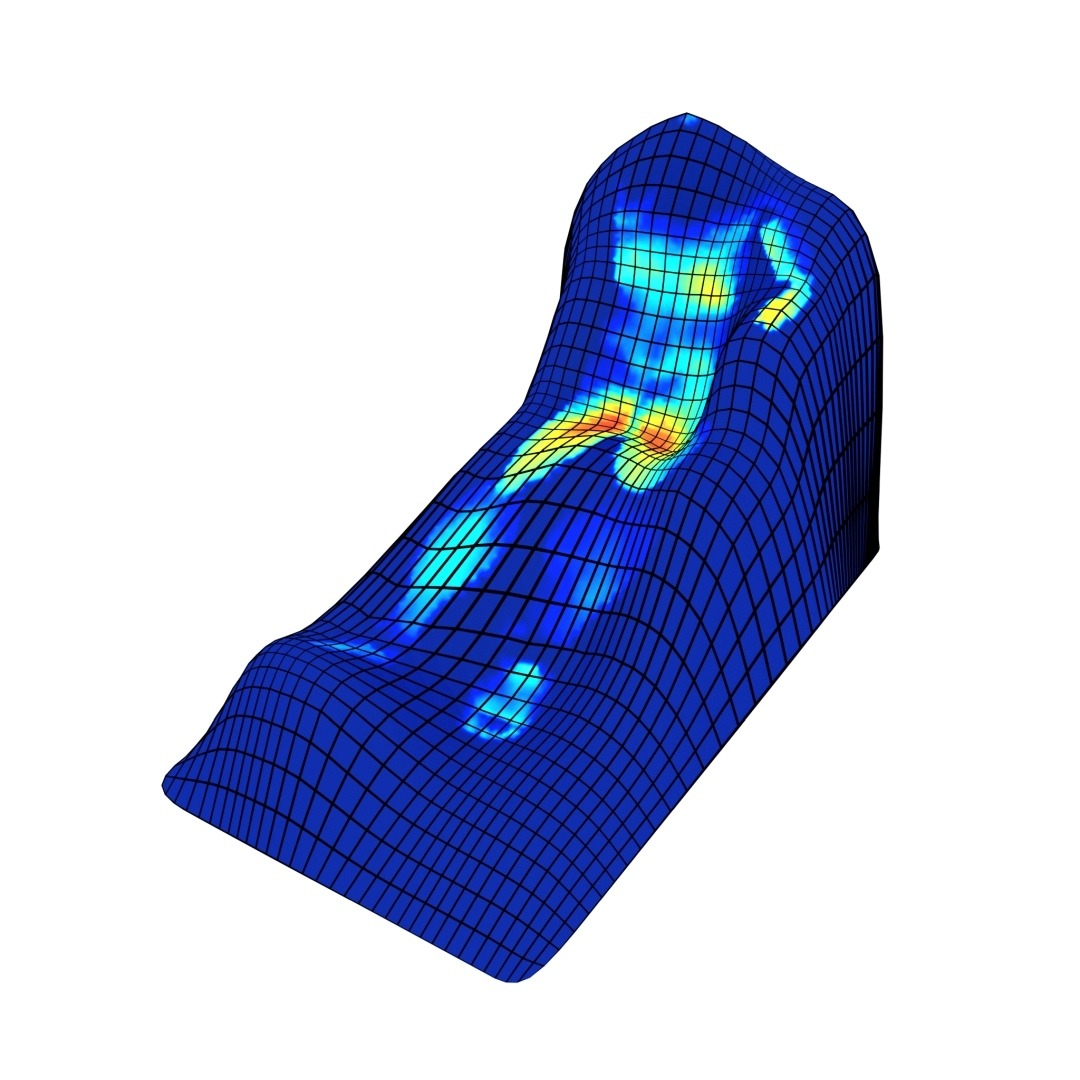}
		\captionsetup{justification=centering}
		\caption[caption]{Contact pressure mapped onto surfaces created by applying the fitting algorithm of Leimer \etal~\cite{leimer2018sit} to the control meshes generated by our method. Please note that both the average and maximum pressure values are significantly improved.}
	\end{subfigure}	
	\caption{Mapping of the contact pressure of Poses 1-6 onto their corresponding generated surfaces. Each column corresponds to one pose. Red color indicates high pressure values.}
	\label{fig:results_weighted2}
\end{figure*}

The poses are also selected from the same pose data set used in their work, which was recorded by having a design student wearing a motion capturing suit find poses that were considered comfortable. Figure \ref{fig:results_weighted2} shows the control meshes created with our method on the left, with the fitting algorithm applied to it in the center, and finally the fitting algorithm applied to a flat patch on the right.

One advantage of our method is that we can infer from the construction of our control mesh which body parts should be supported or not. For example, if the creation of an armrest is impossible, it is also unlikely that we can properly support the arm by applying the fitting algorithm. We therefore do not try to fit the surface to the arm of the person. On the other hand, the method of Leimer \etal~will always attempt to do so, unless additional user input specifically designates some body segments to not be supported. This often leads to very thin regions or even self-intersections of the surface. But for our comparison, we choose to support the same body parts in both methods and also use the same algorithm parameters.

As can be seen in Figure \ref{fig:results_weighted2}, using a control mesh that already serves as a suitable support for the given pose improves the fit to the body when using the fitting algorithm, especially in areas of the back and arms. The reason for this is that the fitting algorithm uses a closest-point search as the basis for the assignment between surface and body. Therefore, a flat patch will have less available area for regions on the body that are further away or perpendicular to the ground plane like the back. This results in greater distortions of the surface and a worse fit. Our surface control mesh generation alleviates this problem by ensuring that each region on the body has a larger area of the surface in close proximity to enable a better fit.

To quantify the advantages of our method, we apply our pressure computation method on the poses shown in Figure \ref{fig:results_pressure} and \ref{fig:teaser}, using only the subset of body vertices that lie within a certain distance to the corresponding generated surface. The results can be seen in Table \ref{tab:eval_m} and Figure \ref{fig:results_weighted2}. All of our generated surfaces result in lower values for the average distance from the body vertices to the closest surface vertex, maximum joint moments, average pressure and maximum pressure. The average joint moments are also lower in all but $2$ examples, which are the result of our algorithm optimizing for both moments and pressure.

Although not all possible poses are supported by our surface generation algorithm (see Section \ref{sec:limitations}), we do support a wide variety of sitting poses, including special cases like crossed legs and forward-leaning poses that do not require a backrest (left side of Figure \ref{fig:results_special}). Also, while not treated in a special way, we can also support poses leaning back while standing (right side of Figure \ref{fig:results_special}).

The simple structure of our furniture model also makes it suitable to use as an initial design candidate that can be edited in conventional geometry modeling applications. Among other additional results, Figure \ref{fig:collage} shows an example of a 3-person bench that is created by manually stitching together 3 control meshes generated by our algorithm.

The total computation time generally ranges from $20$ to $30$ seconds, with control mesh generation taking between $7$ and $12$ seconds, mesh optimization taking $12$ to $14$ seconds, and application of the fitting algorithm taking $1$ to $5$ seconds.

\begin{figure}[t!]
	\centering
	\iflowres
	\includegraphics[width=.22\textwidth,trim=0cm 0cm 0cm 0cm,clip]{./renders/replacements/lores_f22_simple}
	\includegraphics[width=.22\textwidth]{./renders/replacements/lores_f35_simple}
	\else
	\includegraphics[width=.22\textwidth,trim=0cm 0cm 0cm 0cm,clip]{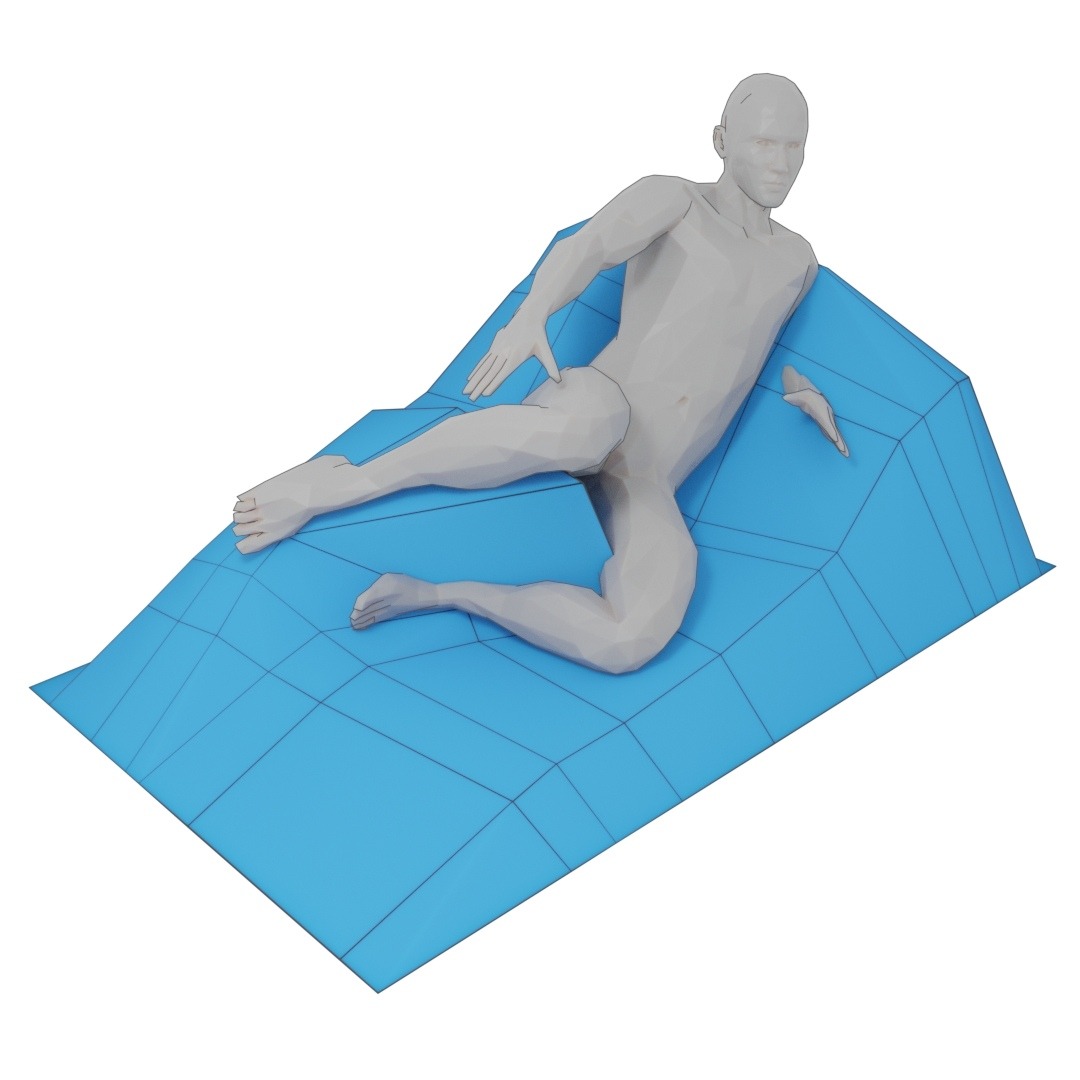}
	\includegraphics[width=.22\textwidth]{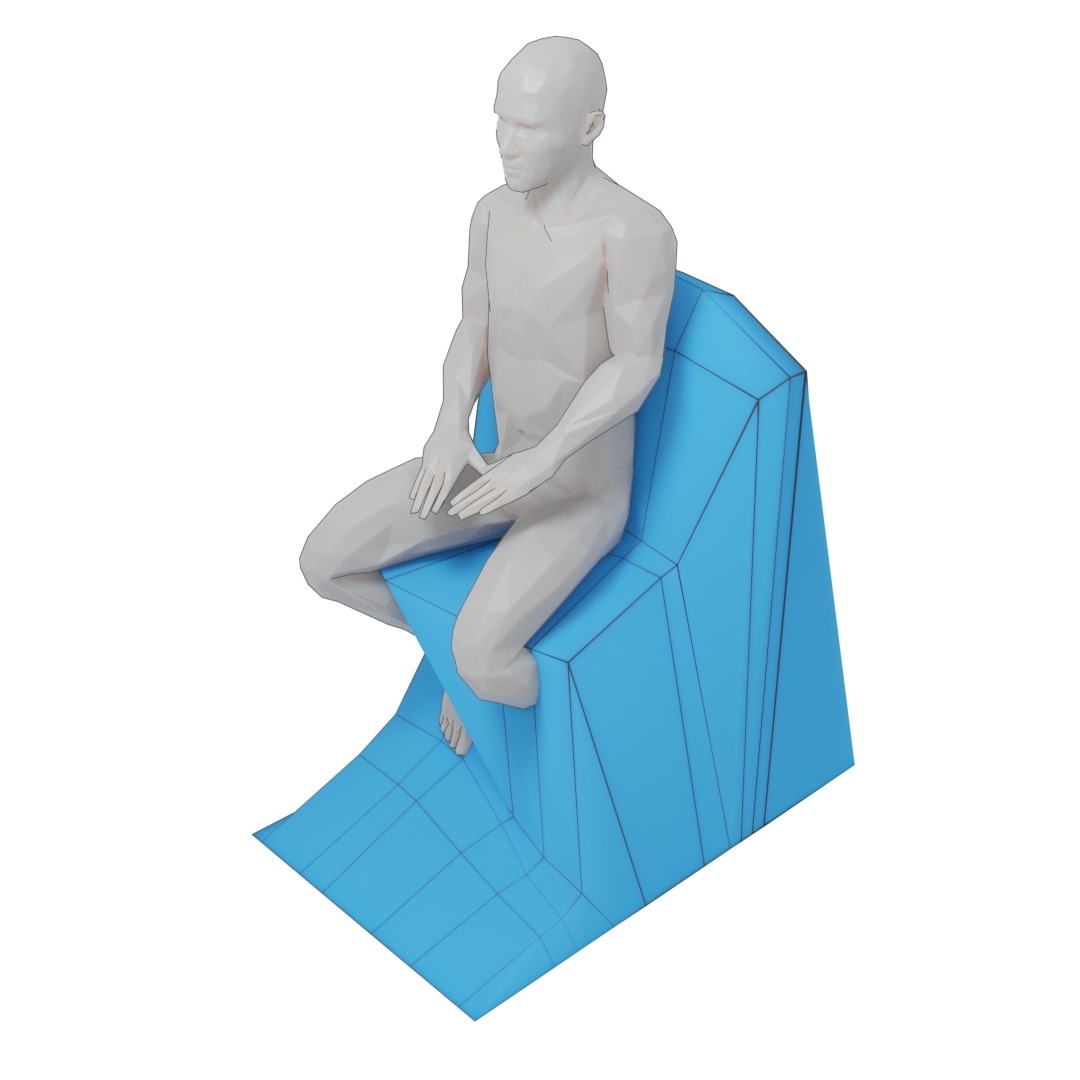}
	\fi
	\setlength{\belowcaptionskip}{-4pt}
	\caption{Our method fails to generate a valid surface from unsupported poses like lying on the side or placing the feet underneath the body.}
	\label{fig:results_unsupported}
\end{figure}

\section{Limitations and Future Work}\label{sec:limitations}
\begin{table*}
	\centering
%	\scriptsize 
	\begin{tabular}{|l|c|c|c|c|c|}
		\hline
		pose & avg. dist. & avg. moment & max. moment & avg. press. & max. press. \\
		\hline
		%F3 - S\&R & 7.10 cm & 3.41 Nm & 10.17 Nm & 479.85 Pa & 4252.50 Pa  \\
		%F3 - ours & 4.27 cm & 3.67 Nm & 9.99 Nm & 450.24 Pa & 3421.38 Pa  \\
		%\rule{0pt}{2.5ex}
		P1 - S\&R & 6.42 cm & 4.02 Nm & 11.57 Nm & 449.81 N/m$^2$ & 3995.31 N/m$^2$  \\
		P1 - ours & 4.00 cm & 4.23 Nm & 11.47 Nm & 433.85 N/m$^2$ & 3598.34 N/m$^2$  \\
		\rule{0pt}{2.5ex}P2 - S\&R & 7.92 cm & 4.08 Nm & 10.79 Nm & 454.52 N/m$^2$ & 4356.73 N/m$^2$  \\
		P2 - ours & 4.09 cm & 4.08 Nm & 10.63 Nm & 438.51 N/m$^2$ & 3889.42 N/m$^2$  \\
		\rule{0pt}{2.5ex}P3 - S\&R & 9.96 cm & 4.62 Nm & 12.05 Nm & 478.29 N/m$^2$ & 4186.42 N/m$^2$  \\
		P3 - ours & 6.08 cm & 3.06 Nm & 9.17 Nm & 458.53 N/m$^2$ & 3390.74 N/m$^2$  \\
		\rule{0pt}{2.5ex}P4 - S\&R & 7.36 cm & 3.87 Nm & 10.84 Nm & 454.52 N/m$^2$ & 4208.46 N/m$^2$  \\
		P4 - ours & 4.15 cm & 3.84 Nm & 10.74 Nm & 441.37 N/m$^2$ & 3953.73 N/m$^2$  \\
		\rule{0pt}{2.5ex}P5 - S\&R & 9.19 cm & 4.44 Nm & 11.74 Nm & 478.88 N/m$^2$ & 4275.08 N/m$^2$  \\
		P5 - ours & 5.15 cm & 2.73 Nm & 8.76 Nm & 451.05 N/m$^2$ & 3739.16 N/m$^2$  \\
		\rule{0pt}{2.5ex}P6 - S\&R & 7.58 cm & 4.23 Nm & 12.88 Nm & 455.13 N/m$^2$ & 3833.27 N/m$^2$  \\
		P6 - ours & 4.10 cm & 4.27 Nm & 12.72 Nm & 432.71 N/m$^2$ & 3507.86 N/m$^2$  \\
		\rule{0pt}{2.5ex}P7 - S\&R & 6.60 cm & 2.22 Nm & 5.66 Nm & 453.61 N/m$^2$ & 3351.65 N/m$^2$  \\
		P7 - ours & 3.17 cm & 2.07 Nm & 4.89 Nm & 422.26 N/m$^2$ & 2654.30 N/m$^2$  \\
		\hline
	\end{tabular}
	\setlength{\belowcaptionskip}{-6pt}
	\caption{Quantitative comparison between the results using the method of Leimer \etal~(S\&R) \cite{leimer2018sit} and our method for the poses shown in Figure \ref{fig:results_weighted2} (P1-P6) and the pose shown in Figure \ref{fig:teaser} (P7). We measure the average distance from the body vertices to the closest surface vertex, average and maximum joints moments, as well as average and maximum contact pressure.}
	\label{tab:eval_m}
\end{table*}

While the developed framework fulfills our goals to a satisfying degree, we acknowledge a number of limitations and weaknesses. %The goal for this framework is to create suitable seating surfaces for a variety of sitting poses.
While the algorithm covers various difficult special cases, there is a number of common sitting poses that are currently not supported.
Poses like sitting in a sideways orientation or having the feet tucked underneath the body (see Figure \ref{fig:results_unsupported}), would require changing the fundamental structure of the surface template to avoid intersections of the surface with the body or with itself and are thus not supported by our algorithm. Improvements to the special case detection and processing steps in the framework could increase the overall robustness of the algorithm and expand the potential input set of poses.

\begin{figure*}[!t]
	\begin{subfigure}[t]{0.96\textwidth}
		\centering
		\iflowres
		\includegraphics[width=0.24\textwidth]{./renders/replacements/lores_f1_collage}
		\includegraphics[width=0.24\textwidth]{./renders/replacements/lores_f12_collage}
		\includegraphics[width=0.24\textwidth]{./renders/replacements/lores_f24_collage}
		\includegraphics[width=0.24\textwidth]{./renders/replacements/lores_f30_collage}
		\else
		\includegraphics[width=0.24\textwidth]{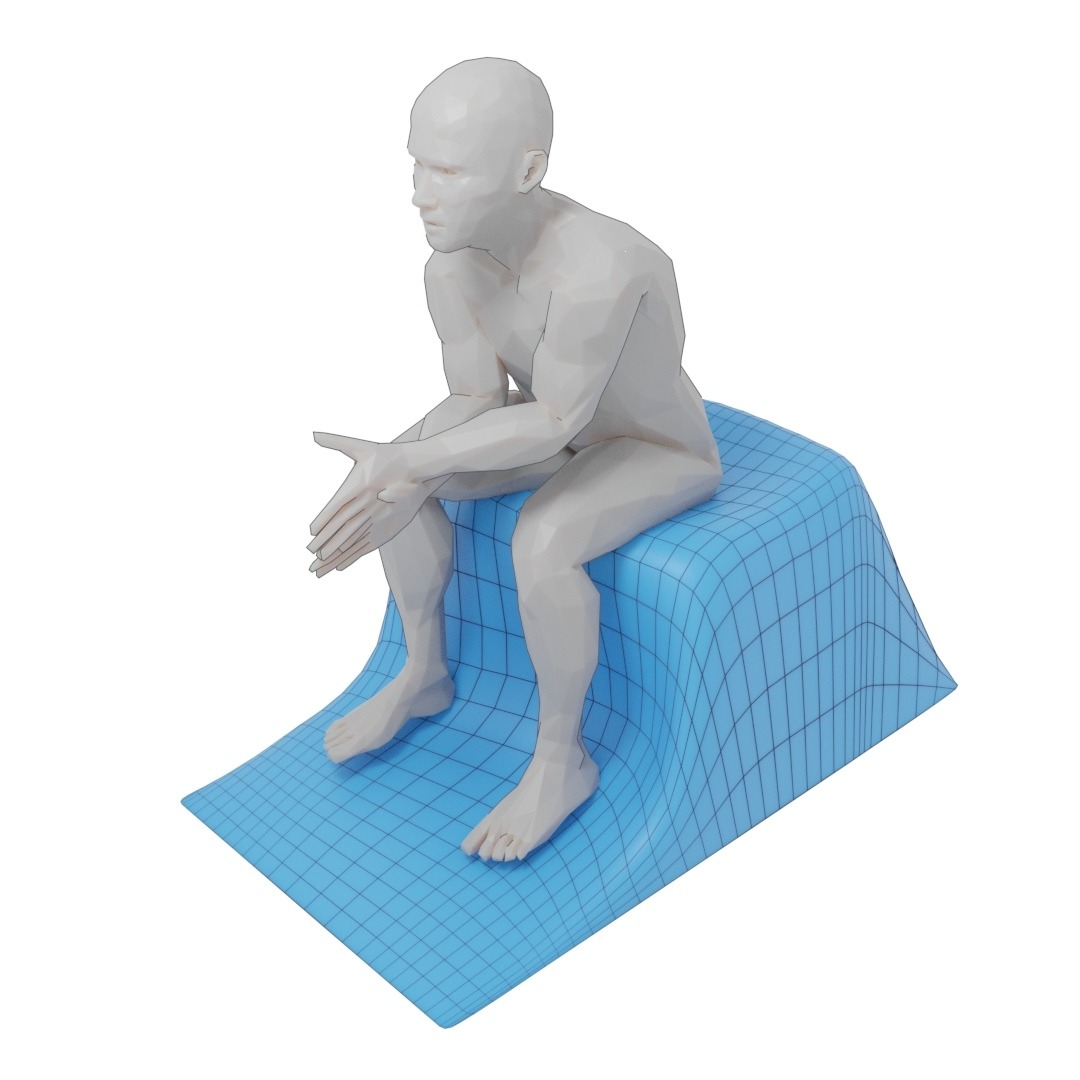}
		\includegraphics[width=0.24\textwidth]{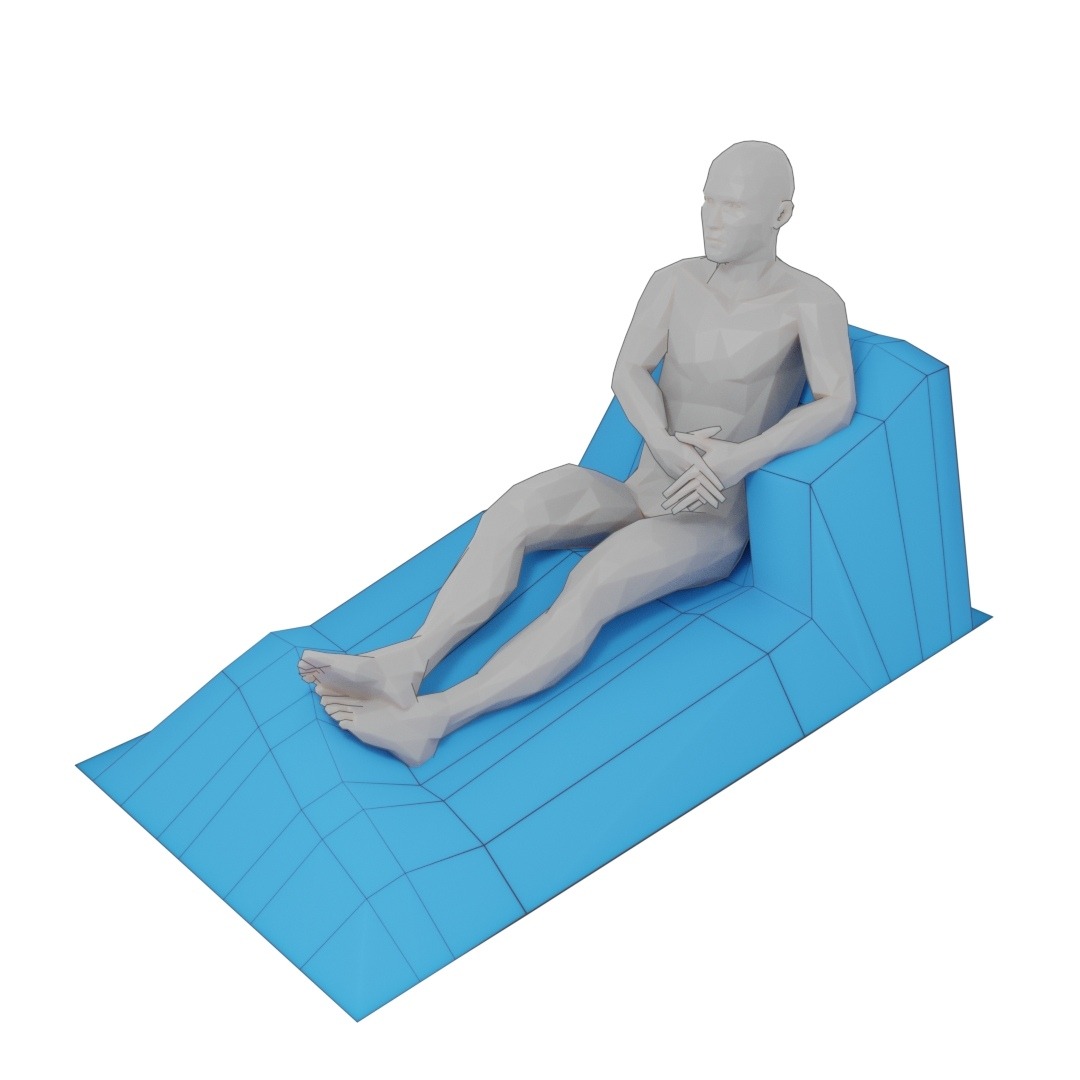}
		\includegraphics[width=0.24\textwidth]{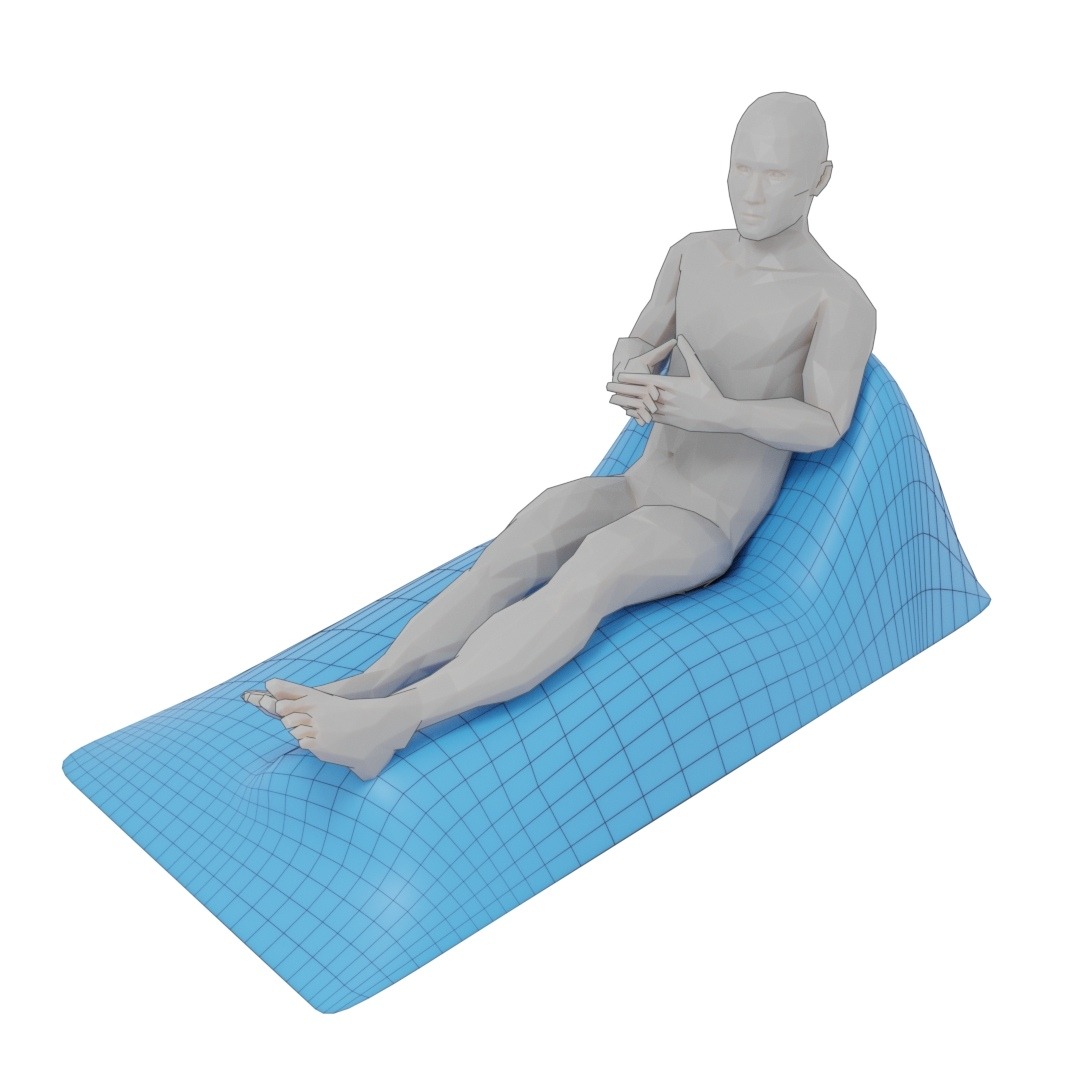}
		\includegraphics[width=0.24\textwidth]{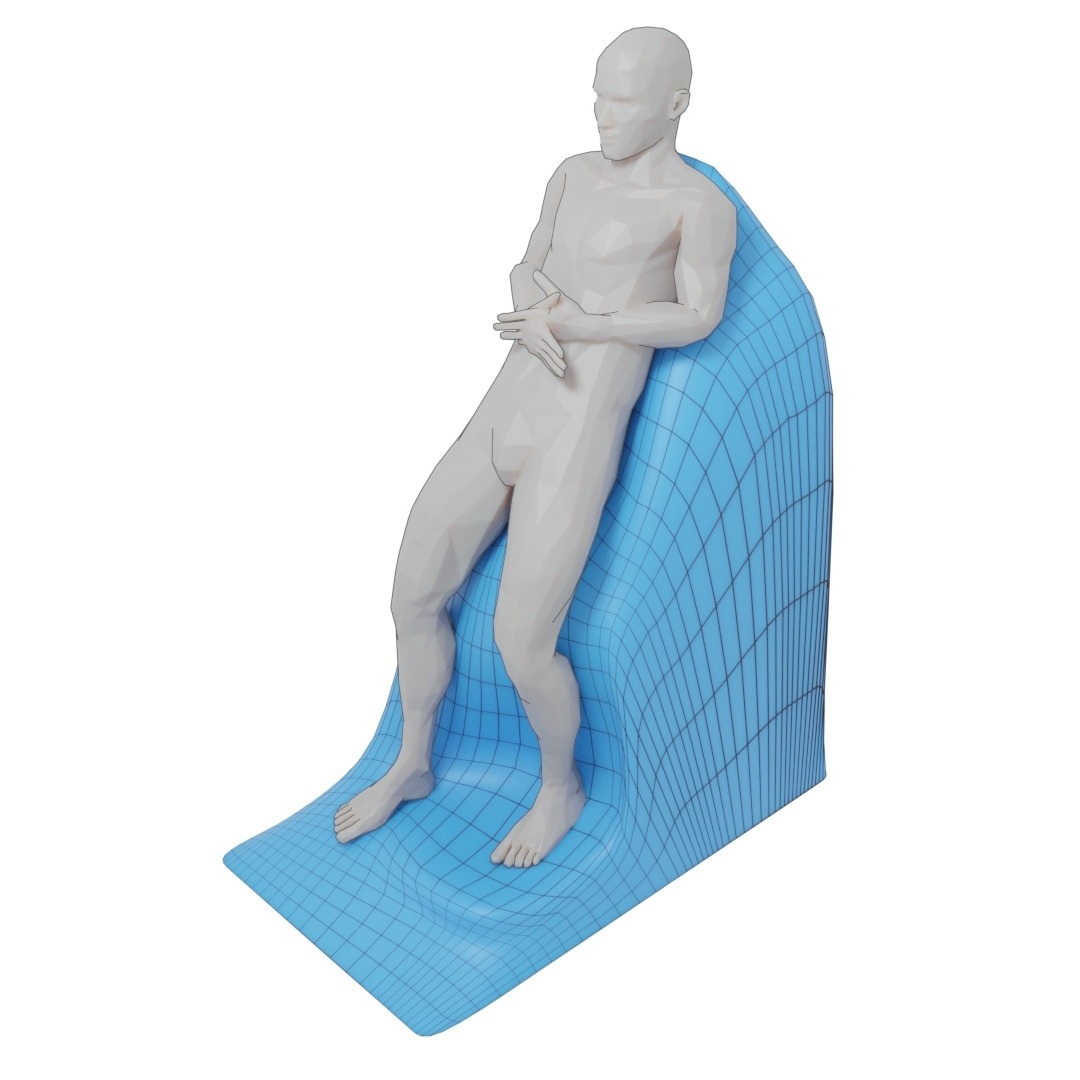}
		\fi
	\end{subfigure}
	\begin{subfigure}[t]{0.96\textwidth}
		\centering
		\iflowres
		\includegraphics[width=0.24\textwidth]{./renders/replacements/lores_f34_collage}
		\includegraphics[width=0.24\textwidth]{./renders/replacements/lores_f46_collage}
		\includegraphics[width=0.24\textwidth]{./renders/replacements/lores_f37_collage}
		\includegraphics[width=0.24\textwidth]{./renders/replacements/lores_f52_collage}
		\else
		\includegraphics[width=0.24\textwidth]{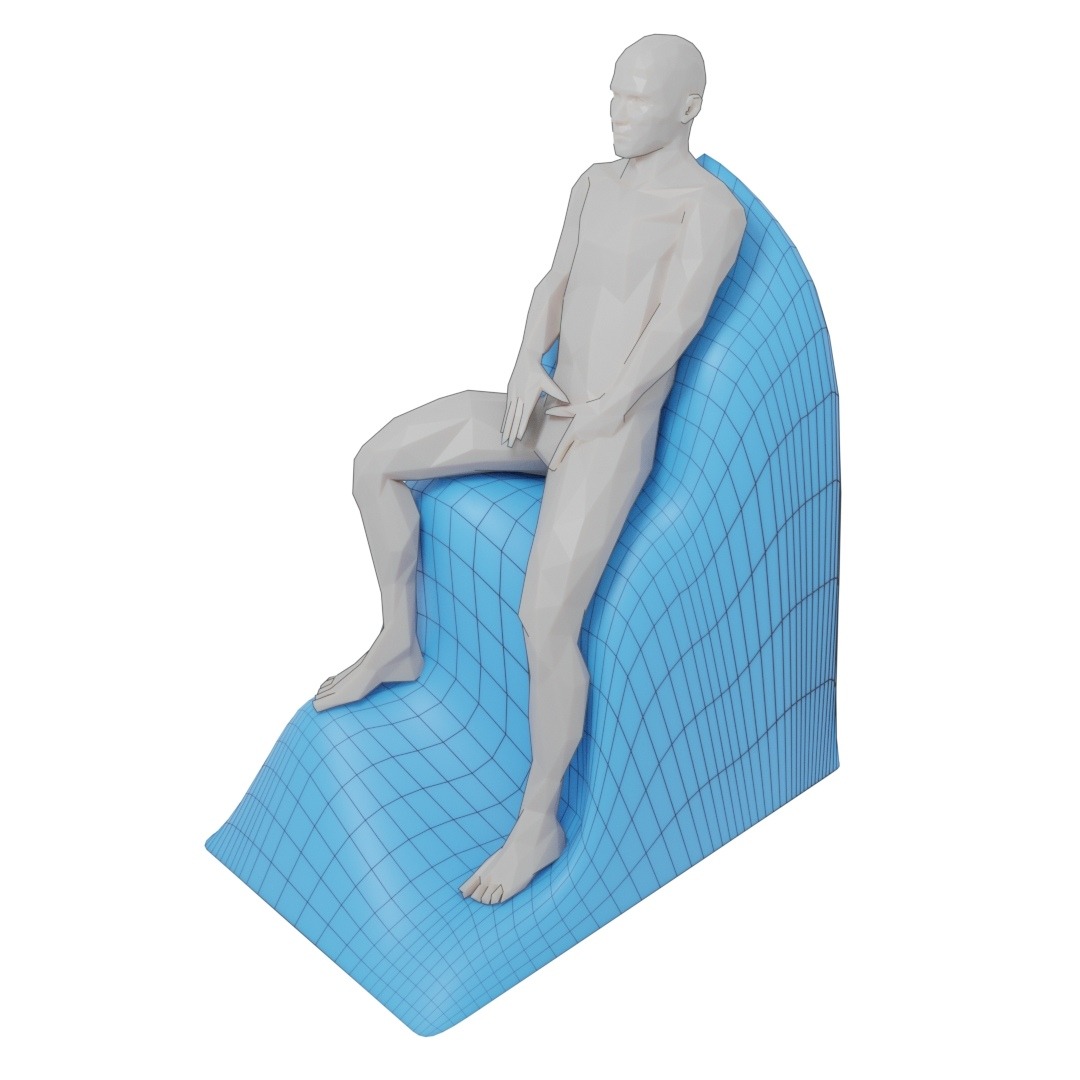}
		\includegraphics[width=0.24\textwidth]{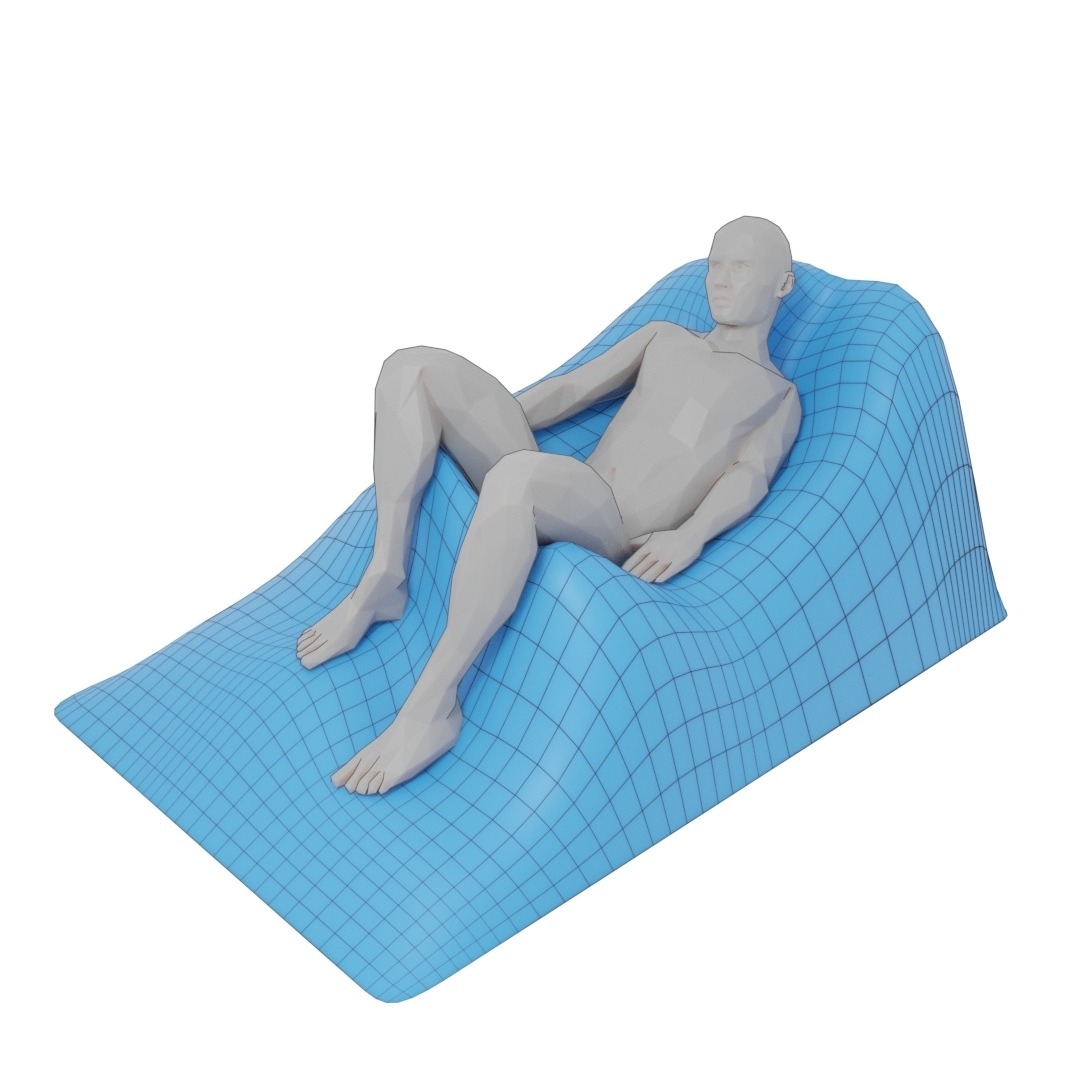}
		\includegraphics[width=0.24\textwidth]{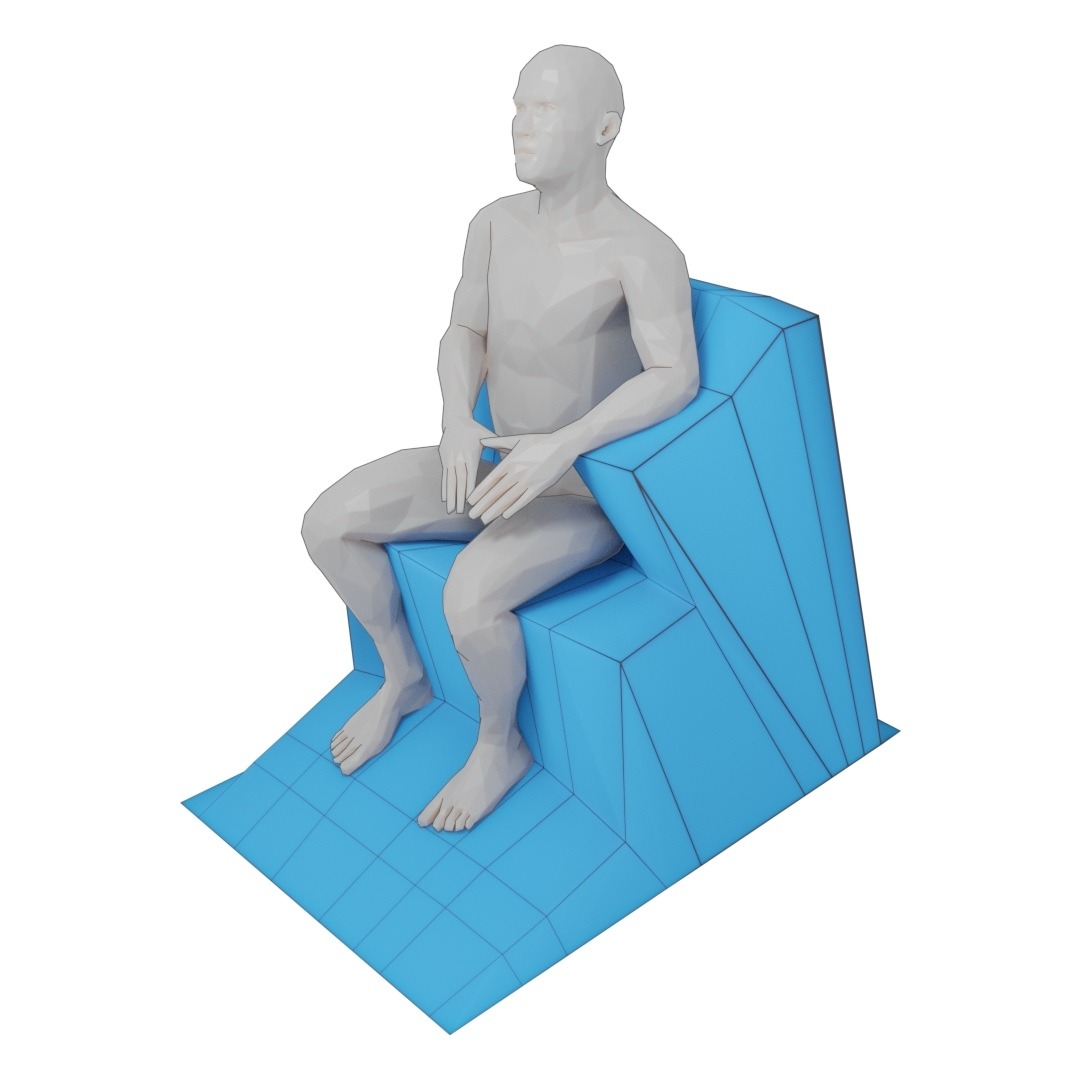}
		\includegraphics[width=0.24\textwidth]{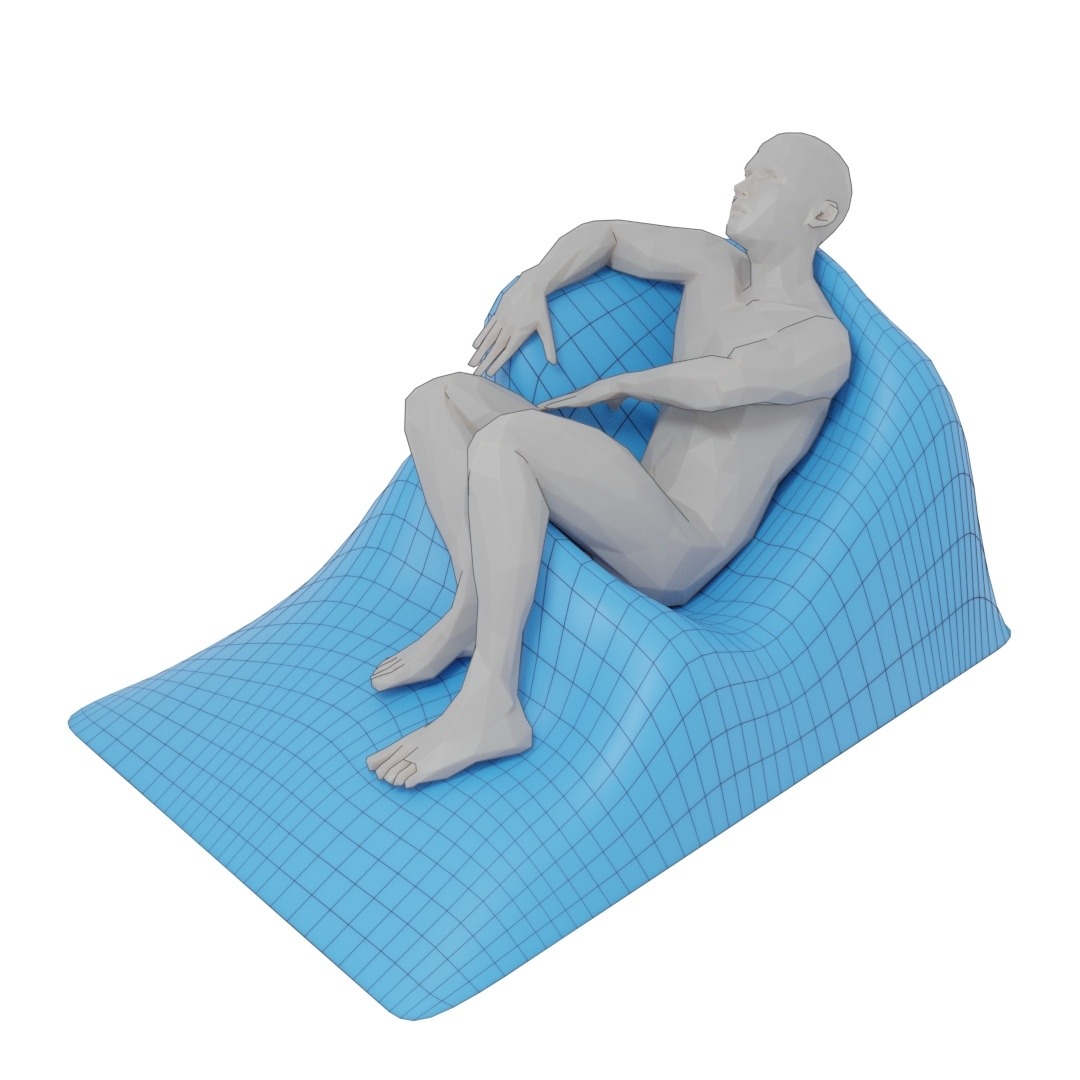}
		\fi
	\end{subfigure}
	\begin{subfigure}[t]{0.96\textwidth}
		\centering
%		\iflowres
%		\includegraphics[width=0.42\textwidth]{./renders/replacements/lores_multi_simple}
%		\includegraphics[width=0.42\textwidth]{./renders/replacements/lores_multi_fitted}
%		\else
%		\includegraphics[width=0.42\textwidth]{./renders/replacements/multi_simple}
%		\includegraphics[width=0.42\textwidth]{./renders/replacements/multi_fitted}
%		\fi
		\iflowres
		\includegraphics[width=0.49\textwidth]{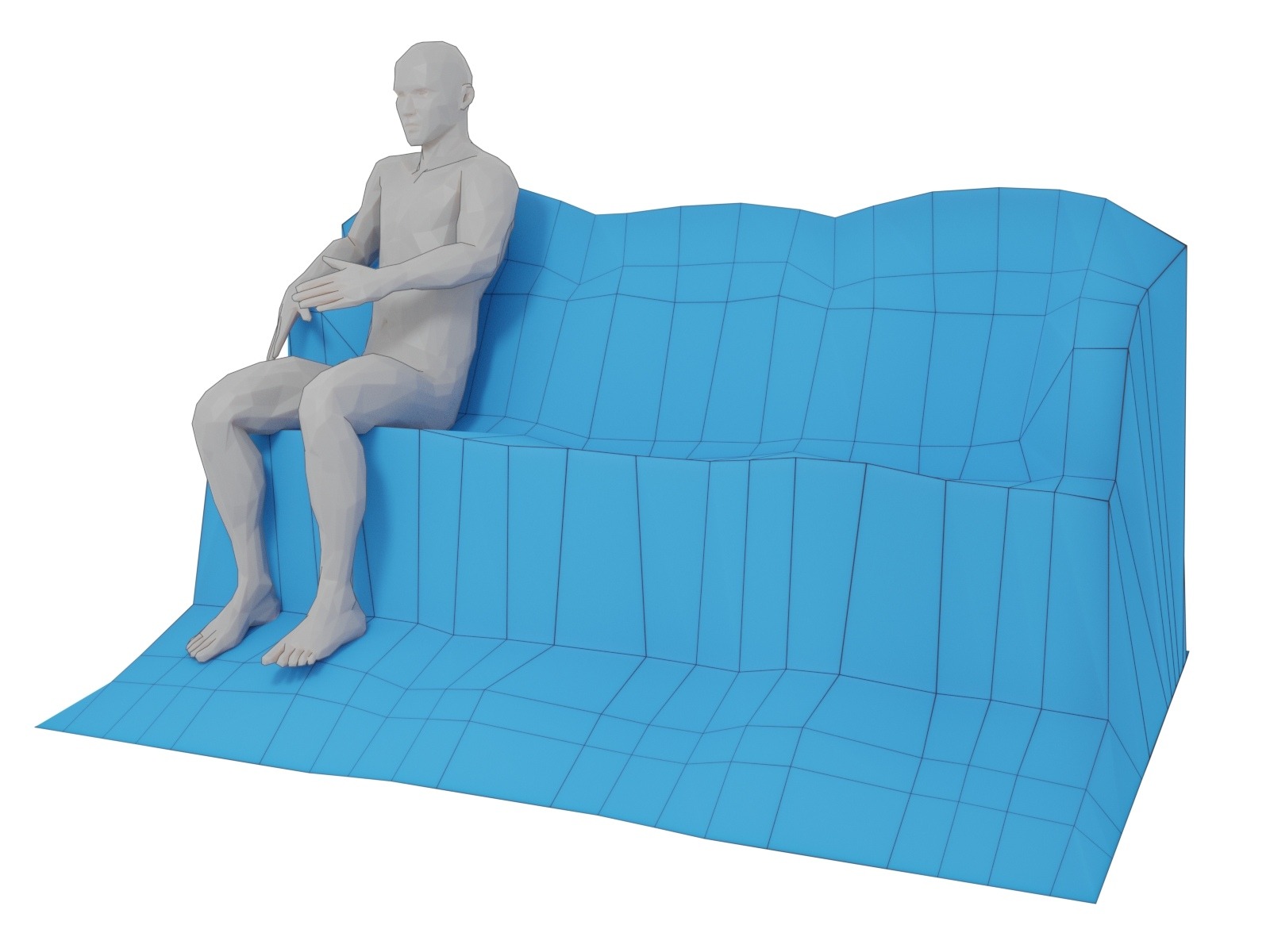}
		\includegraphics[width=0.49\textwidth]{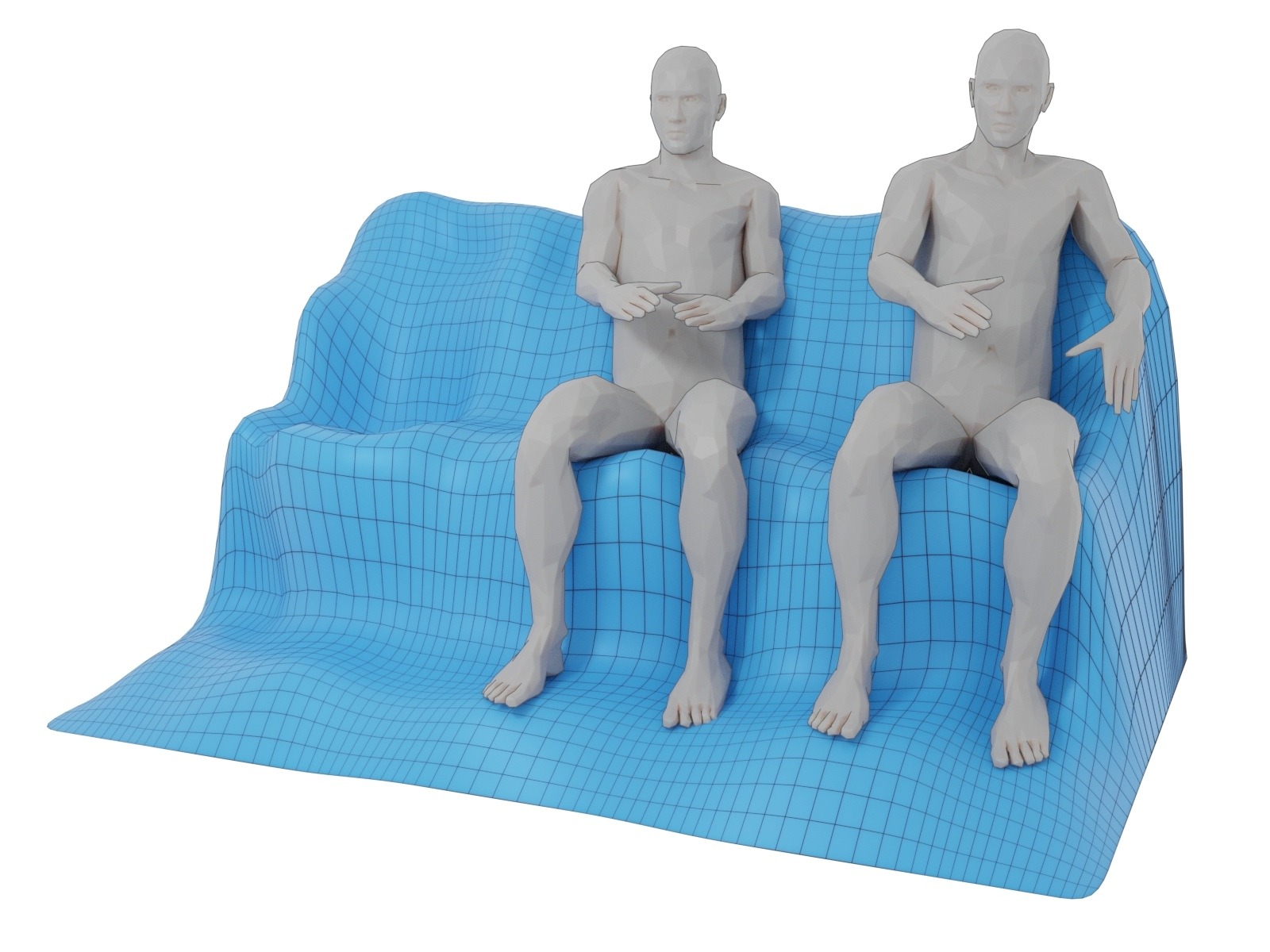}
		\else
		\includegraphics[width=0.49\textwidth]{./renders/replacements/multi_simple_B}
		\includegraphics[width=0.49\textwidth]{./renders/replacements/multi_WF_B}
		\fi
	\end{subfigure}	
	\caption{Additional results generated using our method. Some examples show the control mesh created by our algorithm, while others additionally have the fitting algorithm of Leimer \etal~\cite{leimer2018sit} applied. The surface in the bottom row is created by manually editing and combining multiple control meshes.}
	\label{fig:collage}
\end{figure*}

Our algorithm can only be used to create a seating surface for a single person. If a surface that allows seating for multiple persons is desired, the surfaces generated by our algorithm have to be manually edited. An example of such a manually modified surface can be seen in the bottom row of Figure \ref{fig:collage}. In the future, it would be interesting to extend our approach to support the generation of surfaces for multiple input poses simultaneously.

Finally, the generic template only accounts for geometric consistency and functional quality. However, since our computational model of sitting delivers physical quantities which are close to reality, a fabrication-aware structural optimization of the furniture could be considered. We leave this extension for future work. 

\section{Conclusions}\label{sec:conclusions}

We presented an automated computational design framework for the generation of functional body supporting furniture that optimizes for comfortable body-support in a given pose. The generated results have been shown to be plausible and can be used as is for the creation of smooth body-supporting subdivision surfaces, or can be used as an initial control meshes for further interactive design by the user. 

As a measure of comfort we combine two categories considered as objective in the ergonomics literature: pressure distribution and moments acting on the body. Additionally, we incorporated a friction component and proposed a computational method that handles it in interactive time in adequate accuracy as compared to sophisticated FEM methods. 

%Additionally, we proposed an algorithm for an automatic generation of generic template meshes that utilizes the given comfort measure to optimally support the body. 
Our method is meant for computed aided design of personalized furniture, where it can be used by professionals in order to create an initial design as well as by inexperienced users. Such designs can be than used for fabrication with modern digital manufacturing methods. 

\section*{Acknowledgments}
This research was funded by the  Austrian  Science  Fund  (FWF P27972-N31 and FWF P32418-N31) and  the  Vienna  Science and Technology Fund (WWTF ICT15-082).

%%Vancouver style references.
%\bibliographystyle{cag-num-names}
\bibliographystyle{cag-num-names-nourldoi}
\bibliography{poses-paper,diplomarbeit}

\clearpage

\appendix

\section{Reaction Force Computation}

The equilibrium constraints of our reaction force computation model can be formulated as the matrix 
\begin{equation*}
\mathbf{C} = \begin{bmatrix}
\mathbf{I}^{F}_{b}   & \textbf{0} 		   & {\upsilon}^R\,{\alpha}_b^R\,\mathbf{T}^{R} \\[5pt]
\left[ \mathbf{a}\right]^{F}_{b} & \mathbf{I}^{M}_{b}  & {\upsilon}^R\,{\alpha}_b^R\,\left[ \mathbf{a}\right]^{R}_{b}\,\mathbf{T}^{R} \\[5pt]
\mathbf{I}^{F}_{j}   			 & \mathbf{0}          & \mathbf{0}   \\[5pt]
\mathbf{0}   					 & \mathbf{I}^{M}_{j}  & \mathbf{0}   \\
\end{bmatrix},
\label{eq:constrmatrix1}
\end{equation*}
with $\mathbf{I}^{F}_{b}$ (respective $\mathbf{I}^{M}_{b}$) being the $3 \times 3$ identity matrix if the force $\mathbf{f}$ (respective moment $\mathbf{m}$) corresponding to the column is active inside the body segment $b$ corresponding to the row, and $\mathbf{0}$ otherwise. Similarly, $\mathbf{I}^{F}_{j}$ (respective $\mathbf{I}^{M}_{j}$) is the $3 \times 3$ identity matrix if the force $\mathbf{f}$ (respective moment $\mathbf{m}$) corresponding to the column is active on the joint $j$ corresponding to the row, and $\mathbf{0}$ otherwise. $\mathbf{T}^{R}$ denotes the transformation matrix from tangent space to world space. Finally, $\left[ \mathbf{a}\right]^{F}_{b}$ (respective $\left[ \mathbf{a}\right]^{R}_{b}$) denotes the skew-symmetric matrix
\begin{equation*}
\left[ \mathbf{a}\right]_{b} = \begin{bmatrix}
0 & a_3 & -a_2 \\
-a_3 & 0 & a_1 \\
a_2 & -a_1 & 0
\end{bmatrix},
\end{equation*}
if the force $\mathbf{f}$ (respective reaction force $\mathbf{r}$) corresponding to the column is active inside the body segment $b$, or $\mathbf{0}$ otherwise. 
The scalars ${\alpha}^{R}_{b}$ are the actual linear blend skinning weights of each vertex-body segment connection, and the scalar values  ${\upsilon}^{R}$ are additional user provided weights, which allow to further control the importance or unimportance of particular surface regions. In particular, we use them to exclude body parts like face, chin, or parts of the abdomen.   

The vector $\mathbf{a}$ denotes the moment arm vector---the vector pointing from the COM of $b$ to the point affected by the force $\mathbf{f}$ (or reaction force $\mathbf{r}$) corresponding to the column. %Similarly, the matrix $\left[ \mathbf{v}\right]^R_B$ denotes the skew symmetric matrix of the vector pointing from the COM of the whole body to the reaction force $R$.
Finally, the right side of the system, $\textbf{z}$, is a $(6 n_B + 6 n_J)$ column vector:
\begin{equation*}
\mathbf{z} = \begin{bmatrix}
-\mathbf{g}_b \\%[-2mm]
%\vdots \\[2mm]
\mathbf{0} \\%[-2mm]
%\vdots \\[2mm]
\mathbf{0} \\%[-2mm]
%\vdots \\[2mm]
\mathbf{0} \\%[-2mm]
%\vdots
\end{bmatrix} \,.
\end{equation*}

The terms of the energy function can be formulated as matrix
\begin{equation*}
\mathbf{A} = \begin{bmatrix}
\mathbf{0} & \mathbf{I}^{M} & \mathbf{0} \\
\mathbf{0} & \mathbf{0} & \lambda \, {\upsilon}^R \, \mathbf{W}^R \, \mathbf{T}^{R} \\
\end{bmatrix},
\label{eq:objmatrix1}
\end{equation*}
where $\mathbf{I}^{M}$ is the identity matrix with $3 n_M$ rows and $\lambda \geq 0$ is a weight that assigns more or less importance to the minimization of the moments, which can be interpreted as the stiffness of the joints. $\mathbf{W}^R$ contains the weights that determine how the reaction forces are distributed on the surface and also acts as a regularizer, as without it the best solution is likely to have a very small number of extreme reaction forces.

\section{Surface Optimization Gradient}\label{sec:appendix:fitting}

In order to improve the efficiency of the solving the surface optimization problem, we evaluate the gradient of the corresponding energy function.

\textit{Laplacian smoothing distance}: To compute the gradient value for the Laplacian smoothing distance metric, we first rewrite the corresponding term as:
\begin{equation*}
S_L = \sum_i^{n^v} \sum_k^{x,y,z} \left(\mathbf{v}_{ik} -  \frac{\sum\limits_{j \in N_1(i)} \mathbf{v}_{jk} w_j}{\sum\limits_{j \in N_1(i)} w_j} \right)^2  \lambda_S^l 
\end{equation*}
For each vertex, each dimension can be computed separately as the squared difference from the average position of its neighbors.
To evaluate the gradient, we need to compute the partial derivatives for each variable of the objective function (i.e. the x, y and z coordinates of each vertex position).
For the x coordinate of an arbitrary vertex (with its 1-ring neighborhood $N_1(i)$), its respective part of the gradient value is computed as.
\begin{equation*}
\frac{\partial S_L}{\partial x} = 2 \lambda_S^l  \left(x - \frac{\sum\limits_{j \in N_1(i)} x_j w_j}{\sum\limits_{j \in N_1(i)} w_j} \right)
\end{equation*}
For each variable, we also have to consider its occurrence as neighbor of another variable. Therefore, for each variable's gradient value, we also accumulate the following term:
\begin{equation*}
\frac{\partial S_L}{\partial x_{n1} } = \frac{- 2\lambda_S^l w_{n1}  \left(x - \frac{\sum\limits_{j \in N_1(i)} x_j w_j}{\sum\limits_{j \in N_1(i)} w_j} \right)}{\sum\limits_{j \in N_1(i)} w_j}
\end{equation*}

\textit{Angle based differences:} To compute the gradient value of a variable corresponding to the angle based differences, we consider its occurrences in the respective computations.
For each face, its four interior angles are considered. To compute an angle, the corresponding vertex and the edges to its two adjacent vertices are required. Each vertex is adjacent to multiple faces in the surface mesh.
This means, to compute the (angle based) gradient value for a variable, we need to consider the angles of all adjacent faces of the vertex. For each face, a vertex position is relevant for three interior angles. The respective parts of the gradient values are computed as follows (for a single variable):
\begin{equation*}
\frac{\partial \alpha_{abc}}{\partial a_x}
= -\frac{\frac{\mathbf{e}_{bc,x}}{m_{bc} m_{ab}}
	- \frac{\mathbf{e}_{ab,x} S_{abc}}
	{m_{bc} {m_{ab}}^3}}{\sqrt{1-\langle\mathbf{t}_{ab}, \mathbf{t}_{bc}\rangle}}
\end{equation*}
\begin{equation*}
\frac{\partial \alpha_{dab}}{\partial a_x}
= -\frac{\frac{\mathbf{e}_{da}+\mathbf{e}_{ab}}{m_{ab} m_{bc}}
	- \frac{\mathbf{e}_{da,x} S_{dab}}
	{{m_{ab}}^3 m_{da}}
	+ \frac{\mathbf{e}_{ab,x} S_{dab}}
	{m_{ab} {m_{da}}^3}
}{\sqrt{1-\langle\mathbf{t}_{da}, \mathbf{t}_{ab}\rangle}}
\end{equation*}
\begin{equation*}
\frac{\partial \alpha_{cda}}{\partial a_x}
=-\frac{\frac{-\mathbf{e}_{cd,x}}{m_{cd} m_{da}}
	- \frac{\mathbf{e}_{da,x}  S_{cda}}{m_{cd} {m_{da}}^3}}{\sqrt{1-\langle \mathbf{t}_{cd}, \mathbf{t}_{da}\rangle}}
\end{equation*}
\begin{center}
	\begin{math}
	S_{abc} = \sum_{i}^{x,y,z} (b_i-a_i) (b_i-c_i)
	\end{math}
\end{center}
\begin{itemize}
	\item $a, b, c$ and $d$ are the corner vertices of the face.
	\item $\mathbf{e}_{ab} = (\mathbf{b}-\mathbf{a})$ refers to the edge between vertices $a$ and $b$.
	\item $m_{ab} = \left \| e_{ab} \right \|_2$ is the magnitude, i.e. the euclidean norm of the edge vector.
	\item $\mathbf{t}_{ab} = \frac{\mathbf{e}_{ab}}{m_{ab}}$ is the normalized edge direction vector.	
\end{itemize}
The gradient values for the face interior angles are accumulated for each variable and utilized separately for the \textit{regular faces} and \textit{maximum angles} error metrics.

%TODO: Data term gradients
\textit{Vertex distance:} The vertex distance error metric from the objective function's data term can be written as
\begin{equation*}
D_V =  \lambda_D^V \sum_{i=1}^{n^V}  \left( (\mathbf{v}_{ix}-\tilde{\mathbf{v}}_{ix})^2 + (\mathbf{v}_{iy}-\tilde{\mathbf{v}}_{iy})^2 + (\mathbf{v}_{iz}-\tilde{\mathbf{v}}_{iz})^2 \right) w_i
\end{equation*}
The corresponding part of the gradient value for each variable is simply computed from its partial derivative:
\begin{equation*}
\frac{\partial D_V}{\partial \mathbf{v}_{ix}} = 2  \lambda_D^V w_i (\mathbf{v}_{ix}-\tilde{\mathbf{v}}_{ix})
\end{equation*}

\textit{Plane distance}: The data term's plane distance metric is computed for the four corner vertices for each face that corresponds to a supporting plane in the original model.
Vertices that are adjacent to more than one of these planes, have multiple plane distance values.
Therefore, the gradient values for each variable has to be accumulated for each face.
A variable's value corresponding to a single face $F_j$ (with $\mathbf{v}_i \in F_j$) is computed as:
\begin{equation*}
\frac{\partial D_P}{\partial \mathbf{v}_{ix}} = 2  \lambda_D^P w_j \mathbf{n}^{P_j}_x \left \langle \mathbf{v}_i - \mathbf{c}^{P_j}, \mathbf{n}^{P_j} \right \rangle
\end{equation*}

The gradient values for all error metrics are further scaled by the corresponding global factors for the data and smoothing terms.
The objective function's gradient evaluation results in vector of length $3 n^V $, where each value corresponds to the sum of the error metric gradient values for an individual variable.

\end{document}